\documentclass[a4paper,11pt]{article}
\pdfoutput=1 % if your are submitting a pdflatex (i.e. if you have
\usepackage{jheppub} % for details on the use of the package, please
\usepackage{orcidlink}
\usepackage{adjustbox}
\usepackage[T1]{fontenc} % if needed
\graphicspath{{pict/}}
\usepackage[running]{lineno}

\usepackage[linkcolor=blue]{hyperref}

\newcommand{\beq}{\begin{equation}}
\newcommand{\eeq}{\end{equation}}

\newcommand{\ph}{\phantom{0}}

\newcommand{\ee}{\ensuremath{e^+ e^-}}
\newcommand{\bb}{\ensuremath{b\bar{b}}}

\newcommand{\nn}{\nonumber}

\newcommand{\Ds}{\ensuremath{D_s^+}}
\newcommand{\D}{\ensuremath{D^{0}}}
\newcommand{\B}{\ensuremath{B}}

\newcommand{\Bs}{\ensuremath{B_s^0}}
\newcommand{\BsB}{\ensuremath{\bar{B}_s^0}}

\newcommand{\DsX}{\ensuremath{D_s^{\pm} \, X}}
\newcommand{\DnX}{\ensuremath{D^0\hspace{-0.2em}/\bar{D}^0 \, X}}
\newcommand{\DX}{\ensuremath{D/\bar{D} \, X}}
\newcommand{\BsBsX}{\ensuremath{B_s^0\bar{B}_s^0 \, X}}
\newcommand{\BBX}{\ensuremath{B\bar{B} \, X}}

\newcommand{\Br}{\ensuremath{\mathcal{B}}}
\newcommand{\ecm}{\ensuremath{E_{\rm {c.m.}}}}
\newcommand{\xp}{\ensuremath{x_p}}
\newcommand{\fs}{\ensuremath{f_{\rm s}}}

\newcommand{\U}{\ensuremath{\Upsilon}}
\newcommand{\Ufo}{\ensuremath{\Upsilon(4S)}}
\newcommand{\Ufi}{\ensuremath{\Upsilon(5S)}}

\newcommand{\ifb}{\ensuremath{\,\mathrm{fb}^{-1}}}
\newcommand{\ipb}{\ensuremath{\,\mathrm{pb}^{-1}}}

\newcommand{\gev}{\,{\ensuremath{\mathrm{\mbox{GeV}}}}}
\newcommand{\mevcc}{\,{\ensuremath{\mathrm{\mbox{MeV}}/c^2}}}
\newcommand{\mev}{\,{\ensuremath{\mathrm{\mbox{MeV}}}}}

\newcommand{\imax}{\ensuremath{\,{i}_{\rm max}}}
\newcommand{\fnB}{\ensuremath{f_{B\!\!\!\!/}}}
\newcommand{\fBBX}{\ensuremath{f_{B\bar{B}X}}}

\title{\boldmath Measurement of the $\ee \to \Bs\BsB X$ cross
  section in the energy range from $10.63$ to $11.02\,\gev$ using
  inclusive $\Ds$ and $\D$ production}

\preprint{\vbox{ \hbox{   }
					    	\hbox{Belle Preprint 2023-09 }
                        	\hbox{KEK Preprint 2023-11} 
                     }}

\collaboration{The Belle Collaboration}
  \author{V.~Zhukova\,\orcidlink{0000-0002-8253-641X},\hbox{$\dagger$}\note[$\dagger$]{Corresponding author.}} % 2387
  \author{R.~Mizuk\,\orcidlink{0000-0002-2209-6969},} % 2483          
  \author{I.~Adachi\,\orcidlink{0000-0003-2287-0173},} % 2590
  \author{H.~Aihara\,\orcidlink{0000-0002-1907-5964},} % 2223
  \author{S.~Al~Said\,\orcidlink{0000-0002-4895-3869},} % 6823
  \author{D.~M.~Asner\,\orcidlink{0000-0002-1586-5790},} % 4684
  \author{H.~Atmacan\,\orcidlink{0000-0003-2435-501X},} % 2538
  \author{V.~Aulchenko\,\orcidlink{0000-0002-5394-4406},} % 8183
  \author{T.~Aushev\,\orcidlink{0000-0002-6347-7055},} % 3747
  \author{R.~Ayad\,\orcidlink{0000-0003-3466-9290},} % 3766
% ...
\emailAdd{zhukova.valentina07@gmail.com}
\emailAdd{roman.miziuk@gmail.com}

\abstract{We report the first measurement of the inclusive $\ee \to
  \bb \to \DsX$ and $\ee \to \bb \to \DnX$ cross sections in the
  energy range from $10.63$ to $11.02\,\gev$. Based on these results, we
  determine $\sigma(e^+ e^- \to \BsBsX)$ and $\sigma(e^+ e^- \to \BBX)$
  in the same energy range. We measure the fraction of \Bs\ events at
  $\U(10860)$ to be $\fs=(22.0^{+2.0}_{-2.1})\%$. We determine also
  the ratio of the \Bs\ inclusive branching fractions $\mathcal{B}(\Bs
  \to \DnX)/\mathcal{B}(\Bs \to \DsX)=0.416 \pm 0.018 \pm 0.092$. The
  results are obtained using the data collected with the Belle
  detector at the KEKB asymmetric-energy \ee\ collider.}


\begin{document} 
\maketitle
\flushbottom

\section{Introduction}
\label{sec:intro}
Hadronic states in the bottomonium spectrum lying above the
open-bottom threshold demonstrate properties at odds with the standard
quark model scheme. In particular, the structures $Z(10610)$ and
$Z(10650)$, observed by Belle in 2012~\cite{Belle:2011aa}, are charged
and contain at least four quarks. The mass splittings for the
high-lying vector bottomonia do not follow the quark model
expectations either. The rates of their transitions to lower
bottomonia with the emission of light hadrons are much higher compared
to the expectations for ordinary bottomonium, in violation of the
Okubo-Zweig-Iizuka rule~\cite{Meng:2007tk,Simonov:2008ci}, and their
$\eta$ transitions are not suppressed relative to the dipion
transitions, which violates Heavy Quark Spin
Symmetry~\cite{Kaiser:2002bm,Voloshin:2012dk}. For a review, see,
e.g. Ref.~\cite{Bondar:2016hva}. Studies of various cross sections
above the open-bottom threshold can help us to understand the
properties of the resonances lying in this energy region.

The total hadronic cross section in the bottomonium energy region was
previously measured by both Belle and BaBar
collaborations~\cite{Belle:2015aea,BaBar:2008cmq}. It has a nontrivial
shape, with peaks near the $\Upsilon(4S,10860,11020)$ resonances,
valley near $\Upsilon(10753)$, and dips near the $B\bar{B}^{*}$,
$B^{*}\bar{B}^{*}$, and $B_s^{*}\bar{B}_s^{*}$ thresholds. To some
extent, the total $b\bar{b}$ cross section has already been decomposed
into exclusive cross sections up to the energy 11.02 GeV. The Belle
experiment measured the energy dependence of the cross sections
$\ee\to B\bar{B}$, $B\bar{B}^*$, $B^*\bar{B}^*$,
$B_s^{(*)}\bar{B}_s^{(*)}$, $\Upsilon(nS)\pi^+\pi^-$ $(n=1,2,3)$, and
$h_b(mP)\pi^+\pi^-$
$(m=1,2)$~\cite{Belle:2021lzm,Abdesselam:2016tbc,Belle:2011aa,Belle:2015tbu}. The
major missing contribution is the $B^{(*)}\bar{B}^{(*)}\pi$ channels;
it can be estimated using the difference between the total cross
section and the sum of the measured exclusive channels.

A combined fit of the available measurements was performed in
Ref.~\cite{Husken:2022yik} using a coupled-channel approach. For the
first time, the decay branching fractions of the $\Upsilon(10753)$,
$\Upsilon(10860)$, and $\Upsilon(11020)$ resonances were determined
rigorously. Also, pole positions (masses and widths) of the \U\ states
and energy dependence of the scattering amplitudes between all
considered channels were extracted. It was noted in
Ref.~\cite{Husken:2022yik} that the accuracy of the data needs to be
improved. This is especially true for the $B_s^{(*)}\bar{B}_s^{(*)}$
channel, in which one can not discriminate the models that predict
different behavior of the cross section near the threshold.

The previous measurement of the $B_s^{(*)}\bar{B}_s^{(*)}$ final
states was performed using full reconstruction of one $\Bs$. The
efficiency of the full reconstruction was relatively low, which
resulted in large statistical uncertainties in the results.

Here we use an inclusive approach: first, we measure
$\sigma(\ee\to\bb\to\DsX)$ and $\sigma(\ee\to\bb\to\DnX)$, then
$\sigma(\ee\to\BsBsX)$ and $\sigma(\ee\to\BBX)$ are determined based
on the above measurements. The cross sections are measured in the
energy range from $10.63$ to $11.02\,\gev$. Since isospin-violating
channels $B_s^{(*)}\bar{B}_s^{(*)}\pi^0$ are strongly suppressed, the
relation
\beq
\sigma(\ee \to \BsBsX)=\sigma(\ee \to B_s^{(*)} \bar{B}_s^{(*)})
\label{eq::relation_in_and_ex}
\eeq
is valid up to the $\Bs\BsB\pi^0\pi^0$ threshold that opens at
$11.004\,\gev$, thus, for most of the energy range studied in this
paper.

For brevity, in the following we denote $\U(10860)$ as \Ufi\ and
$\U(11020)$ as $\U(6S)$.

\section{Belle detector and data samples}
\label{det}

The analysis is based on data collected by the Belle
detector~\cite{Abashian:2000cg,Belle:2012iwr} at the KEKB asymmetric-energy
\ee\ collider~\cite{Kurokawa:2001nw,Abe:2013kxa}.

The Belle detector is a large-solid-angle magnetic spectrometer that
consists of a silicon vertex detector (SVD), a 50-layer central drift
chamber (CDC), an array of aerogel threshold Cherenkov counters (ACC),
a barrel-like arrangement of time-of-flight scintillation counters
(TOF), and an electromagnetic calorimeter (ECL) composed of CsI(Tl)
crystals located inside a superconducting solenoid coil that provides
a 1.5 T magnetic field. An iron flux-return located outside of the
coil is instrumented to detect $\,\ensuremath{K^0_L}$ mesons and to
identify muons (KLM). Two different inner detector configurations were
used. For the first sample of 156 \ifb, a 2.0 cm radius beam pipe and
a 3-layer silicon vertex detector were used; for the latter sample of
833 \ifb, a 1.5 cm radius beam pipe, and a 4-layer silicon vertex
detector (SVD2) and a small-cell inner drift chamber were used. This
analysis is based only on data collected with the SVD2
configuration. A detailed description of the detector can be found,
for example, in Ref.~\cite{Abashian:2000cg,Belle:2012iwr}.

We use energy scan data with approximately $1\,\ifb$ per point: six
points collected in 2007 and 16 points collected in 2010. We use also
the $\Ufi$ on-resonance data with a total integrated luminosity
of $121\,\ifb$ collected at five points with energies from
$10.864\,\gev$ to $10.868\,\gev$. 
The center-of-mass (c.m.) energies of these data samples are calibrated 
using the $\ee\to\mu^+\mu^-$ and 
$\ee\to\Upsilon(nS)\pi^+\pi^-$ ($n=1,2,3$) processes~\cite{Belle:2019cbt}. 
We combine the data samples
with similar energies so that finally we obtain 23 energy points. The
energies and integrated luminosities of these 23 data samples are
presented in Table~\ref{tab::all_cross_section} below.
We also use the SVD2 part of the $\Ufo$ data sample with an integrated
luminosity of $571\,\ifb$ and the data sample collected $40\,\mev$
below the $B\bar{B}$ threshold (c.m.\ energy $10.52\,\gev$) with an
integrated luminosity of $74\,\ifb$. 

The signal $\ee \to \bb$ and the continuum $\ee\to q\bar{q}\ (q = u,
d, s, c)$ events are generated using EvtGen ~\cite{Lange:2001uf}. The
size of the Monte-Carlo (MC) samples corresponds to an integrated
luminosity six times that of the data.  The detector response is
simulated using GEANT3~\cite{Brun:1987ma}. The MC simulation
includes run-dependent variations in the detector performance and
background conditions.

\section{Analysis strategy}

The method used in this paper was developed by the CLEO
collaboration~\cite{CLEO:2005pyn} and then applied by
Belle~\cite{Belle:2006jvm} for one energy point near the \Ufi.  We
slightly modify the method to mitigate low accuracy in inclusive $\Bs$
branching fractions.

We measure the inclusive $\ee\to\bb\to\DsX$ and $\ee\to\bb\to\DnX$
cross sections at various energies above the $B\bar{B}$ threshold by
subtracting the continuum contribution from the total $\ee\to\DsX$ and
$\ee\to\DnX$ cross sections. To perform the subtraction, we use
distributions in the normalized momentum $\xp$, which is defined as
$\xp=p/\sqrt{(\ecm/2)^2-m^2}$,\footnote{We are using $c=1$ units.}
where $p$ is the $D$-meson ($D$ corresponds to \Ds\ or \D) momentum
measured in the c.m.\ frame, $\ecm$ is the c.m.\ energy, and $m$ is
the $D$-meson mass.
The $\xp$ spectra of $\Ds$ mesons at the $\Ufi$ energy in the
simulated $\bb$ and continuum events are shown in
Fig.~\ref{fig::mc_spectra}. 
\begin{figure}[h!]
\centering 
\begin{tabular}{cc}
\includegraphics[width=0.45\linewidth]{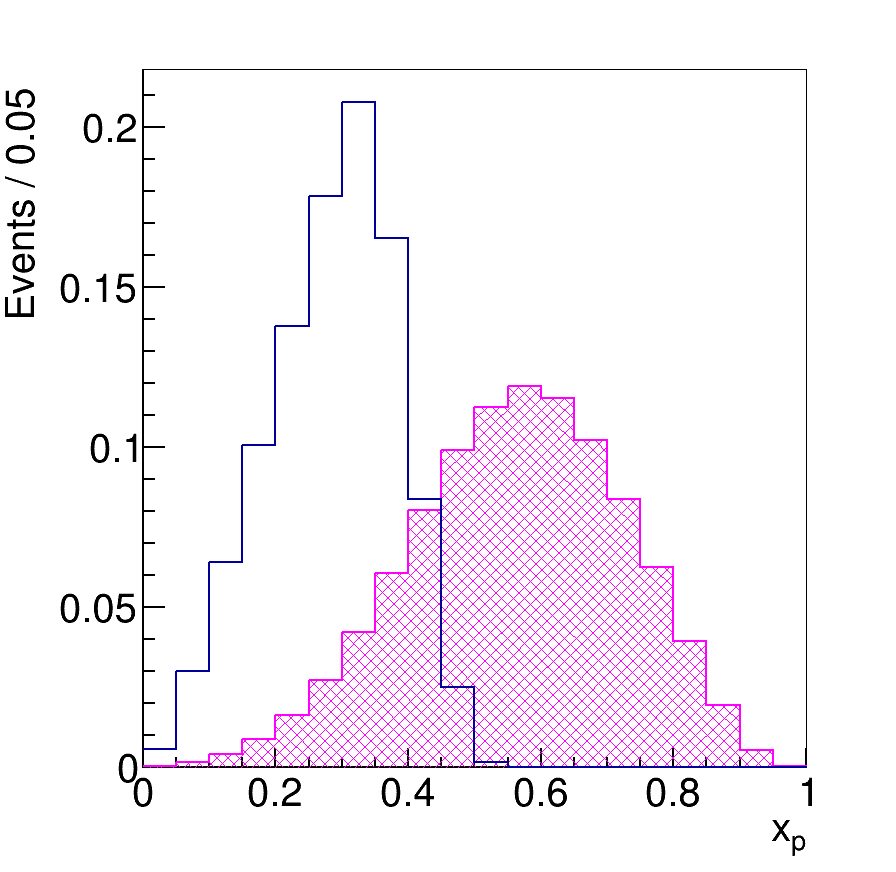} &
\end{tabular}
\caption{The $\xp$ spectra of $\Ds$ mesons at the \Ufi\ energy in the
  simulated $\bb$ (open blue histogram) and continuum (hatched magenta
  histogram) events. Both distributions are normalized to unity.}
\label{fig::mc_spectra}
\end{figure}
The $\bb$ events are restricted to the lower half of the $\xp$ range,
while the continuum events are enhanced in the high $\xp$ region. We
determine the shape of the continuum contribution using the data
collected below the $B\bar{B}$ threshold, normalize the contribution
using the high $\xp$ region, and subtract. The events remaining after
the subtraction are corrected for the efficiency in the $\xp$ bins,
and their sum is used to determine $\sigma(\ee\to\bb\to\DX)$.

In the considered energy range, the \bb\ events are of three types: with
$B$ mesons ($B$ corresponds to $B^+$ or $B^0$), with $\Bs$, and with
bottomonium; the latter contributes at the level of a few per
cent. Neglecting $D$ meson production in bottomonium decays, we write
\beq
\begin{split}
 \begin{array}{lccc}
   \sigma(\ee\to\bb\to\DsX) & = & 2\,\sigma(\ee\to\BsBsX) & \Br(\Bs\to\DsX)\\
   & + & 2\,\sigma(\ee\to\BBX) & \Br(B\to \DsX),
 \end{array} \\
 \begin{array}{lcccc}
   \sigma(\ee\to\bb\to\DnX) & = & 2\,\sigma(\ee\to\BsBsX) & \Br(\Bs\to\DnX)\\
   & + & 2\,\sigma(\ee\to\BBX) & \Br(B\to\DnX).
 \end{array}
 \label{eq:general}
\end{split} 
\eeq
The multiplicity of $D$ mesons is up to two in $B$ decays and up to
four in $\bb$ events; the branching fractions $\Br(B_{(s)}\to\DX)$ and
the cross sections $\sigma(\ee\to\bb\to\DX)$ correspond to the average
multiplicity of the $D$ mesons in the considered processes.
From the system of equations (\ref{eq:general}), we find the
ratio
\begin{align}
  C & \equiv \frac{\Br(\Bs\to\DnX)}{\Br(\Bs\to\DsX)} \nn \\ 
  & = \frac{\sigma(\ee\to\bb\to\DnX) - 
    2\;\sigma(\ee\to\BBX)\;\Br(B\to\DnX)} 
  {\sigma(\ee\to\bb\to\DsX) - 
    2\;\sigma(\ee\to\BBX)\;\Br(B\to\DsX)}.
\label{eq:general_brs_ratio}
\end{align}

We determine the ratio $C$ by measuring $\sigma(\ee\to\bb\to\DsX)$ and
$\sigma(\ee\to\bb\to\DnX)$ at the \Ufi\ energy and using the value of
$\sigma(\ee\to\BBX)$ at this energy reported in
Ref.~\cite{Belle:2021lzm}. Then we re-write the system of equations
\eqref{eq:general} as
\beq
\begin{split}
 \begin{array}{lccc}
   \sigma(\ee\to\bb\to\DsX) & = & 2\,\sigma(\ee\to\BsBsX) & \Br(\Bs\to\DsX)\\
   & + & 2\,\sigma(\ee\to\BBX) & \Br(B\to \DsX),
 \end{array} \\
 \begin{array}{lcccc}
   \sigma(\ee\to\bb\to\DnX) & = & 2\,C\,\sigma(\ee\to\BsBsX) & \Br(\Bs\to\DsX)\\
   & + & 2\,\sigma(\ee\to\BBX) & \Br(B\to\DnX).
 \end{array}
 \label{eq:general_2}
\end{split} 
\eeq
We define $X=\sigma(\ee\to\BsBsX)\,\Br(\Bs\to\DsX)$,
$Y=\sigma(\ee\to\BBX)$ and solve the system of
equations~\eqref{eq:general_2} with respect to $X$ and $Y$:
\begin{align}
  X & = \frac{B\,U-A\,W }{2(B - A\,C)}, \nn \\[-2mm]
  \label{eq:solv_general12}\\[-2mm]
  Y & = \frac{W - C\,U }{2(B - A\,C)}\nn, 
\end{align}
where we introduced notations
\begin{align}
  &U  = \sigma(\ee \to \bb \to \DsX),\nn\\
  &W  = \sigma(\ee \to \bb \to \DnX),\label{eq:definition}\\
  &A  = \Br(\B \to \DsX),\nn\\
  &B  = \Br(\B \to \DnX).\nn
\end{align}
To study energy dependence of the $\ee\to\BsBsX$ cross section, it is
convenient to consider the product
$\sigma(\ee\to\BsBsX)\;\Br(\Bs\to\DsX)$, since in this case a rather
large uncertainty in $\Br(\Bs\to\DsX)$ will affect only the overall
normalization. 

Based on the $\Ufo$ data, we measure $\Br(B\to\DsX)$ and
$\Br(B\to\DnX)$, and use them in Eqs.~\eqref{eq:general_brs_ratio} and
\eqref{eq:solv_general12} to reduce systematic uncertainties.

\section{Event selection}
\label{select}

All charged tracks are required to be consistent with originating from
the interaction point (IP): we require $dr<0.5 \, {\mathrm{cm}}$ and
$|dz|<2\,{\mathrm{cm}}$, where $dr$ and $|dz|$ are the impact
parameters perpendicular to and along the beam direction,
respectively, with respect to the IP. Information from the TOF, the
number of the photoelectrons from the ACC, and the $dE/dx$ measurement
in the CDC are combined to form a likelihood $\mathcal{L}_h$ for a
hadron hypothesis $h$~\cite{Nakano:2002jw}. Charged kaon candidates
are required to have a likelihood ratio ${\mathcal{P}}_{K/\pi} =
\mathcal{L}_K / (\mathcal{L}_K + \mathcal{L}_\pi)>0.6$. Charged pion
candidates are required to have ${\mathcal{P}}_{K/\pi}<0.9$. The
efficiency for kaon (pion) identification is about 90\% (97\%) with a
misidentification rate of a pion as a kaon (a kaon as a pion) of about
8\% (20\%).

The \Ds\ and \D\ candidates are reconstructed using only the clean
$\Ds \to \phi \pi^+$ and $\D \to K^- \pi^+$ decay
channels.\footnote{Throughout this paper, charge conjugation is always
  included.} Since there might be several $D$ mesons in an event, we
do not apply best candidate selection. The $\phi$ mesons are
reconstructed from $K^+K^-$ pairs. The invariant mass of the two kaons
should be within $\pm 19$\mevcc\ from the nominal $\phi$ mass
(Fig.~\ref{fig::mphi}).
\begin{figure}[h!]
 \centering
 \includegraphics[width=0.49\linewidth]{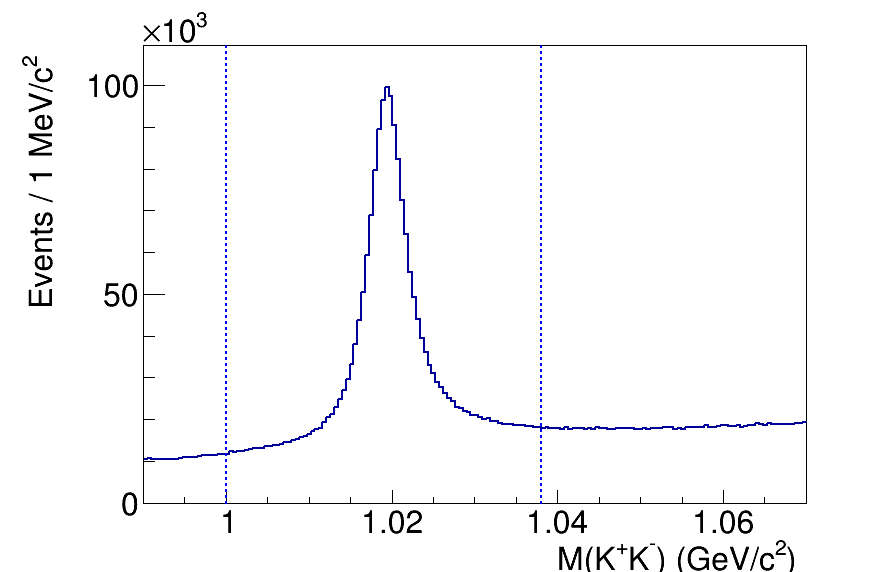} 
 \caption{The $K^+ K^-$ mass distribution in the data without the
     helicity angle requirement. The dashed vertical lines indicate
   the signal region.}
  \label{fig::mphi}
\end{figure}
The helicity angle $\theta_{\rm hel}$ is defined as the angle between
the $\Ds$ and $K^+$ momenta in the $\phi$ rest frame; a requirement
$|\cos(\theta_{\rm hel})|>0.25$ is applied.

\section{Analysis of the \Ufo\ and \Ufi\ data samples}
\label{sec::5s_and_4s}

In this section, we describe the analysis of the \Ufo\ and \Ufi\ data
samples. Here our goal is to measure the cross sections
$\sigma(\ee\to\bb\to\DX)$, the branching fractions $\Br(B\to\DX)$, the
$\Bs$ production fraction $f_s$, and the ratio
$\Br(\Bs\to\DnX)\;/\;\Br(\Bs\to\DsX)$.

\subsection{Measurement of $\sigma(\ee \to \bb \to DX)$}

We fit the mass distributions of the \Ds\ and \D\ candidates in bins
of \xp. The signals are described by a sum of four Gaussians with
parameters determined from the MC simulation. We introduce a shift and
a broadening factor, common to all Gaussians, that are floated in each
\xp\ bin. The background is described by a second-order polynomial. We
use binned likelihood fits. Examples of the fits to the \Ufi\ data for
\xp\ bins (0.25,\,0.3) and (0.65,\,0.7) are shown in
Fig.~\ref{fig:mD_fit}. The p-values of the fits, quoted in
Fig.~\ref{fig:mD_fit}, are calculated assuming Gaussian errors in each
bin.
\begin{figure}[h!]
  \centering 
  \begin{tabular}{cc}
    \includegraphics[width=.45\textwidth]{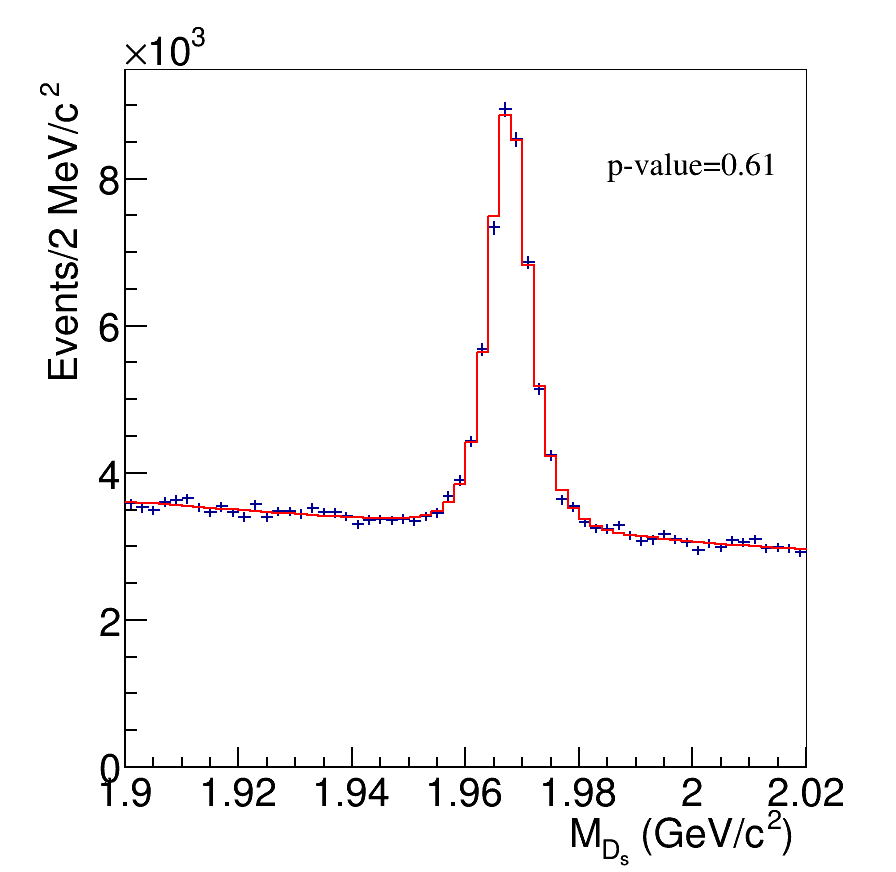} &
    \includegraphics[width=.45\textwidth]{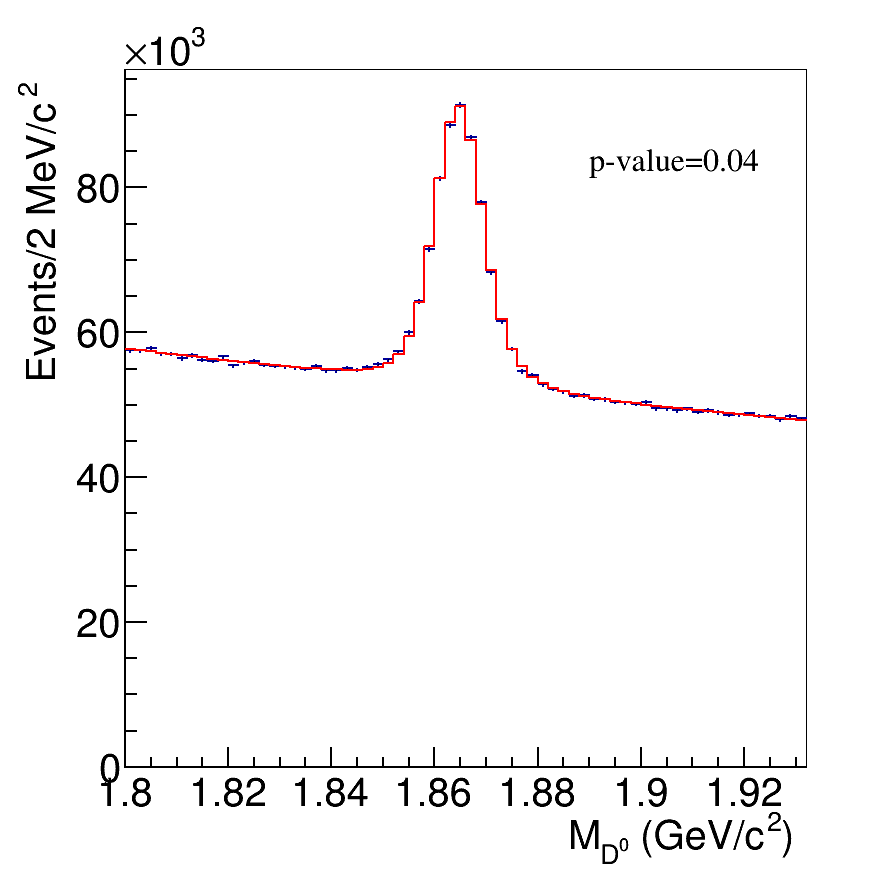}\\
    \includegraphics[width=.45\textwidth]{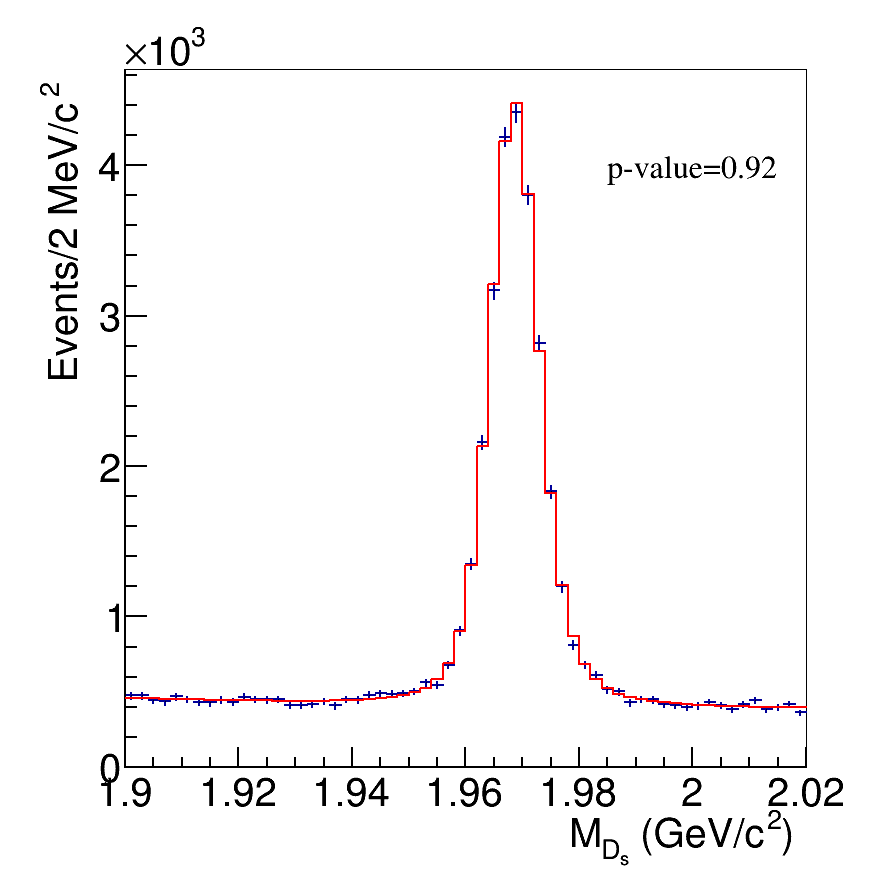} &
    \includegraphics[width=.45\textwidth]{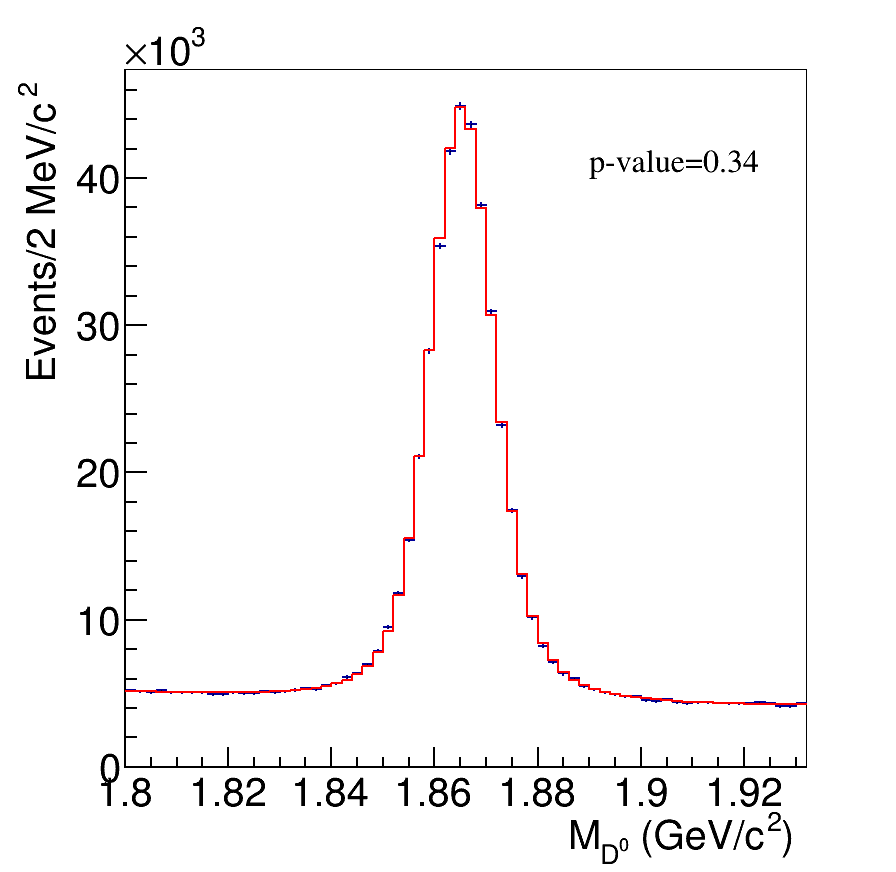}
  \end{tabular}
  \caption{\label{fig:mD_fit} The mass distributions of the
    \Ds\ (left) and \D\ (right) candidates in the $0.25 < x_p <0.3$
    (top) and $0.65< x_p <0.7$ (bottom) regions. Points with error bars
    are the \Ufi\ data, and histograms are the fit results. }
\end{figure}

The dependence of the \Ds\ and \D\ yields on \xp\ for the \Ufi, \Ufo,
and continuum data samples is shown in Fig.~\ref{fig::xp}.
\begin{figure}[tbp]
  \centering % \begin{center}/\end{center} takes some additional vertical space
  \begin{tabular}{cc}
    \includegraphics[width=.45\textwidth]{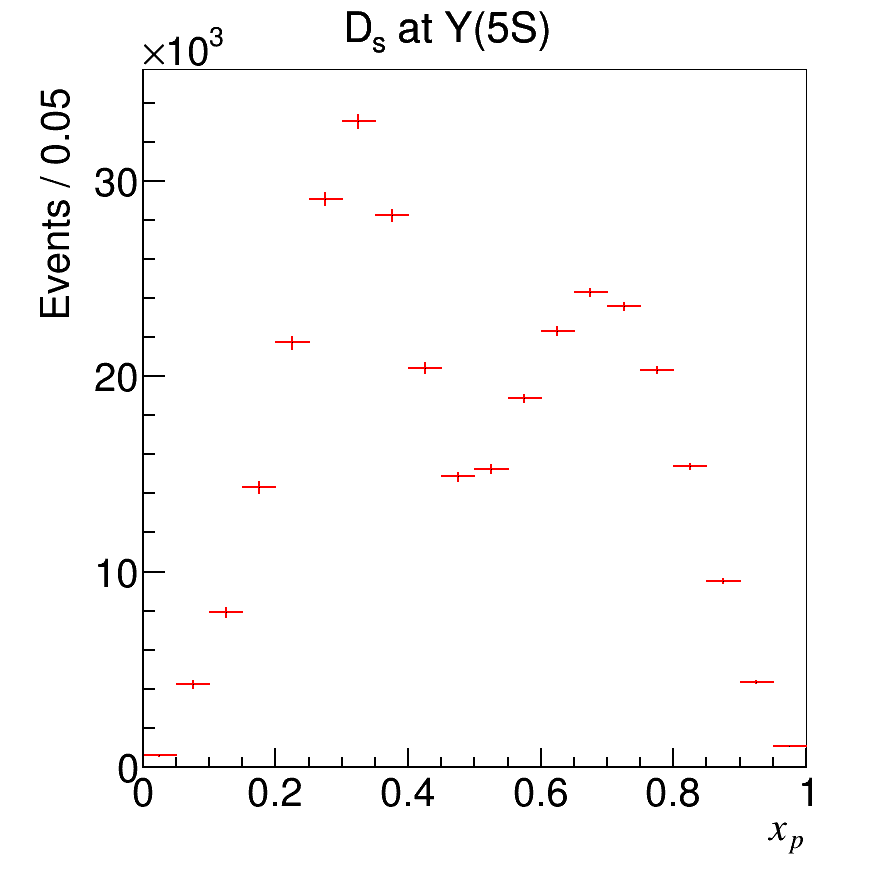} &
    \includegraphics[width=.45\textwidth]{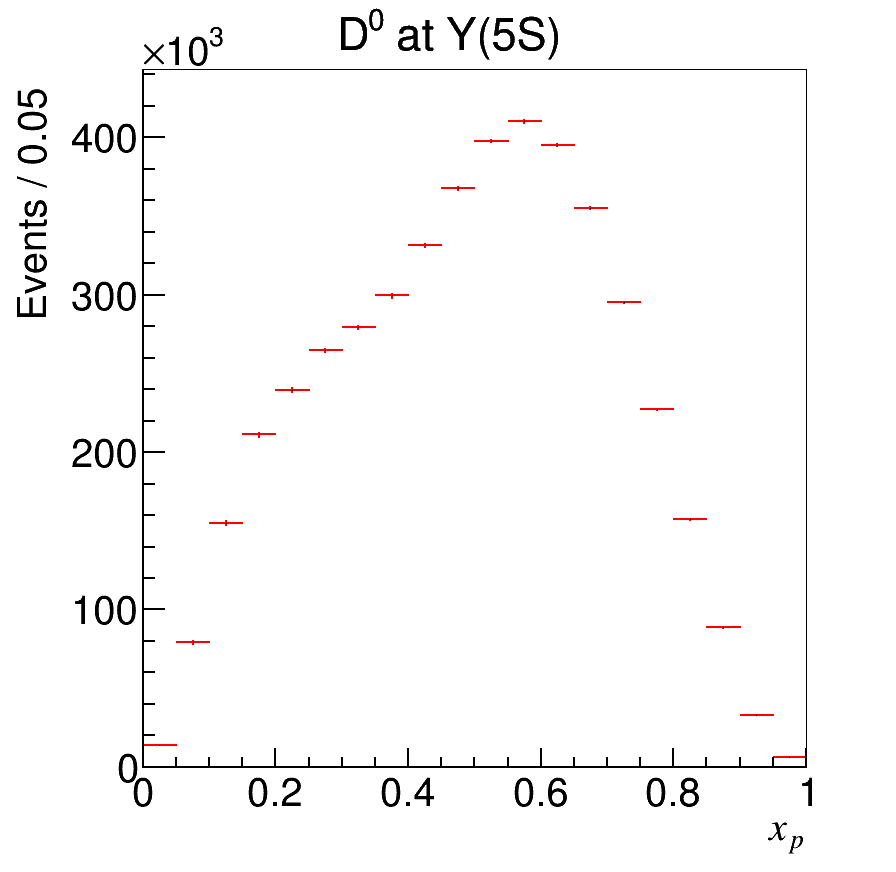}\\
    \includegraphics[width=.45\textwidth]{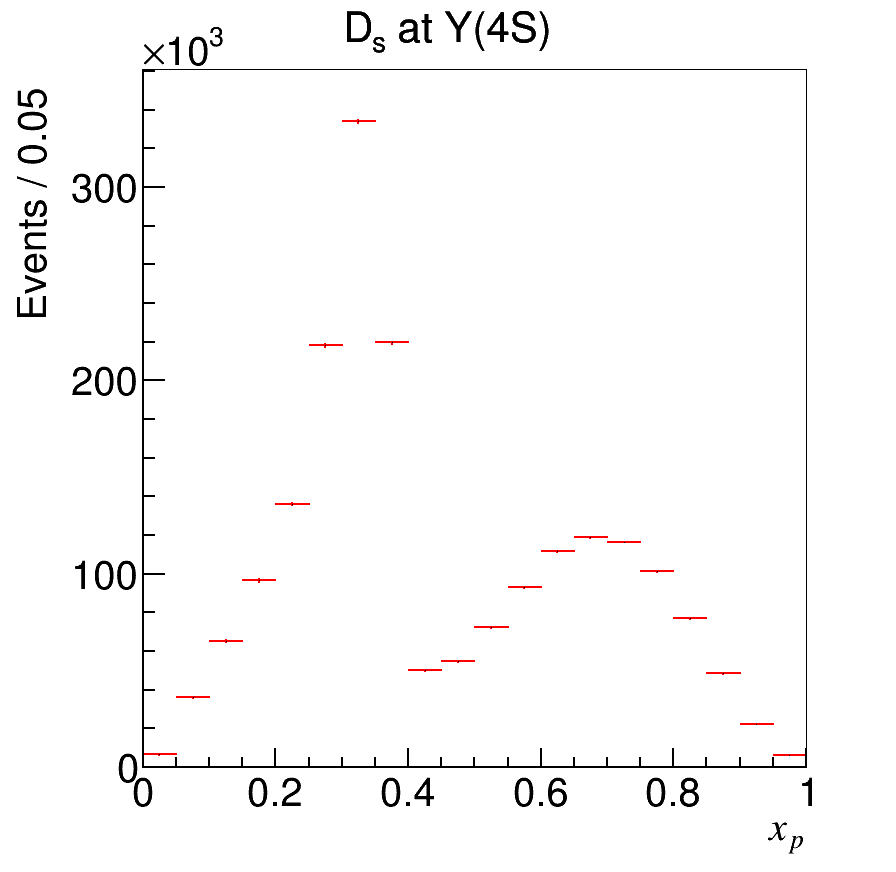} &
    \includegraphics[width=.45\textwidth]{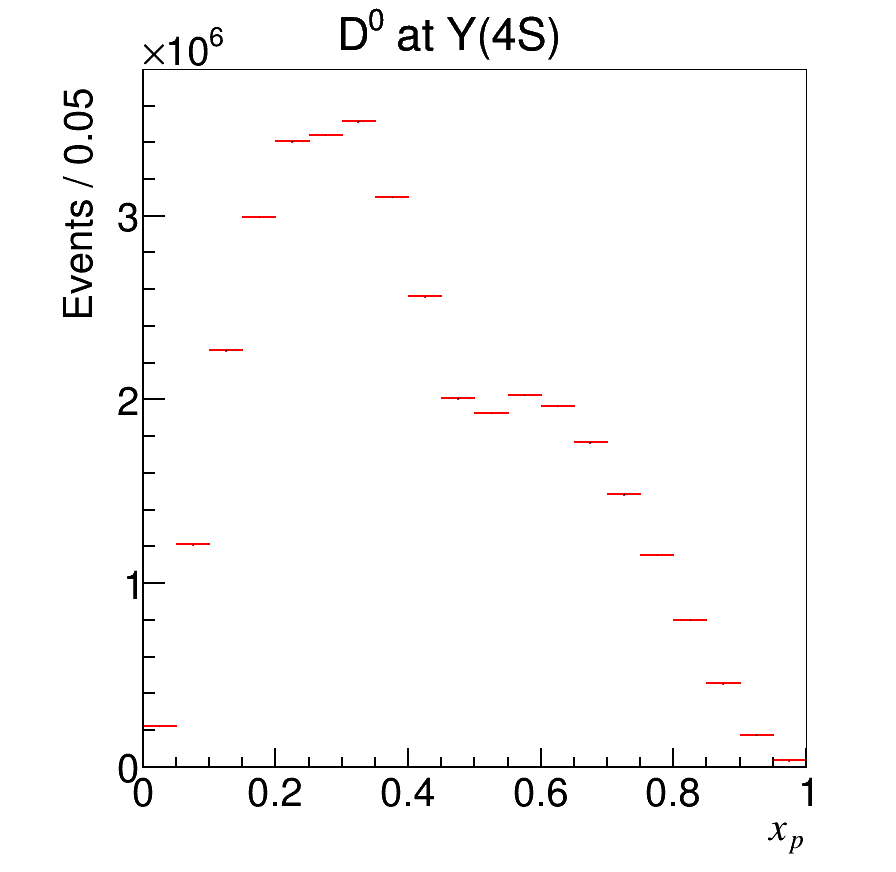}\\
    \includegraphics[width=.45\textwidth]{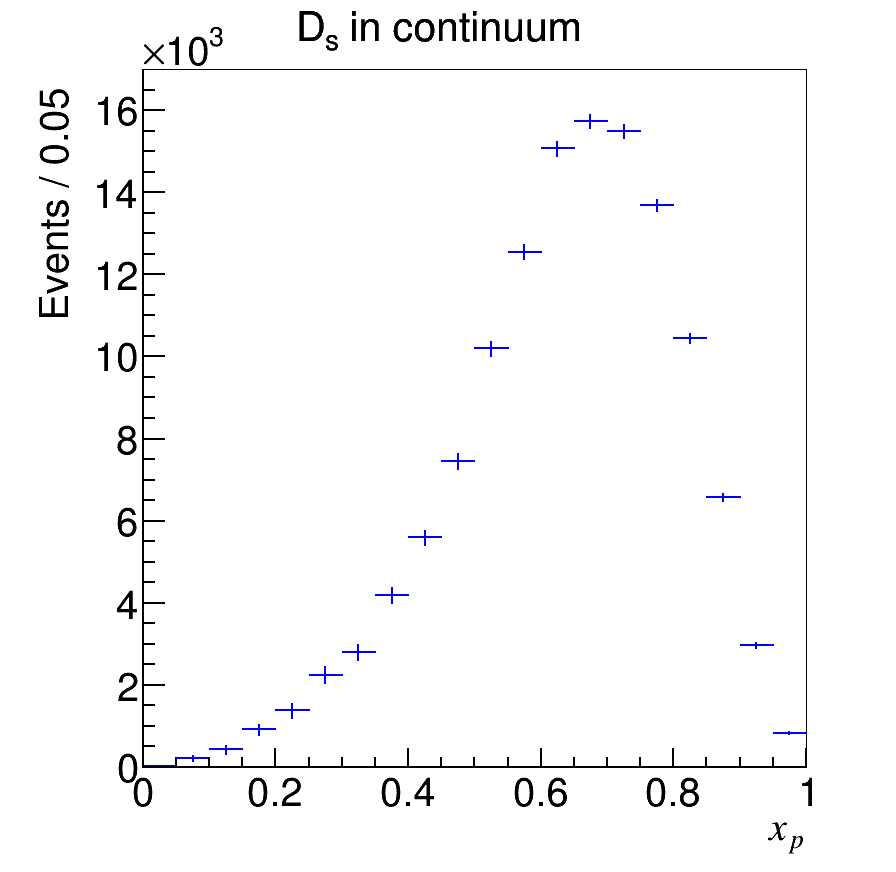} &
    \includegraphics[width=.45\textwidth]{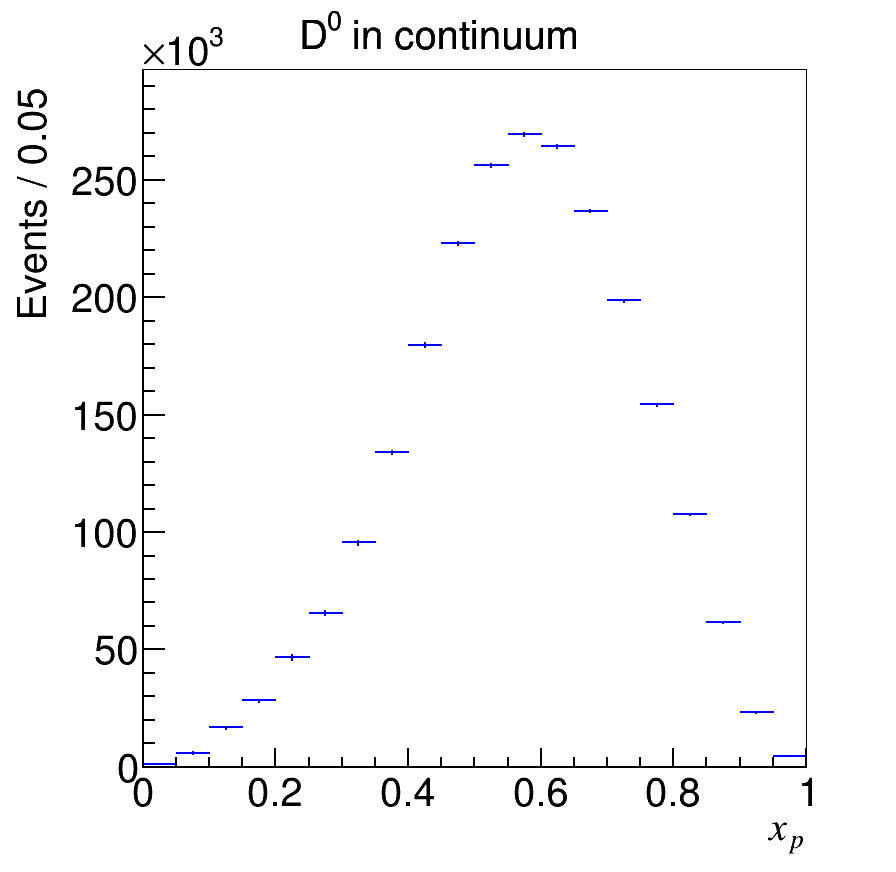}\\
  \end{tabular}
  \caption{\label{fig::xp} The yield of \Ds\ (left) and \D\ (right) in
    bins of \xp\ for the data samples collected at the
    \Ufi\ (top), \Ufo\ (middle) and below the $B\bar{B}$ threshold (bottom).}
\end{figure}
There is a clear enhancement at low \xp\ in the \Ufi\ and \Ufo\ data
due to the production of the $b\bar{b}$ events. We subtract the
continuum contribution using the \xp\ spectra for the data collected
below the $B\bar{B}$ threshold. The shape of the continuum spectrum
changes noticeably between $\ecm=10.52\,\gev$ and the $\Ufi$ energy,
primarily due to the evolution of fragmentation with energy. We
determine corrections with the help of the event generator developed
for Belle II that integrates KKMC and
Pythia~\cite{ref::b2_generators}. The KKMC generator is used to
simulate initial state radiation and the Pythia generator is used to
simulate $c$-quark fragmentation. The correction factors for \Ds\ and
\D, defined as the ratio of the continuum \xp\ spectra at the
  \Ufi\ energy and at $\ecm=10.52$ \gev, are shown in
Fig.~\ref{fig::xp_corr_5s}. In the \Ufo\ case, we find that no
correction is needed since the \Ufo\ energy is close to 10.52 GeV.
\begin{figure}[h!]
 \centering
 \includegraphics[width=0.49\linewidth]{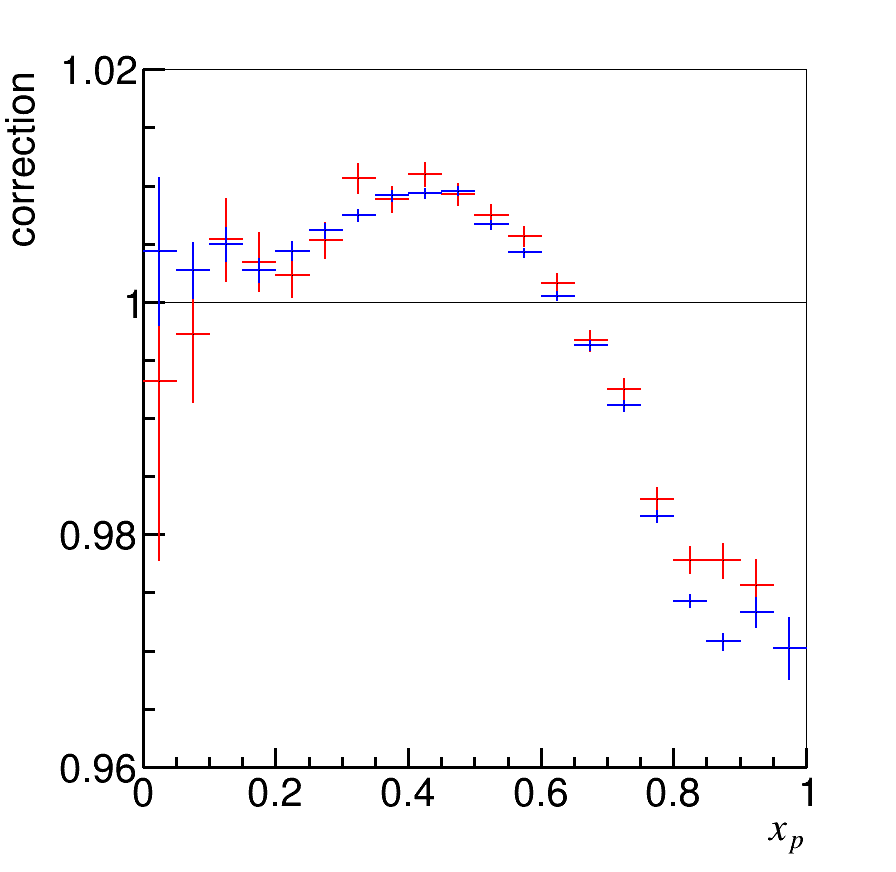} 
 \caption{The ratio of the continuum \xp\ spectra at the \Ufi\ energy
   and $\ecm=10.52\,\gev$ obtained using MC simulation. Red and blue
   points correspond to \Ds\ and \D, respectively.}
 \label{fig::xp_corr_5s}
\end{figure}

From the MC simulation, we find that $b\bar{b}$ events contribute only
at lower \xp\ values: the highest bin with a \bb\ contribution,
$i_{\rm max}$, and the corresponding upper bin edge $x_p^{\rm max}$,
are shown for \Ds\ and \D\ in different data samples in
Table~\ref{tab::xp}.
%$x_p<x_p^{\rm max}$, with the values of $x_p^{\rm max}$ given in Table~\ref{tab::xp}.
%
\begin{table}[ht!]
 \caption{The values of $x_p^{\rm max}$, $i_{\rm max}$, $k$ and
   $k^{\rm est}$ for \Ds\ and \D\ in different data samples; see
   the main text for the definition of these
   quantities.}\label{tab::xp}
  \begin{center}
  \begin{tabular}{ccccc}
    \hline
    & \Ds\ at \Ufi & \D\ at \Ufi & \Ds\ at \Ufo & \D\ at \Ufo\\ 
    \hline
    $i_{\rm max}$ & 11 & 12 & 10 &11 \\
    $x_p^{\rm max}$ & 0.55 & 0.60 & 0.50 & 0.55\\
    $k$ & $1.510 \pm 0.004$ & $1.499 \pm 0.001$ & $ 7.410 \pm 0.008$ & $ 7.460 \pm 0.002$\\
    $ k^{\rm est}$ & \multicolumn{2}{c}{1.516} & \multicolumn{2}{c}{7.430}  \\
    \hline
  \end{tabular}
  \end{center}
\end{table}
Thus, we use the $x_p>x_p^{\rm max}$ region for the normalization of
the continuum \xp\ distribution and fit the \Ufi\ and \Ufo\ data in
this range using the (corrected) \xp\ spectrum of the data below the
$B\bar{B}$ threshold as the fitting function. The results of these
fits are shown in Fig.~\ref{fig::fit_xp}.

\begin{figure}[h!]
 \centering
 \begin{tabular}{cc}
   \includegraphics[width=0.49\linewidth]{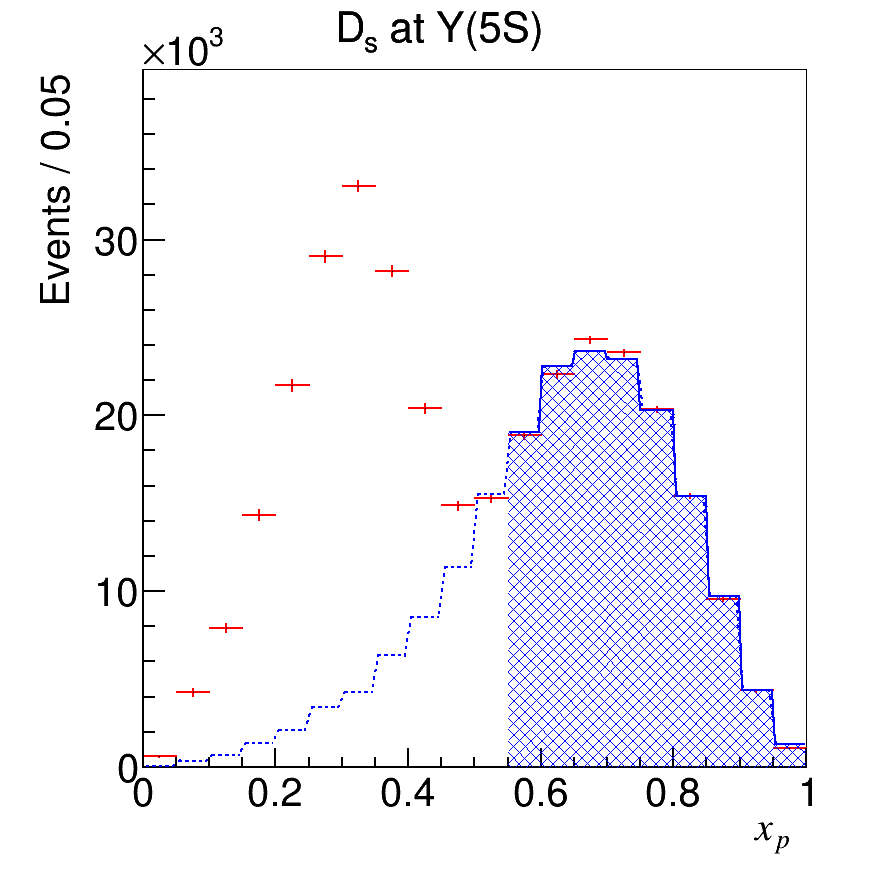} &
   \includegraphics[width=0.49\linewidth]{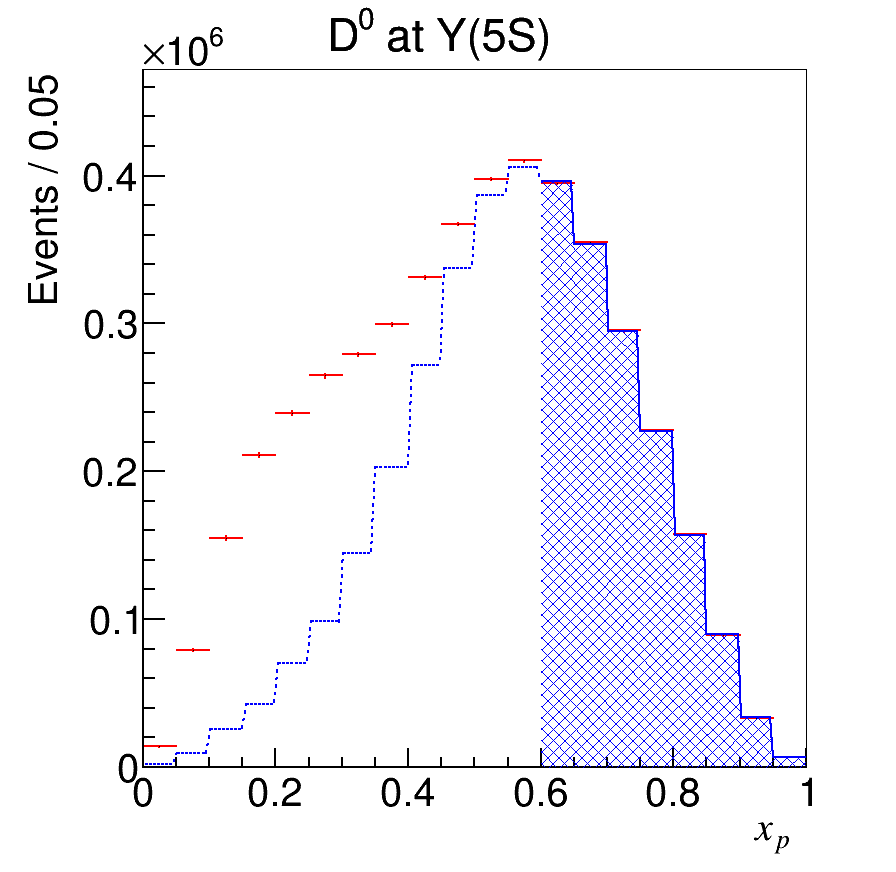} \\
   \includegraphics[width=0.49\linewidth]{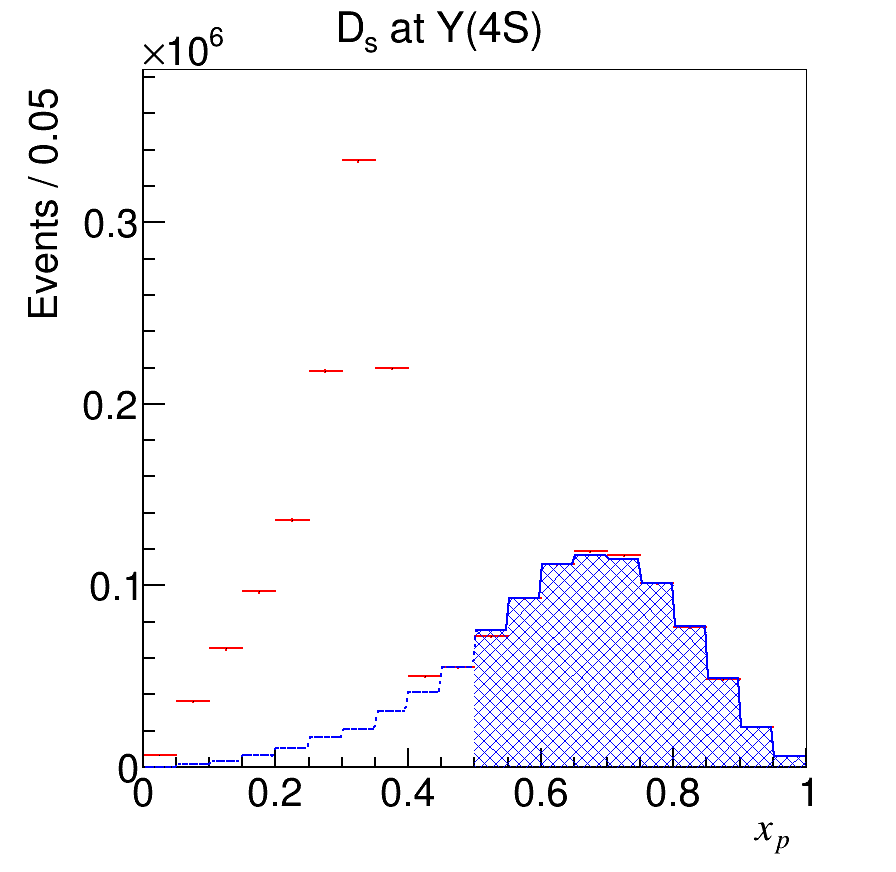}  &
   \includegraphics[width=0.49\linewidth]{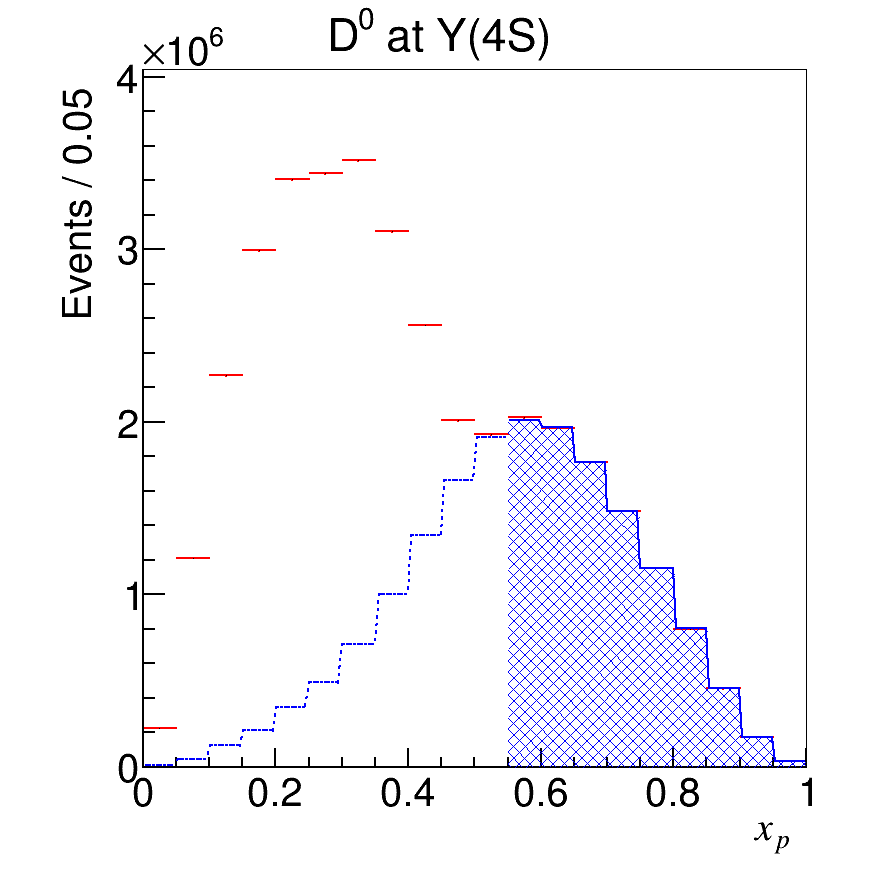} 
 \end{tabular} 
 \caption{The yield of \Ds\ (left) and \D\ (right) in bins of \xp\ for
   the \Ufi\ (top) and \Ufo\ (bottom) data. Points with error bars
   show the data, solid hatched histograms show the
   fit results, and open dashed histograms show the extrapolation of
   the continuum component into the $b\bar{b}$ signal region.}
 \label{fig::fit_xp}
\end{figure}

The normalization factors $k$ for the continuum
contribution obtained from the fits are listed in Table~\ref{tab::xp}. 
These factors can be roughly estimated as
\beq
k_i^{\rm est}=\frac{\mathcal{L}_{i}}{\mathcal{L}_{\rm cont}} \;
\left(\frac{E_{\rm cont}}{E_{i}} \right)^2, %\quad i=\Ufi, \Ufo,
\eeq
where $i$ runs over $\Ufi$ and $\Ufo$, ``cont'' denotes data sample 
collected below the $B\bar{B}$ threshold, $E$ and ${\mathcal L}$ are the 
corresponding energy and integrated luminosity.
The values of $k_i^{\rm est}$ are in reasonable agreement with the fit
results, as shown in Table~\ref{tab::xp}.
The \xp\ spectra after the continuum subtraction are shown in
Fig.~\ref{fig::xp_after_subtr}. The points in the subtraction region
are consistent with zero, which indicates that the continuum spectra
are determined correctly.

\begin{figure}[h!]
 \centering
 \begin{tabular}{cc}
 \includegraphics[width=0.49\linewidth]{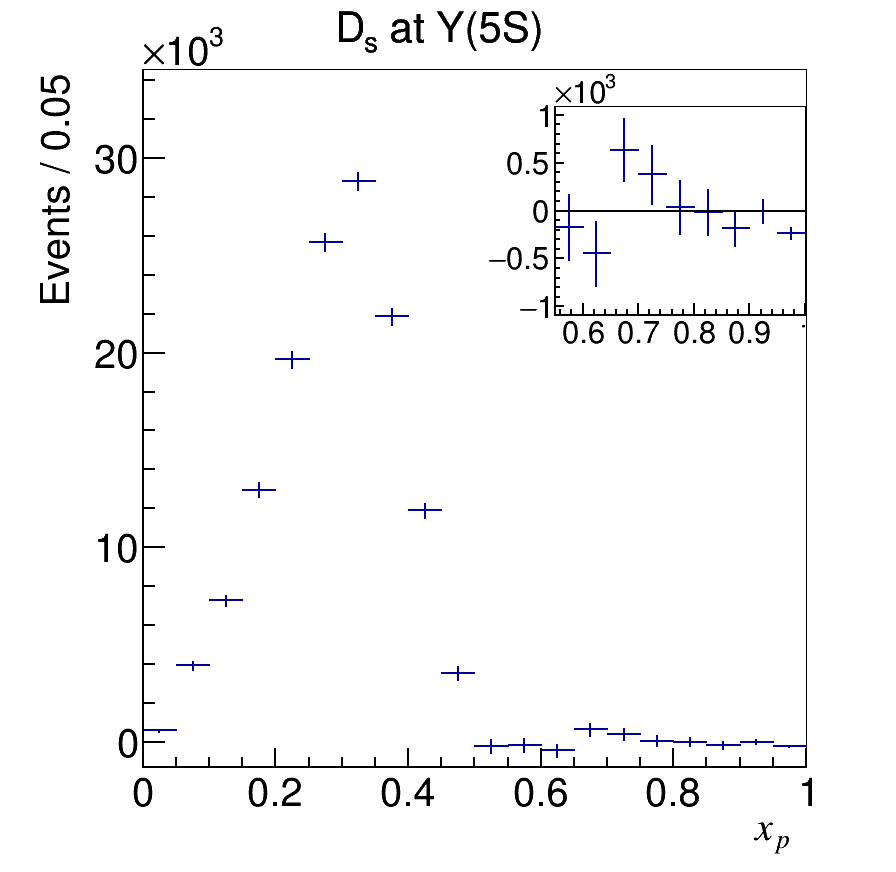} &
 \includegraphics[width=0.49\linewidth]{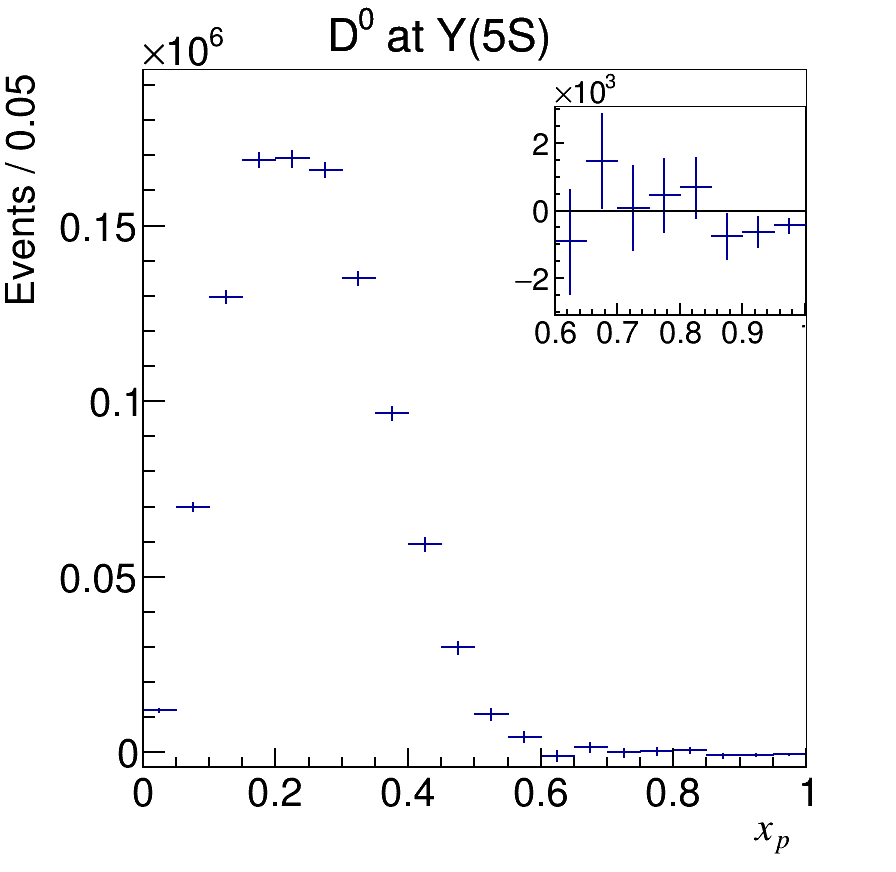}\\
 \includegraphics[width=0.49\linewidth]{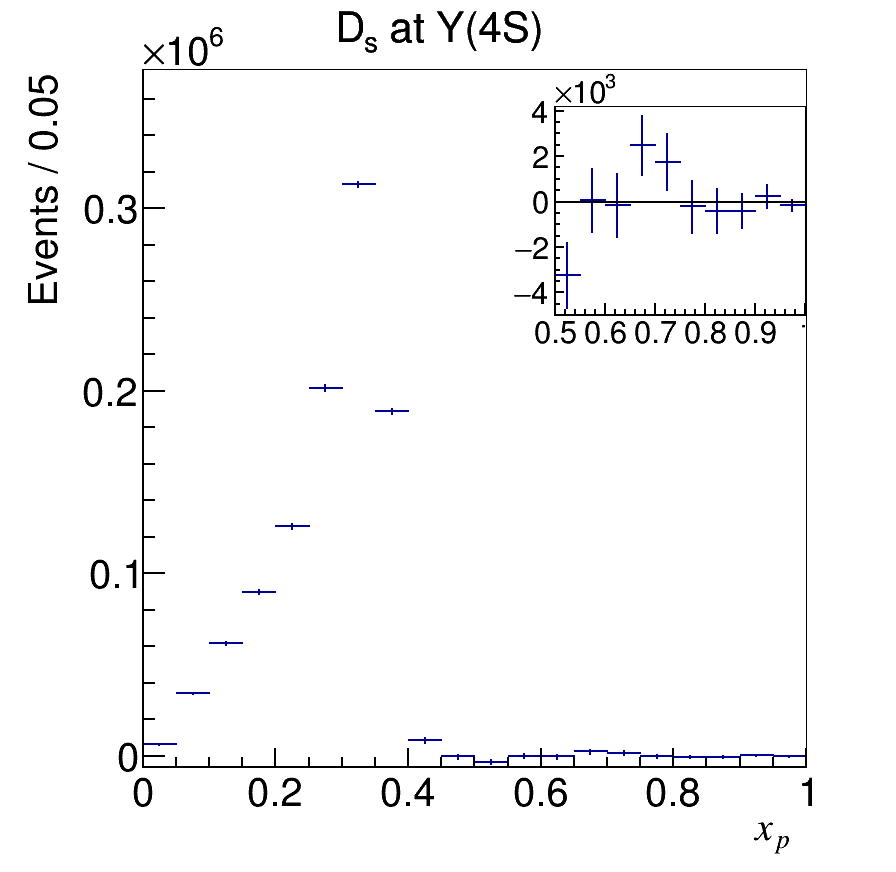} &
 \includegraphics[width=0.49\linewidth]{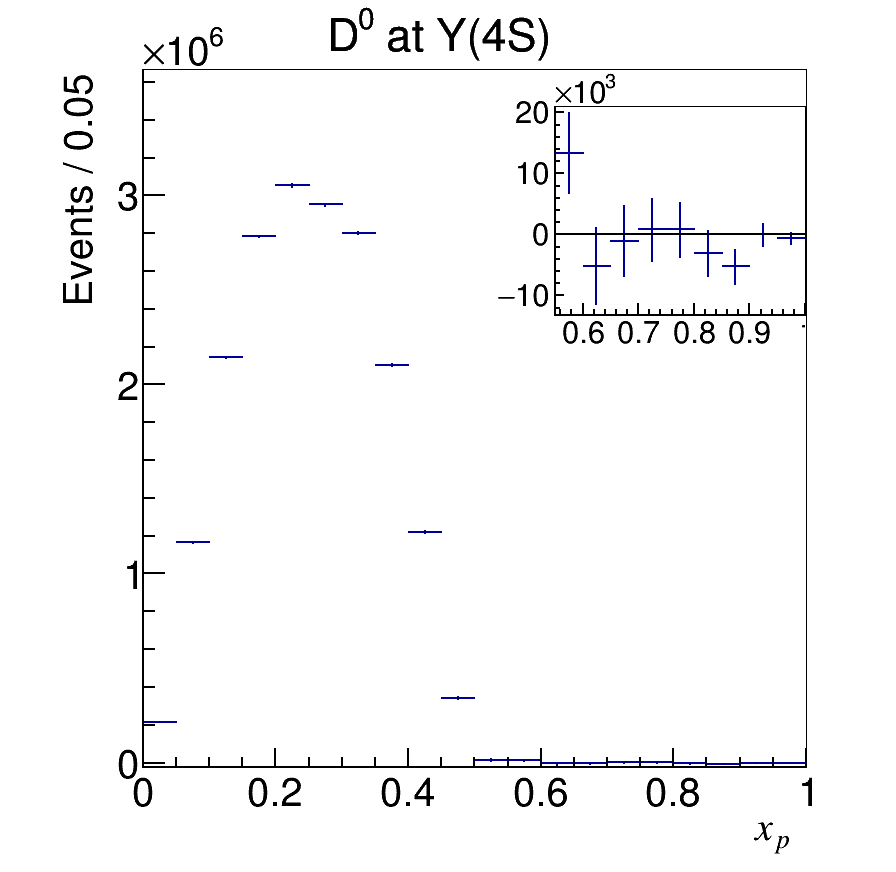}\\
 \end{tabular}
 \caption{The distribution of the \xp\ of \Ds\ (left) and \D\ (right)
  in the \Ufi\ (top) and \Ufo\ (bottom) data after subtracting the
   continuum contribution. Insets show the high \xp\ region with
     an expanded vertical scale.}
 \label{fig::xp_after_subtr}
\end{figure}

The $D$ meson reconstruction efficiency as a function of \xp\ is shown
in Fig.~\ref{fig::eff}. It takes into account the known difference
between data and simulation for particle identification
efficiency; the corresponding momentum and polar angle dependent
correction factors are determined using the $D^{*+}\to \D (\to K^- \pi^+) \pi^+$
decays~\cite{Nakano:2002jw}. The presented \Ds\ reconstruction
efficiency includes the efficiency of the $\phi$ mass and helicity
angle requirements. 
\begin{figure}[h!]
 \centering
 \begin{tabular}{cc}
 \includegraphics[width=0.49\linewidth]{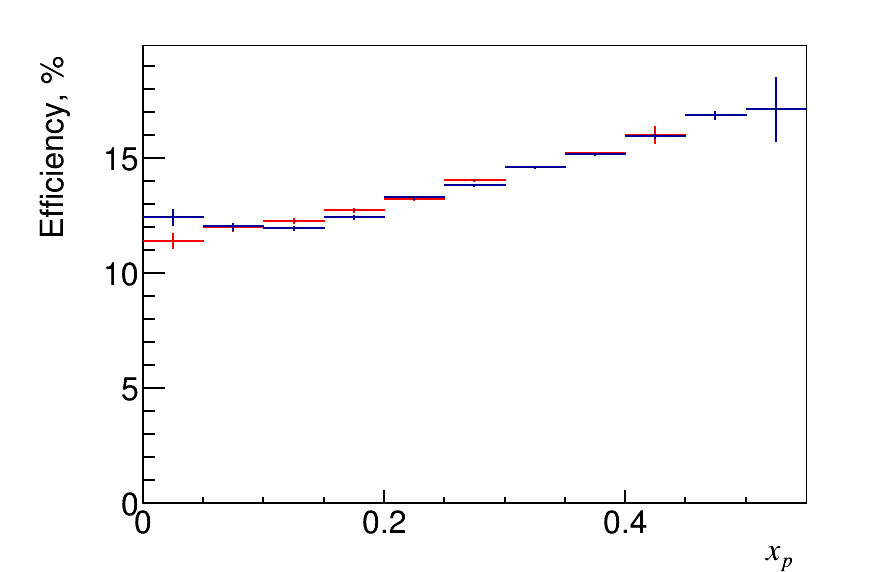} &
 \includegraphics[width=0.49\linewidth]{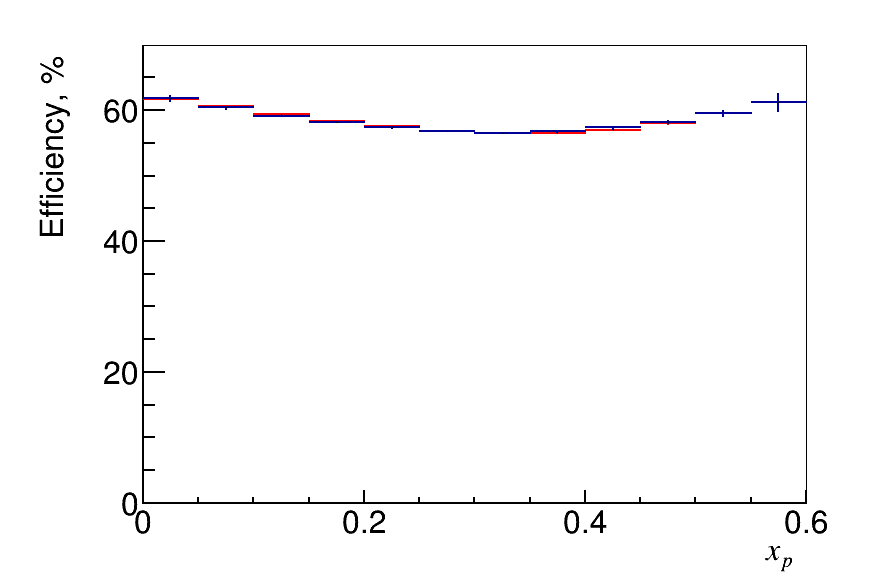}\\
 \end{tabular}
 \caption{The \Ds\ (left) and \D\ (right) meson reconstruction
   efficiency at the \Ufi\ (blue points) and \Ufo\ (red points) as
   a function of \xp.}
 \label{fig::eff}
\end{figure}

We introduce a correction factor
$r_{\phi}=\varepsilon_{\phi}^{\rm data}/\varepsilon_{\phi}^{\rm MC}$
to account for the difference in $M(K^+K^-)$ distribution between the
data and the MC simulation.
To determine $\varepsilon_{\phi}$, we fit the mass distributions for
the \Ds\ candidates that satisfy the $\phi$ mass and helicity angle
requirements and that are rejected by them. Based on the corresponding
signal yields, $N_{\phi}$ and $N_{\overline{\phi}}$, we find
\beq
\varepsilon_{\phi}=\frac{N_{\phi}}{N_{\phi}+N_{\overline{\phi}}}.
\eeq
%We use the \xp\ interval $0.2 < \xp < 0.95$, where background is low
Here we use events with $0.2 < \xp < 0.95$, where background is low
and all bins are well-populated. To estimate systematic
uncertainty, we vary the considered \xp\ interval. The result is
\beq
r_{\phi}=0.981 \pm 0.005 \pm 0.004.
\label{formula::r_phi}
\eeq
Here and throughout this paper if two uncertainties are shown, the
first is statistical and the second is systematic.

The inclusive visible $\ee\to\DX$ cross sections are calculated as
\beq
\sigma(\ee\to\bb\to\DsX) =\sum_{i=1}^{\imax} \frac{N_i(\Ds)-k(\Ds)
 \; n_i(\Ds)}{\mathcal{L} \; \mathcal{E}_i(\Ds) \; r_{\phi} \; \Br(\Ds
 \to K^+ K^- \pi)}
\label{formula::ds_cross_sec}
\eeq
and
\beq
\sigma(\ee\to\bb\to\DnX) =\sum_{i=1}^{\imax} \frac{N_i(\D)-k(\D) \; n_i(\D)}{\mathcal{L} \; \mathcal{E}_i(\D) \; \Br(\D \to K \pi) },
\label{formula::d0_cross_sec}
\eeq
where $i$ runs over the \xp\ bins, the values of $\imax$ are given in
Table~\ref{tab::xp}, $N_i$ and $n_i$ are the numbers of the $D$ mesons
in the $i$-th \xp\ bin in the on-resonance and continuum spectra,
respectively, $\mathcal{E}_i(D)$ is the $D$ reconstruction efficiency
in the $i$-th bin, $\mathcal{L}$ is the integrated luminosity of the
\Ufi\ or \Ufo\ data samples, $\Br(\Ds\to K^+ K^-\pi)={(5.38\pm0.10)}\%$
and
$\Br(\D\to{K}^\mp\pi^\pm)=(3.95\pm0.03)\%$~\cite{ParticleDataGroup:2022pth};
$k$ is given in Table~\ref{tab::xp}, and $r_{\phi}$ is given in
Eq.~\eqref{formula::r_phi}. The cross section values measured at the
\Ufi\ and \Ufo\ are listed in Table~\ref{tab::cross_sec_at_Y(nS)}.
\begin{table}[ht!]
    \caption{The $\ee\to\bb\to\DsX$ and $\ee\to\bb\to\DnX$ cross
      sections (in pb) measured at the \Ufi\ and
      \Ufo.}\label{tab::cross_sec_at_Y(nS)}
  \begin{center}
  \begin{tabular}{ccc}
    \hline
    &  $\sigma(\ee \to \bb \to \DsX)$ & $\sigma(\ee \to \bb \to \DnX)$ \\ 
    \hline
    \Ufi  & $151.8 \pm 1.0\pm 5.5$ &  $\phantom{1}379.7 \pm 1.6 \pm 10.0$ \\
     \hline
    \Ufo& $248.6 \pm 0.6\pm 9.2$ &  $ 1468.5 \pm 0.9\pm 36.6$ \\
    \hline
  \end{tabular}
  \end{center}
\end{table}
Their statistical uncertainties are calculated as
\beq
\sqrt{ \sum_{i=1}^{\imax} \left( \sigma_i\;\frac{\Delta N_i}{N_i-k\;n_i}\right)^2 + \left( \Delta k \; {
\sum_{i=1}^{\imax} \frac{\sigma_i\;n_i}{N_i-k\;n_i}}\right)^2},
\label{eq::stat_error_ds_5s}
\eeq where ${\sigma_i}$ is the inclusive cross section in the $i$-th
momentum bin and $\Delta X$ is the statistical uncertainty of
the quantity $X$.

Below we list various contributions to the systematic uncertainty in
the cross sections; corresponding summary is presented in
Table~\ref{tab::syst_sum}.
\begin{itemize}
\item We vary the fit model for the mass spectra of the $D$
  candidates. In particular, we (1) introduce one additional shift for
  one narrow Gaussian; (2) introduce additional shift and broadening
  factor for one narrow Gaussian; (3) change the background function
  from 2nd- to 3rd-order polynomial. Variations (1) and (3) result in
  negligibly small changes in the cross section. The uncertainties
  related to variation (2) are shown in Table~\ref{tab::syst_sum}.
\item Contribution of the statistical error in the continuum
  \xp\ spectrum is calculated as \beq \frac{1}{\sigma}\;\sqrt{
    \sum_{i=1}^{\imax} \left( \sigma_i \;\frac{ \Delta n_i \;
      k}{N_i-k\;n_i}\right)^2}.
\label{eq::stat_err_in_cont}
\eeq
\item The systematic uncertainty related to the continuum spectrum
  correction is estimated as half of the change in the cross section
  obtained with and without this correction.
\item The contribution of the MC statistical error is calculated as
\beq
\frac{1}{\sigma}\;\sqrt{ \sum_{i=1}^{\imax} \left( \sigma_i \; \frac{\Delta{\cal{E}}_i}{{\cal{E}}_i} \right)^2}.
\label{eq::stat_err_in_MC}
\eeq
\item We account for the 0.6\% uncertainty in $r_{\phi}$.
\item The systematic uncertainty of the track reconstruction
  efficiency, estimated using partially
    reconstructed $D^{*+} \to \D \pi^+$, $\D \to \pi^+ \pi^- K_S^0$
    and $K_S^0 \to \pi^+ \pi^-$ events, is 0.35\% per track; thus we have 1.1\%\ for
  \Ds\ and 0.7\%\ for \D.
\item The uncertainty of the $K/\pi$ identification efficiency is due
  to a possible difference between MC
  and data. This difference is studied using $D^{*+} \to
    D^0(K^-\pi^+)\pi^+$ decays --- see Chapter 5.4 of Ref.~\cite{BaBar:2014omp}. The uncertainty is calculated
  as 2.3\% for $\Ds \to K^- K^+ \pi^+$ and 1.4\% for $\D \to K^-
  \pi^+$.
\item The uncertainty in the integrated luminosity is 1.4\%. 
\item The uncertainty in the world average $\Br(\Ds \to K^+ K^- \pi^+)$
is 1.9\% and in $\Br(\D \to K^+ \pi^+)$ is
0.8\% \cite{ParticleDataGroup:2022pth}.
\end{itemize}
The total systematic uncertainty is calculated by adding the various
contributions in quadrature.

\begin{table}[h]
  \caption{Systematic uncertainties in the $\ee \to \bb \to \DsX$ and
  $\ee \to \bb \to \DnX$ cross sections at \Ufi\ and \Ufo\ (in \%).}
\label{tab::syst_sum}
%\begin{ruledtabular}
\begin{tabular}{lcccc}
\hline
 \centering 
Source & \Ds\ at \Ufi\ & \D\ at \Ufi\ & \Ds\ at \Ufo\ & \D\ at \Ufo\ \\
\hline
Fit model & 0.6 & 0.3 & 1.0 & 1.1\\
Cont. $\xp$ spectrum stat. unc. & 0.6 & 0.4 & 0.4 & 0.1\\
Cont. $\xp$ spectrum correction & 0.3 & 1.3 & - & -\\
MC statistical unc. & 0.2 & 0.1 & 0.1 & 0.0\\
$r_{\phi}$ & 0.6 & - & 0.6 & -\\
Tracking & 1.1 & 0.7 & 1.1 & 0.7\\
$K/\pi$ identification & 2.3 & 1.4 & 2.3 & 1.4 \\
Integrated luminosity & 1.4 & 1.4 & 1.4 & 1.4\\
Branching fraction & 1.9 & 0.8 & 1.9 & 0.8\\
\hline
Total & 3.6 & 2.6 & 3.7 & 2.5\\
\hline
\end{tabular}
%\end{ruledtabular}
\end{table}

Cross sections $\sigma(\ee\to\bb\to\DX)$ for various $\xp$
bins are presented in Appendix~\ref{app::xp_spectra}.

\subsection{Determination of $\Br(\B \to \DX)$}

The $B \to \DX$ branching fractions are found as 
\beq
\Br(\B\to\DX)=\frac12\;\frac{\sigma(\ee\to\bb\to\DX)|_{\Ufo}}{\sigma(\ee\to\bb)|_{\Ufo}},
\label{eq:b_to_dx}
\eeq
where we use the cross sections measured at the \Ufo. The total cross
section $\sigma(\ee \to \bb)$ is calculated as
\beq
\sigma(\ee \to \bb)|_{\Ufo}=\frac{N^{\Ufo}_{B\bar{B}}}{\mathcal{L}}=(1102 \pm 24)\ {\rm pb},
\eeq
where $N^{\Ufo}_{B\bar{B}}=(619.6 \pm 9.4)\times 10^6$ is the total
number of the $B\bar{B}$ pairs in the \Ufo\ SVD2
data~\cite{Belle:2012iwr} and $\mathcal{L}=562\,\ifb$ is the total
integrated luminosity of this data sample. The number
$N^{\Ufo}_{B\bar{B}}$ is obtained by counting the hadronic events at
the \Ufo\ and subtracting the continuum contribution determined using
the data below the $B\bar{B}$ threshold. The transitions from \Ufo\ to
lower bottomonia have a total branching fraction of
0.26\%\ \cite{ParticleDataGroup:2022pth} and are neglected.

Using the cross section values presented in Table~\ref{tab::cross_sec_at_Y(nS)}, we find 
\begin{eqnarray}
  \Br(\B \to \DnX) &=  & (66.63 \pm 0.04 \pm 1.77) \%,
  \label{eq:br_b_to_d0} \\
  \Br(\B \to \DsX) &= & (11.28 \pm 0.03 \pm 0.43) \%.
  \label{eq:br_b_to_ds} 
\end{eqnarray}
The systematic uncertainty due to the integrated luminosity is the
same in the numerator and denominator of Eq.~\eqref{eq:b_to_dx} and,
therefore, cancels.

The world-average results obtained by a similar method are
$(61.6 \pm 2.9)\%$ for $\D$ and $(8.3 \pm 0.8)\%$ for
$\Ds$~\cite{ParticleDataGroup:2022pth}. Our uncertainties are lower
than those of the world-average values; there is a $3.2\,\sigma$
tension in the $\Ds$ channel.
One can also use for comparison the measurements performed with a full
reconstruction of one $B$ meson in the event~\cite{BaBar:2006wbf}. In
this case, one has to add branching fractions for $B^+$ and $B^0$. The
results are $(71.6 \pm 4.6)\%$ and $(10.4^{+1.3}_{-1.8})\%$; the
agreement with our measurements is better.

\subsection{Production fractions at the $\Ufi$}
 \label{sec::prod_frac}

We determine the average number of the $D$ mesons produced at the \Ufi\ as 
\beq
\Br(\Ufi \to D/\bar{D}X)=\frac{\sigma(\ee \to \bb \to DX)|_{\Ufi}}{\sigma(\ee \to \bb)|_{\Ufi}}.
\eeq
Using the values from Table~\ref{tab::cross_sec_at_Y(nS)} and 
$\sigma(\ee\to\bb)|_{\Upsilon(5S)}=(340\pm16)\,\ipb$~\cite{Belle:2012tsw},
we find
\begin{eqnarray}
   \Br(\Ufi \to \DnX) &= & (111.7 \pm 0.5 \pm 6.0) \%,
  \label{eq:br_Y5S_to_d0} \\
  \Br(\Ufi \to \DsX) &= & (44.7 \pm 0.3 \pm 2.7) \%.
  \label{eq:br_Y5S_to_ds} 
\end{eqnarray}
These results agree with the previous measurements $(108 \pm 8)\%$ for
\D\ and $(46 \pm 6)\%$ for \Ds~\cite{Belle:2006jvm}, and supersede
them.

The fraction of $\BsBsX$ events produced at the \Ufi\ is defined as
\beq
\fs=\frac{\sigma(\ee\to\BsBsX)|_{\Ufi}}{\sigma(\ee \to \bb)|_{\Ufi}},
\label{eq::fs_def}
\eeq 
where $\sigma(\ee \to \BsBsX)$ can be found from the
first equation of \eqref{eq:general}
\beq
\sigma(\ee\to\BsBsX)=\frac{\sigma(\ee\to\bb\to\DsX)/2 -
  \sigma(\ee\to\BBX)\;\Br(B\to\DsX)}{\Br(\Bs\to\DsX)}. 
\eeq
Then, using Eq.~\eqref{eq:b_to_dx} for $\Br(\B \to \DsX)$, we
find 
\beq
\fs=\frac{\sigma(\ee\to\bb\to\DsX)|_{\Ufi} -
  \sigma(\ee\to\BBX)|_{\Ufi}\frac{\sigma(\ee\to\bb\to\DsX)|_{\Ufo}}
        {\sigma(\ee\to\bb)|_{\Ufo}}}
{2\;\Br(\Bs\to\DsX)\;\sigma(\ee\to\bb)|_{\Ufi}}
\label{eq:fs_general}.
\eeq
Using $\sigma(\ee\to\bb\to\DsX)$ from
Table~\ref{tab::cross_sec_at_Y(nS)},
$\Br(\Bs\to\DsX)=(60.2\pm5.8\pm2.3)\%$~\cite{Belle:2021qxu}, and
$\sigma(\ee\to\BBX)=(255.5\pm7.9)$~pb~\cite{Belle:2021lzm}, we obtain: 
\beq \fs = (23.0 \pm 0.2 \pm 2.8)\%.
\label{eq::fs}
\eeq
While estimating the systematic uncertainty in \fs, we take into
account that the systematic uncertainties of the quantities entering
Eq.~\eqref{eq:fs_general} are correlated and to a large extent cancel.
\begin{itemize}
\item The uncertainty due to the integrated luminosity is the same in
  all $\sigma$'s in Eq.~(\ref{eq:fs_general}) and, therefore, cancels
  in $f_s$.
\item Both quantities $\sigma(\ee\to\BBX)|_{\Ufi}$ and
  $\sigma(\ee\to\bb)|_{\Ufo}$ contain the same uncertainty due to
  $N^{\Ufo}_{B\bar{B}}$~\cite{Belle:2021lzm} --- it cancels in their
  ratio.
\item The uncertainties due to the reconstruction efficiency and the
  \Ds-meson branching fraction are fully correlated between the
  inclusive \Ds\ cross sections in the numerator. They are treated as
  common correlated errors for the resulting value of $f_s$, and,
  thus, the total uncertainty in the difference is considerably
  reduced.
\end{itemize}
The individual contribution from each quantity and the correlated contributions
are listed in Table~\ref{tab::syst_fs}. We sum all presented errors
in quadrature to obtain the total systematic uncertainty.

\begin{table}[h!]
  \caption{Systematic uncertainty in $\fs$.}
    \label{tab::syst_fs} 
  %\begin{ruledtabular}
  \begin{center}
    \begin{tabular}{lc}
      \hline
      \centering
      Source & Systematic uncertainty (\%) \\
      \hline
      $\sigma(\ee \to \bb \to \DsX)|_{\Ufi}$      &  \ph1.4     \\
      $\sigma(\ee \to \bb \to \DsX)|_{\Ufo}$      &  \ph0.7      \\
      $\sigma(\ee \to \BBX)|_{\Ufi}$              &   \ph1.4      \\
      $\Br(\Bs \to \DsX)$                        &   10.5      \\
      $\sigma(\ee \to \bb)|_{\Ufi}$                &   \ph4.5       \\
      Correlated contributions                    & \\ 
      \ \ --  tracking                            &  \ph1.1        \\
      \ \ --  $K/\pi$ identification              &   \ph2.3       \\
      \ \ --  $r_{\phi}$                           &  \ph0.6       \\
      \ \ --  $\Br(\Ds \to K^+K^-\pi^+)$         & \ph1.9       \\
      \hline
      Total & 12.0\\
      \hline
    \end{tabular}
    \end{center}
    %\end{ruledtabular}
\end{table}

To improve the accuracy in $f_s$, we use the relation
\beq
\fs + \fBBX + \fnB = 1,
\label{eq:fs_constraint}
\eeq
where
$\fBBX=\sigma(\ee\to\BBX)/\sigma(\ee\to\bb)=(75.1\pm4.0)\%$~\cite{Belle:2021lzm}
is the fraction of the $\BBX$ events at \Ufi\ and $\fnB$ is the
fraction of $\bb$ events without open-bottom mesons in the final
state. The $\fnB$ fraction is due to the transitions to lower
bottomonia with the emission of light hadrons. In
Ref.~\cite{Belle:2021lzm}, it was estimated that the known bottomonium
channels sum up to \beq
\fnB^\mathrm{known} =(4.9\pm 0.6)\%.
\label{eq:f_bottomonium}
\end{equation}
We perform a fit to three measurements: $f_s$, $\fBBX$ and $\fnB$,
applying one constraint -- Eq.~\eqref{eq:fs_constraint}. The free
parameters of this fit are the fitted values
of the production fractions.
%of $f_s$, $\fBBX$ and $\fnB$.
Since potentially not all bottomonium channels are known, we use
Eq.~\eqref{eq:f_bottomonium} as a constraint from below. The
production fractions contain a factor $1/\sigma(\ee\to\bb)|_{\Ufi}$
which results in a correlated uncertainty of
4.5\%\ (Table~\ref{tab::syst_fs}). The presence of
$\sigma(\ee\to\BBX)$ on the right-hand side of
Eq.~\eqref{eq:fs_general} results in an anti-correlated uncertainty in
\fs\ and \fBBX\ of 1.4\%\ and 2.4\%, respectively. The above value for
\fs\ is taken from Table~\ref{tab::syst_fs}; the value for \fBBX\ is
obtained taking into account that the uncertainty in
$\sigma(\ee\to\BBX)$ in Eq.~\eqref{eq:fs_general} partially cancels.
The correlated uncertainties are taken into account using the 
method described in Ref.~\cite{HFLAV:2022pwe}. From the fit, we find
\beq
\fs = (22.0^{+2.0}_{-2.1})\%.
\eeq
This result for \fs\ supersedes the previous Belle measurement
$\fs = (17.2 \pm 3.0)\% $~\cite{Belle:2012tsw} obtained with a
model-dependent estimate $\Br(\Bs \to \DsX)= (92 \pm 11)\%$;
it also supersedes the result $\fs = (28.5 \pm 3.2 \pm 3.7)\%$
reported in Ref.~\cite{Belle:2021qxu}.

\subsection{Determination of $\Br(\Bs\to\DnX)\;/\;\Br(\Bs\to\DsX)$}

The measurements presented in Eqs.~\eqref{eq:br_b_to_d0} and
\eqref{eq:br_b_to_ds}, and in Table~\ref{tab::cross_sec_at_Y(nS)} are
substituted in Eq.~\eqref{eq:general_brs_ratio}; we find
\beq
\frac{\Br(\Bs \to \DnX)}{\Br(\Bs \to \DsX)}=0.416 \pm 0.018 \pm 0.092.
\label{eq::brs_ratio}
\eeq 
As in the case of \fs, here we consider the correlations between
the systematic uncertainties of the quantities in
Eq.~\eqref{eq:general_brs_ratio}.
\begin{itemize}
\item The uncertainty due to integrated luminosity cancels in the ratio of
  the cross sections.
\item The uncertainty due to $N^{\Ufo}_{B\bar{B}}$ cancels in the
  product of $\Br(B\to\DX)$ and $\sigma(\ee\to\BBX)$
  (we note that $\Br(B\to\DX)$ is inversely proportional to
  $N^{\Ufo}_{B\bar{B}}$).
\item The uncertainties due to the reconstruction efficiency and the
  $D$-meson branching fractions are completely correlated between the
  two terms in the numerator of Eq.~\eqref{eq:general_brs_ratio}. This
  correlation considerably reduces the uncertainty in the difference
  of the two terms. The same is true about the denominator.
\item The uncertainty due to tracking efficiency partly cancels
  between the numerator and the denominator (two tracks in the
  numerator and three tracks in the denominator).
\item We conservatively assume that the uncertainties due to the
  particle identification are not correlated between the numerator and
  the denominator because the corresponding momentum spectra of kaons
  and pions are different.
\end{itemize}
%See description in Sec.~\ref{sec::prod_frac} for details.
In Table~\ref{tab::syst_brs_ratio} we list first the uncorrelated
contributions from the quantities in Eq.~\eqref{eq:general_brs_ratio},
and then the correlated contributions; the total uncertainty is the
sum in quadrature of all listed errors.
\begin{table}[h]
  \caption{Systematic uncertainty in $\Br(\Bs \to \DnX)/\Br(\Bs \to \DsX)$.}
    \label{tab::syst_brs_ratio} 
  %\begin{ruledtabular}
  \begin{center}
    \begin{tabular}{lc}
      \hline
      \centering
      Source & Systematic uncertainty (\%) \\
      \hline
      $\sigma(\ee \to \bb \to \DnX)$          &    13.3     \\
      $\sigma(\ee \to \bb \to \DsX)$          &  \ph1.4     \\
      $\Br(\B \to \DnX)$                      &    11.2     \\
      $\Br(\B \to \DsX)$                      &  \ph0.8     \\
      $\sigma(\ee \to \BBX)$                  &    13.7     \\
      Correlated contributions         & \\ 
      \ \ --  tracking                        &  \ph0.4     \\
      \ \ --  $K/\pi$ identification          &  \ph2.7     \\
       \ \ --  $r_{\phi}$                      &  \ph0.6      \\
      \ \ --  $\Br(\Ds \to K^+K^-\pi^+)$       &  \ph1.9    \\
      \ \ -- $\Br(\D \to K^-\pi^+)$            &  \ph0.8      \\
      \hline
      Total & 22.2\\
      \hline
    \end{tabular}
    \end{center}
    %\end{ruledtabular}
\end{table}

The fractions of $B^+B^-$ and $B^0\bar{B}^0$ events at the \Ufo\ and
\Ufi\ are different: at \Ufo\ the ratio of production fractions
$f^{+-}/f^{00} = 1.065\pm 0.012\pm 0.019 \pm
0.047$~\cite{Belle:2022hka}, while at \Ufi\ this ratio is expected to
be close to one since \Ufi\ is far from the $B\bar{B}$ production
thresholds and no isospin violation is expected. Given that the
branching fractions $\Br(B^+\to\DnX)=(87.6\pm4.1)\%$ and
$\Br(B^0\to\DnX)=(55.5\pm3.2)\%$ are considerably
different~\cite{ParticleDataGroup:2022pth}, we expect $\Br(B\to\DnX)$
at the \Ufi\ to be $(0.71\pm0.54)\%$ lower than at the \Ufo, here $B$
denotes the relevant mixture of $B^+$ and $B^0$. The effect is small
and is neglected.

\section{Energy scan data}

The analysis strategy described previously in
Sec.~\ref{sec::5s_and_4s} for the \Ufi\ and \Ufo\ data is now applied
at each energy point. We fit the mass distributions of the $\Ds$ and
$\D$ candidates in each \xp\ bin. As in the \Ufi\ and \Ufo\ analysis,
the signal function is a sum of the four Gaussians with parameters
obtained from fitting the MC sample. The shift and the broadening
factor, introduced to describe the signal in the data, are common for
all the Gaussians. They are fixed to the values obtained from fitting
the \Ufi\ data sample for all energy points, except the three with the
largest luminosity. At these three points, near the \Ufi\ resonance,
the shift and broadening factor are allowed to vary freely. The
background is fitted by a second-order polynomial.

We use the \xp\ spectra for the data collected below the $B\bar{B}$
threshold, shown in Fig.~\ref{fig::xp} (bottom), to subtract the
continuum contribution at each energy point. First, the continuum
$\xp$ spectrum for the \Ds\ and \D\ mesons is corrected for the energy
difference between $\ecm$ = 10.52 GeV and the energy of the relevant
point. As before, these corrections are obtained using the Belle II
event generator. Then, the shape of the corrected continuum $\xp$
spectrum is used in the fit to the high-momentum region of the $\xp$
distribution to obtain the normalization factor $k$ for the continuum
contribution at the current energy. The results of these fits are
provided in Appendix~\ref{app::xp_fit_for_scan}.

The inclusive $\ee\to\bb\to\DsX$ and $\ee\to\bb\to\DnX$ cross sections
are calculated according to formulas~\eqref{formula::ds_cross_sec} and
\eqref{formula::d0_cross_sec} with $\imax=11$ and $\imax=12$,
respectively (Table~\ref{tab::xp}). The obtained values of the cross
sections are listed in Table~\ref{tab::all_cross_section} and shown in
Fig.~\ref{fig::inclusive_ee_bb_dx}. The errors in this figure are
statistical only; they are calculated according to
Eq.~\eqref{eq::stat_error_ds_5s}.

\begin{figure}[htbp]
  \centering
 \begin{tabular}{cc}
 \includegraphics[width=0.49\linewidth]{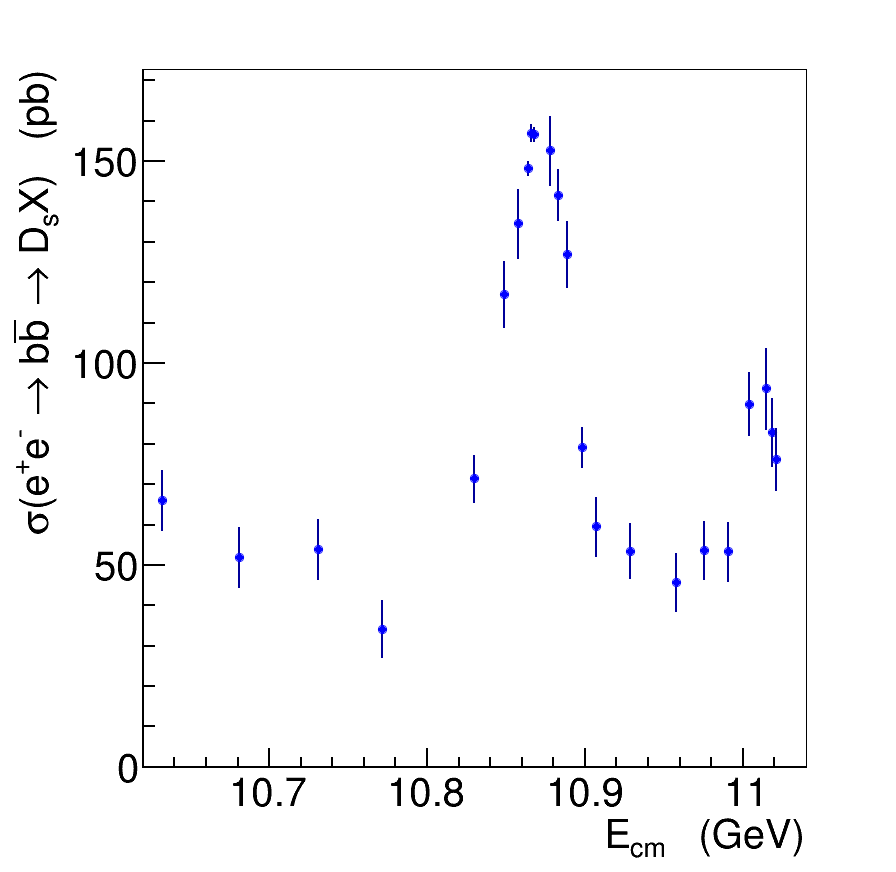} &
 \includegraphics[width=0.49\linewidth]{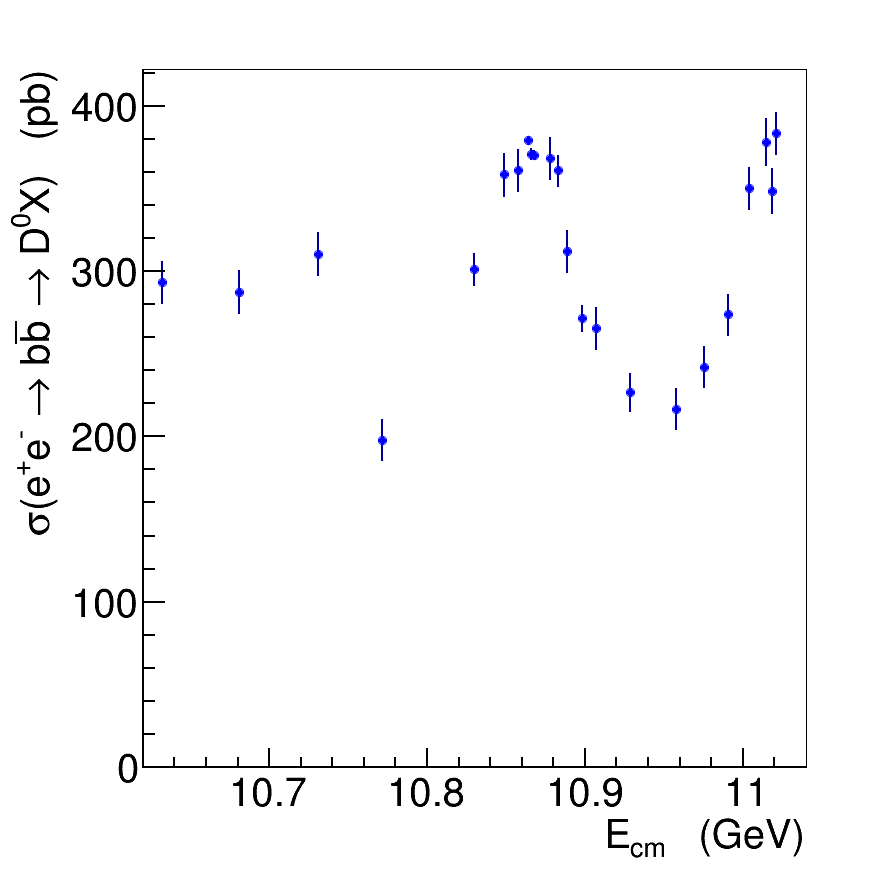} 
 \end{tabular} 
 \caption{The energy dependence of the inclusive $\ee\to\bb\to\DsX$
   (left) and $\ee\to\bb\to\DnX$ (right) cross sections.}
 \label{fig::inclusive_ee_bb_dx}
\end{figure}

We consider the same sources of systematic uncertainty as listed in
Table~\ref{tab::syst_sum}. We assume that the systematic uncertainties
are fully correlated at the various energy points and find three types
of the energy dependence of the systematic uncertainties:
\begin{itemize}
\item The contribution of the statistical uncertainty of the continuum
  \xp\ spectrum is additive and is almost energy-independent
  (Fig.~\ref{fig::uncert_in_cross_sec}, blue points).
\begin{figure}[htbp]
 \centering
 \begin{tabular}{cc}
   \includegraphics[width=0.49\linewidth]{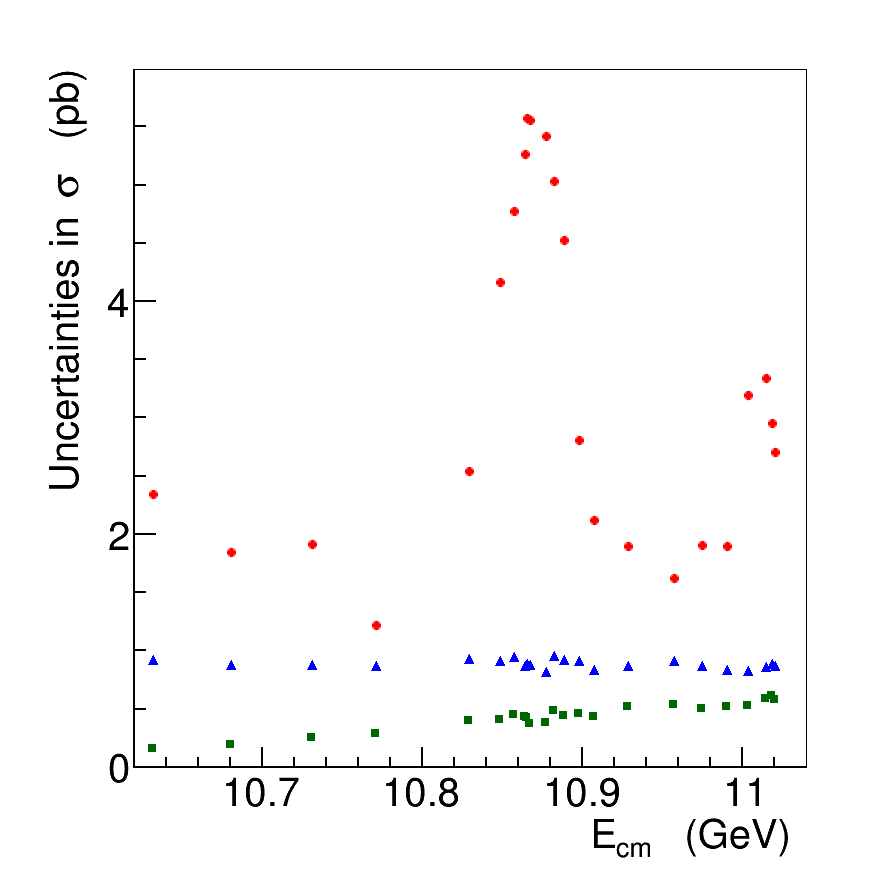} &
   \includegraphics[width=0.49\linewidth]{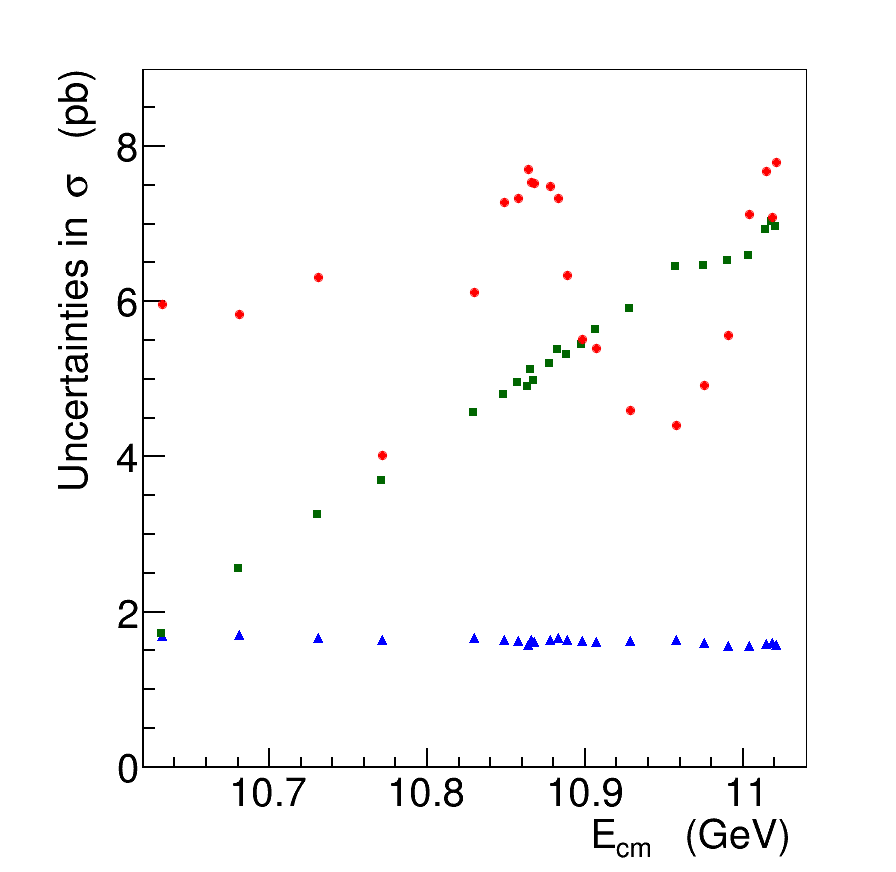}
 \end{tabular} 
 \caption{The energy dependence of the absolute systematic
   uncertainties in the $\ee \to \bb \to \Ds X$ (left) and $\ee \to
   \bb \to \D X$ (right) cross sections. Shown are contributions from
   the statistical uncertainty of the continuum \xp\ spectrum
   (blue triangles), the continuum \xp\ spectrum correction (green
   squares), and the sum of the other sources (red circles). The
   vertical scale is chosen to coincide with the maximal statistical
   uncertainty in the corresponding cross section.}
 \label{fig::uncert_in_cross_sec}
\end{figure}
\item The contribution of the \xp\ spectrum correction is additive and shows a
  linear rise with energy (Fig.~\ref{fig::uncert_in_cross_sec}, green points).
\item Other sources are multiplicative; their relative uncertainties
  are energy-independent. Their shapes repeat those of the cross
  sections themselves (Fig.~\ref{fig::uncert_in_cross_sec}, red
  points).
\end{itemize}
In Table~\ref{tab::all_cross_section} the contributions of these three
types of systematic uncertainties are summed in quadrature.

\begin{table}[htbp]
   \caption{Energies (in GeV), luminosities (in \ifb) for various data
     samples and the results for the $\sigma(\ee\to\bb\to\DsX)$,
     $\sigma(\ee\to\bb\to\DnX)$, $\sigma(\ee \to \BsBsX)\cdot\Br(\Bs\to \DsX)$, and $\sigma(\ee \to \BBX)$ (in
     pb). The first error in the cross section is statistical, the
     second is systematic.}
\label{tab::all_cross_section} 
  \begin{adjustbox}{max width=\textwidth}
  \begin{tabular}{cccccc}
  \hline
\centering 
\ecm & $\mathcal{L}$ & $\sigma(\Ds X)$ & $\sigma(\D X)$ & ~~~$\sigma(\BsBsX)\cdot \Br$ & $\sigma(\BBX)$ \\
\hline
10.6322   &  $\ph$0.989 & $ \ph65.4 \pm 7.4 \pm 2.5 $  & $ 298.5 \pm   12.9 \pm \ph7.2 $ & $ \ph\phantom{-~}8.0 \pm 4.2 \pm 0.7 $  & $ 219.0 \pm 10.7 \pm 3.2 $\\ 
10.6810   &  $\ph$0.949 & $ \ph51.3 \pm 7.4 \pm 2.1 $  & $ 292.2 \pm   13.2 \pm \ph7.3 $ & $ \ph\phantom{-~}1.0 \pm 4.2 \pm 0.8 $  & $ 218.7 \pm 10.9 \pm 3.6 $\\ 
10.7313   &  $\ph$0.946 & $ \ph53.4 \pm 7.4 \pm 2.1 $  & $ 315.7 \pm   13.2 \pm \ph8.0 $ & $            \ph-0.0 \pm 4.2 \pm 0.8 $  & $ 236.9 \pm 10.9 \pm 4.1 $\\ 
10.7712   &  $\ph$0.955 & $ \ph33.9 \pm 7.0 \pm 1.5 $  & $ 201.2 \pm   12.8 \pm \ph6.1 $ & $            \ph-0.1 \pm 3.9 \pm 0.7 $  & $ 151.0 \pm 10.6 \pm 3.8 $\\ 
10.8295   &  $\ph$1.697 & $ \ph70.8 \pm 5.8 \pm 2.7 $  & $ 306.2 \pm \ph9.8 \pm \ph8.5 $ & $   \phantom{-~}10.2 \pm 3.2 \pm 0.9 $  & $ 223.4 \pm \ph8.1 \pm 4.8 $\\ 
10.8489   &  $\ph$0.989 & $   116.0 \pm 8.2 \pm 4.3 $  & $ 364.4 \pm   13.0 \pm \ph9.7 $ & $   \phantom{-~}29.2 \pm 4.6 \pm 1.5 $  & $ 255.2 \pm 10.9 \pm 6.3 $\\ 
10.8574   &  $\ph$0.988 & $   133.3 \pm 8.4 \pm 4.9 $  & $ 366.9 \pm   13.0 \pm \ph9.8 $ & $   \phantom{-~}38.3 \pm 4.7 \pm 1.8 $  & $ 251.4 \pm 10.9 \pm 7.4 $\\ 
10.8642   &      47.648 & $   146.9 \pm 1.7 \pm 5.4 $  & $ 385.3 \pm \ph2.6 \pm   10.1 $ & $   \phantom{-~}43.9 \pm 0.9 \pm 2.0 $  & $ 261.7 \pm \ph2.2 \pm 8.0 $\\ 
10.8658   &      29.107 & $   155.6 \pm 2.1 \pm 5.7 $  & $ 376.8 \pm \ph3.3 \pm   10.1 $ & $   \phantom{-~}49.4 \pm 1.2 \pm 2.2 $  & $ 251.9 \pm \ph2.8 \pm 8.9 $\\ 
10.8676   &      45.284 & $   155.2 \pm 1.7 \pm 5.6 $  & $ 376.3 \pm \ph2.7 \pm   10.0 $ & $   \phantom{-~}49.2 \pm 0.9 \pm 2.1 $  & $ 251.6 \pm \ph2.3 \pm 8.8 $\\ 
10.8778   &  $\ph$0.978 & $   151.2 \pm 8.6 \pm 5.5 $  & $ 374.5 \pm   13.2 \pm   10.1 $ & $   \phantom{-~}47.2 \pm 4.8 \pm 2.1 $  & $ 251.5 \pm 11.0 \pm 8.6 $\\ 
10.8828   &  $\ph$1.848 & $   140.3 \pm 6.2 \pm 5.1 $  & $ 367.0 \pm \ph9.6 \pm   10.0 $ & $   \phantom{-~}42.1 \pm 3.4 \pm 1.9 $  & $ 249.1 \pm \ph8.0 \pm 8.0 $\\ 
10.8889   &  $\ph$0.990 & $   125.8 \pm 8.2 \pm 4.6 $  & $ 317.0 \pm   12.9 \pm \ph9.1 $ & $   \phantom{-~}38.8 \pm 4.6 \pm 1.8 $  & $ 213.6 \pm 10.8 \pm 7.6 $\\ 
10.8983   &  $\ph$2.408 & $ \ph78.4 \pm 4.9 \pm 3.0 $  & $ 276.0 \pm \ph8.2 \pm \ph8.4 $ & $   \phantom{-~}17.0 \pm 2.7 \pm 1.1 $  & $ 196.5 \pm \ph6.8 \pm 5.4 $\\ 
10.9073   &  $\ph$0.980 & $ \ph59.0 \pm 7.3 \pm 2.3 $  & $ 269.8 \pm   12.7 \pm \ph8.5 $ & $ \ph\phantom{-~}7.2 \pm 4.1 \pm 0.8 $  & $ 198.0 \pm 10.5 \pm 5.3 $\\ 
10.9287   &  $\ph$1.149 & $ \ph53.0 \pm 6.7 \pm 2.1 $  & $ 230.3 \pm   11.6 \pm \ph8.0 $ & $ \ph\phantom{-~}7.5 \pm 3.8 \pm 0.9 $  & $ 168.1 \pm \ph9.6 \pm 5.3 $\\ 
10.9575   &  $\ph$0.969 & $ \ph45.3 \pm 7.2 \pm 1.9 $  & $ 220.2 \pm   12.7 \pm \ph8.3 $ & $ \ph\phantom{-~}4.3 \pm 4.0 \pm 0.9 $  & $ 162.5 \pm 10.5 \pm 5.7 $\\ 
10.9753   &  $\ph$0.999 & $ \ph53.2 \pm 7.1 \pm 2.2 $  & $ 246.2 \pm   12.5 \pm \ph8.7 $ & $ \ph\phantom{-~}6.2 \pm 4.0 \pm 1.0 $  & $ 180.9 \pm 10.4 \pm 5.8 $\\ 
10.9904   &  $\ph$0.985 & $ \ph52.9 \pm 7.2 \pm 2.1 $  & $ 278.3 \pm   12.7 \pm \ph9.2 $ & $ \ph\phantom{-~}3.1 \pm 4.0 \pm 0.9 $  & $ 206.9 \pm 10.5 \pm 5.9 $\\ 
11.0039   &  $\ph$0.976 & $ \ph89.0 \pm 7.8 \pm 3.3 $  & $ 356.2 \pm   13.0 \pm   10.5 $ & $   \phantom{-~}15.4 \pm 4.4 \pm 1.1 $  & $ 257.6 \pm 10.8 \pm 6.4 $\\ 
11.0148   &  $\ph$0.771 & $ \ph92.9 \pm 9.9 \pm 3.5 $  & $ 384.5 \pm   14.7 \pm   11.2 $ & $   \phantom{-~}15.0 \pm 5.5 \pm 1.2 $  & $ 279.2 \pm 12.3 \pm 6.7 $\\ 
11.0185   &  $\ph$0.859 & $ \ph82.1 \pm 8.4 \pm 3.1 $  & $ 354.3 \pm   13.9 \pm   10.8 $ & $   \phantom{-~}11.9 \pm 4.7 \pm 1.2 $  & $ 258.4 \pm 11.6 \pm 6.6 $\\ 
11.0208   &  $\ph$0.982 & $ \ph75.5 \pm 7.6 \pm 2.9 $  & $ 390.0 \pm   13.0 \pm   11.4 $ & $ \ph\phantom{-~}5.1 \pm 4.3 \pm 1.1 $  & $ 289.5 \pm 10.8 \pm 6.7 $\\ 
\hline
  \end{tabular}
  \end{adjustbox}
\end{table}

Substituting the obtained values of $\Br(B\to\DnX)$, $\Br(B\to\DsX)$,
and $\Br(\Bs\to\DnX)\;/\;\Br(\Bs\to\DsX)$
(Eqs.~\eqref{eq:br_b_to_d0},~\eqref{eq:br_b_to_ds} and
\eqref{eq::brs_ratio}) in
Eqs.~\eqref{eq:solv_general12}, we find
\begin{align}
    X & = \phantom{-}0.54 \cdot U - 0.09 \cdot W,\nn\\[-2mm]
  \label{eq:num_solve}\\[-2mm]
     Y & = -0.34  \cdot U + 0.81 \cdot W.\nn
\end{align}
The results for $X=\sigma(\ee\to\BsBsX)\cdot\Br(\Bs\to\DsX)$ and
$Y=\sigma(\ee\to\BBX)$
%, calculated using Eqs.~\eqref{eq:num_solve},
are presented in Table~\ref{tab::all_cross_section} and in
Fig.~\ref{fig::inclusive_ee_bx}.
\begin{figure}[h!]
  \centering
  \begin{tabular}{cc}
    \includegraphics[width=0.49\linewidth]{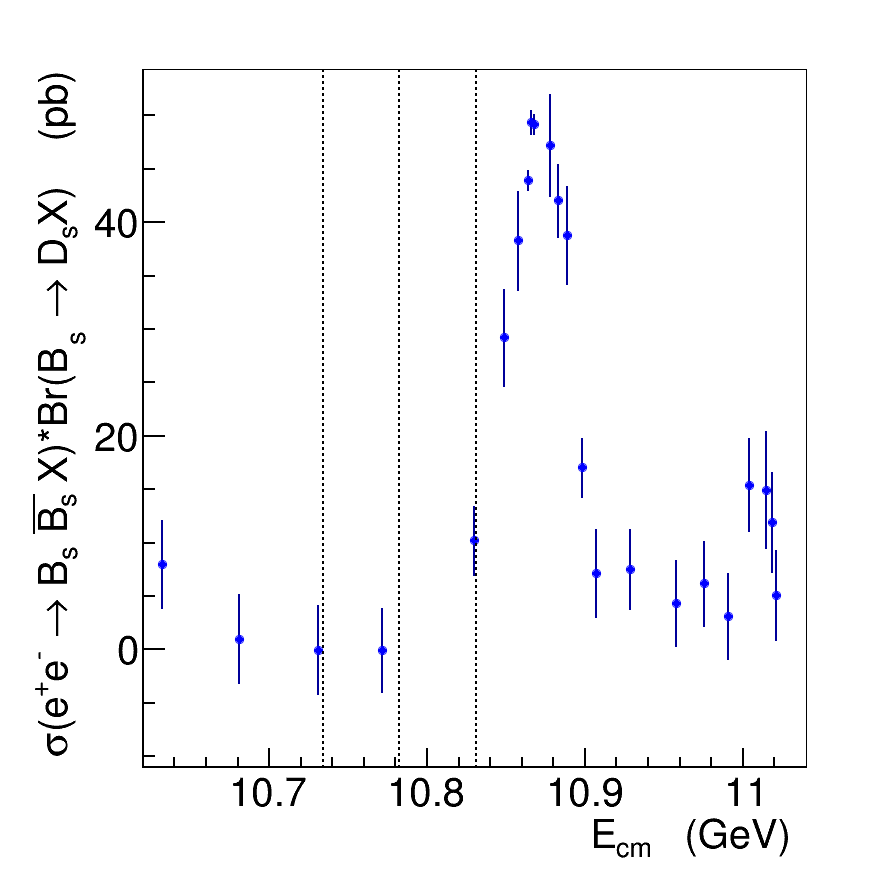} &
    \includegraphics[width=0.49\linewidth]{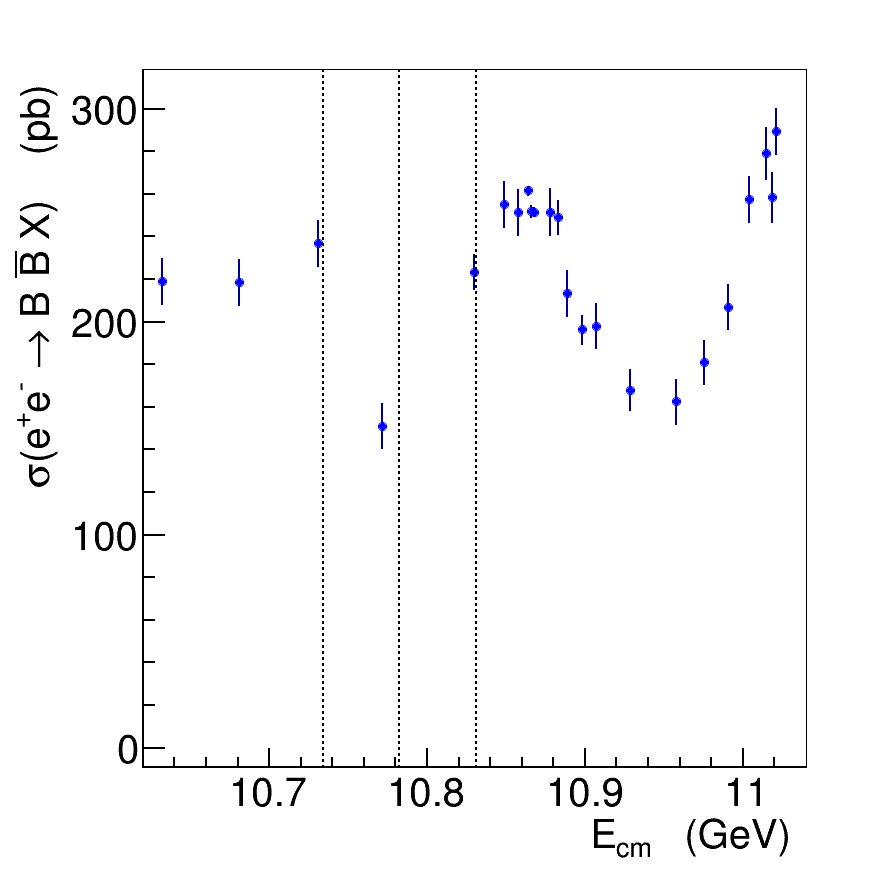} 
  \end{tabular} 
  \caption{The energy dependence of the product
    $\sigma(\ee\to\BsBsX)\cdot\Br(\Bs\to\DsX)$ (left) and the
    $\sigma(\ee\to\BBX)$ (right). Shown are statistical uncertainties
    calculated using Eq.~\eqref{eq:num_solve} based on the statistical
    uncertainties of $U=\sigma(\ee\to\bb\to\DsX)$ and
    $W=\sigma(\ee\to\bb\to\DnX)$. The dashed lines indicate the
    $\Bs\BsB$, $\Bs\bar{B}_s^{*}$ and $B_s^{*}\bar{B}_s^{*}$
    thresholds.}
     \label{fig::inclusive_ee_bx}
\end{figure}
The $\ee\to\BsBsX$ cross section shows a clear peak near the
\Ufi\ energy and a hint of a peak near the $\Upsilon(6S)$, while the
non-resonant contribution is small.

We separate the systematic uncertainties of the quantities entering
Eqs.~\eqref{eq:solv_general12} into correlated and uncorrelated
parts, similarly as for the \fs\ (Table
\ref{tab::syst_fs}). We add all the contributions in quadrature to
obtain the total systematic uncertainty shown in
Table~\ref{tab::all_cross_section}.

\section{Conclusions}

We have measured the inclusive cross sections
$\sigma(\ee\to\bb\to\DsX)$, $\sigma(\ee\to\bb\to\DnX)$,
$\sigma(\ee\to\BBX)$ and the product
$\sigma(\ee\to\BsBsX)\cdot\Br(\Bs\to\DsX)$ in the energy range from
$10.63$ to $11.02\,\gev$. Results are presented in
Table~\ref{tab::all_cross_section}. The energy dependence of the
$\ee\to\BsBsX$ cross section shows a clear peak near the $\Ufi$ energy
and a hint of a peak near the $\Upsilon(6S)$.
The obtained results can be used in a combined analysis of the data in
various final states within coupled-channel approaches to investigate
the nature and properties of the bottomonium and bottomonium-like
states lying above the $B\bar{B}$ threshold.

We have measured the following inclusive branching fractions and
production fractions: 
\begin{alignat}{3}
\Br(\B \to \DsX) & =  (11.28 \pm 0.03 \pm 0.43) \%,\\
\Br(\B \to \DnX) & = (66.63 \pm 0.04 \pm 1.77) \%,\\
\Br(\Ufi \to \DsX) & = (44.7 \pm 0.3 \pm 2.7) \%, \label{eq:Ufi_DsX} \\
\Br(\Ufi \to \DnX) & = (111.7 \pm 0.5 \pm 6.0) \%. \label{eq:Ufi_DnX}
\end{alignat}
There might be several $D$ mesons in $B$ decays and $\bb$ events; the
measurements correspond to the average multiplicities.
The results shown in Eqs.~\eqref{eq:Ufi_DsX} and \eqref{eq:Ufi_DnX}
supersede previous Belle measurements reported in
Ref.~\cite{Belle:2006jvm}.

The fraction of the events containing the \Bs\ mesons at the $\Ufi$ is
found to be
\begin{align}
 (22.0^{+2.0}_{-2.1})\%.
\end{align}
This value supersedes the previous Belle results reported in
Refs.~\cite{Belle:2012tsw} and ~\cite{Belle:2021qxu}.

We also determined the ratio of the \Bs\ branching fractions
\begin{align}
  %\Br(\Bs\to\DnX)\;/\;\Br(\Bs\to\DsX) = 0.415\pm0.018\pm0.092.
\frac{\Br(\Bs\to\DnX)}{\Br(\Bs\to\DsX)} = 0.416\pm0.018\pm0.092.
\end{align}

The inclusive method allows to measure energy dependence of the
$\ee\to\BsBsX$ cross section with relatively high precision even if
relatively low integrated luminosity is available. It can be used
by the Belle~II experiment for exploratory studies of various energy
regions of interest, for example, near the $B_s^{(*)}\bar{B}_s^{(*)}$
production thresholds or to search for $P$-wave $B_{s0}^0$ and
$B_{s1}^0$ states via $\sigma(\ee \to\BsBsX)$ enhancements at the
$B_{sJ}\bar{B}_s^{(*)}$ thresholds~\cite{Bondar:2016hva}. Additional
advantage of the method is that the inclusive $\ee\to\BBX$ cross
section is also determined.

\acknowledgments This work, based on data collected using the Belle
detector, which was operated until June 2010, was supported by the
Ministry of Education, Culture, Sports, Science, and Technology (MEXT)
of Japan, the Japan Society for the Promotion of Science (JSPS), and
the Tau-Lepton Physics Research Center of Nagoya University; the
Australian Research Council including grants DP210101900, % Urquijo
DP210102831, % Sevior DE220100462, % Hsu LE210100098, % Infrastructure
LE230100085; % Infrastructure Austrian Federal Ministry of Education,
Science and Research (FWF) and FWF Austrian Science Fund
No.~P~31361-N36; the National Natural Science Foundation of China
under Contracts No.~11675166, %Wen-Biao Yan No.~11705209; %Yi-Ming Li
No.~11975076; %Chengping Shen No.~12135005; %Chengping Shen
No.~12175041; %Xiaolong Wang No.~12161141008; %Chengping Shen Key
Research Program of Frontier Sciences, Chinese Academy of Sciences
(CAS), Grant No.~QYZDJ-SSW-SLH011; % Chang-Zheng Yuan Project
ZR2022JQ02 supported by Shandong Provincial Natural Science
Foundation; the Ministry of Education, Youth and Sports of the Czech
Republic under Contract No.~LTT17020; the Czech Science Foundation
Grant No. 22-18469S; Horizon 2020 ERC Advanced Grant No.~884719 and
ERC Starting Grant No.~947006 ``InterLeptons'' (European Union); the
Carl Zeiss Foundation, the Deutsche Forschungsgemeinschaft, the
Excellence Cluster Universe, and the VolkswagenStiftung; the
Department of Atomic Energy (Project Identification No. RTI 4002) and
the Department of Science and Technology of India; the Istituto
Nazionale di Fisica Nucleare of Italy; National Research Foundation
(NRF) of Korea Grant Nos.~2016R1\-D1A1B\-02012900,
2018R1\-A2B\-3003643, 2018R1\-A6A1A\-06024970, RS\-2022\-00197659,
2019R1\-I1A3A\-01058933, 2021R1\-A6A1A\-03043957,
2021R1\-F1A\-1060423, 2021R1\-F1A\-1064008, 2022R1\-A2C\-1003993;
Radiation Science Research Institute, Foreign Large-size Research
Facility Application Supporting project, the Global Science
Experimental Data Hub Center of the Korea Institute of Science and
Technology Information and KREONET/GLORIAD; the Polish Ministry of
Science and Higher Education and the National Science Center; the
Ministry of Science and Higher Education of the Russian Federation,
Agreement 14.W03.31.0026, % from 15.02.2018 and the HSE University
Basic Research Program, Moscow; % from 15.04.2021 University of Tabuk
research grants S-1440-0321, S-0256-1438, and S-0280-1439 (Saudi
Arabia); the Slovenian Research Agency Grant Nos. J1-9124 and P1-0135;
Ikerbasque, Basque Foundation for Science, Spain; the Swiss National
Science Foundation; the Ministry of Education and the Ministry of
Science and Technology of Taiwan; and the United States Department of
Energy and the National Science Foundation.  These acknowledgements
are not to be interpreted as an endorsement of any statement made by
any of our institutes, funding agencies, governments, or their
representatives.  We thank the KEKB group for the excellent operation
of the accelerator; the KEK cryogenics group for the efficient
operation of the solenoid; and the KEK computer group and the Pacific
Northwest National Laboratory (PNNL) Environmental Molecular Sciences
Laboratory (EMSL) computing group for strong computing support; and
the National Institute of Informatics, and Science Information NETwork
6 (SINET6) for valuable network support.

\bibliographystyle{JHEP}
\bibliography{draft_ee_bsbsx_v1}

\newpage
\appendix

\section{$\xp$ spectra of \Ds\ and \D\ at the \Ufo\ and \Ufi\ resonances}
\label{app::xp_spectra}
To determine $\sigma(\ee\to\bb\to D/\bar{D} X)$ for various $\xp$
intervals, we use Eqs.~\eqref{formula::ds_cross_sec},
\eqref{formula::d0_cross_sec}, and \eqref{eq::stat_error_ds_5s}
without summing over index $i$. The results are presented in
Table~\ref{tab::number_of_D}. We consider the same sources of the
systematic uncertainty as shown in Table~\ref{tab::syst_sum}.

\begin{table}[htbp]
  \caption{Cross sections $\sigma(\ee\to\bb\to DX)$ for various
    \xp\ intervals at the \Ufi\ and \Ufo\ resonances. The first and
    second uncertainties are statistical for on-resonance and
    continuum data, respectively; the third uncertainty is
    multiplicative systematic.}
\label{tab::number_of_D} 
  \begin{adjustbox}{max width=\textwidth}
  \begin{tabular}{ccccc}
  \hline
\centering 
\xp\ interval & \Ds\ at the \Ufi\ & \D\ at the \Ufi\ & \Ds\ at the \Ufo\ & \D\ at the \Ufo\ \\
\hline
(0.00,0.05) & $ 0.73 \pm 0.08 \pm 0.00 \pm 0.06 $  & $ 4.10 \pm 0.17 \pm 0.00 \pm 0.17 $ & $ 1.88 \pm 0.06 \pm 0.06 \pm 0.11 $  & $ 15.69 \pm 0.09 \pm 0.00 \pm 0.42 $ \\ 
(0.05,0.10) & $ 5.09 \pm 0.27 \pm 0.00 \pm 0.24 $  & $ 24.29 \pm 0.39 \pm 0.00 \pm 0.70 $ & $ 9.70 \pm 0.15 \pm 0.15 \pm 0.40 $  & $ 86.67 \pm 0.21 \pm 0.00 \pm 2.19 $ \\ 
(0.10,0.15) & $ 9.49 \pm 0.34 \pm 0.00 \pm 0.41 $  & $ 45.94 \pm 0.53 \pm 0.01 \pm 1.27 $ & $ 17.02 \pm 0.20 \pm 0.20 \pm 0.68 $  & $ 162.73 \pm 0.29 \pm 0.00 \pm 4.09 $ \\ 
(0.15,0.20) & $ 16.28 \pm 0.40 \pm 0.00 \pm 0.64 $  & $ 60.70 \pm 0.59 \pm 0.01 \pm 1.65 $ & $ 23.77 \pm 0.22 \pm 0.22 \pm 0.92 $  & $ 214.84 \pm 0.32 \pm 0.01 \pm 5.40 $ \\ 
(0.20,0.25) & $ 23.13 \pm 0.40 \pm 0.01 \pm 0.89 $  & $ 61.56 \pm 0.57 \pm 0.02 \pm 1.74 $ & $ 32.08 \pm 0.22 \pm 0.22 \pm 1.23 $  & $ 239.29 \pm 0.33 \pm 0.01 \pm 6.03 $ \\ 
(0.25,0.30) & $ 29.04 \pm 0.39 \pm 0.01 \pm 1.11 $  & $ 60.81 \pm 0.55 \pm 0.03 \pm 1.70 $ & $ 48.40 \pm 0.21 \pm 0.21 \pm 1.82 $  & $ 233.92 \pm 0.31 \pm 0.01 \pm 5.89 $ \\ 
(0.30,0.35) & $ 30.85 \pm 0.36 \pm 0.01 \pm 1.16 $  & $ 49.63 \pm 0.51 \pm 0.04 \pm 1.41 $ & $ 72.22 \pm 0.21 \pm 0.21 \pm 2.67 $  & $ 223.39 \pm 0.30 \pm 0.02 \pm 5.62 $ \\ 
(0.35,0.40) & $ 22.53 \pm 0.32 \pm 0.02 \pm 0.87 $  & $ 35.31 \pm 0.48 \pm 0.06 \pm 1.07 $ & $ 41.72 \pm 0.18 \pm 0.18 \pm 1.57 $  & $ 167.71 \pm 0.27 \pm 0.03 \pm 4.24 $ \\ 
(0.40,0.45) & $ 11.65 \pm 0.27 \pm 0.02 \pm 0.51 $  & $ 21.41 \pm 0.44 \pm 0.08 \pm 0.76 $ & $ 1.85 \pm 0.12 \pm 0.12 \pm 0.30 $  & $ 96.56 \pm 0.24 \pm 0.03 \pm 2.48 $ \\ 
(0.45,0.50) & $ 3.27 \pm 0.23 \pm 0.03 \pm 0.29 $  & $ 10.61 \pm 0.41 \pm 0.10 \pm 0.56 $ & $ -0.07 \pm 0.16 \pm 0.16 \pm 0.41 $  & $ 26.65 \pm 0.21 \pm 0.04 \pm 0.85 $ \\ 
(0.50,0.55) & $ -0.22 \pm 0.21 \pm 0.04 \pm 0.25 $  & $ 3.84 \pm 0.39 \pm 0.11 \pm 0.48 $ & $ - $  & $ 1.08 \pm 0.17 \pm 0.04 \pm 0.45 $ \\ 
(0.55,0.60) & $ - $  & $ 1.53 \pm 0.36 \pm 0.11 \pm 0.44 $ & $ - $  & $ - $ \\ 
 \hline
  \end{tabular}
  \end{adjustbox}
\end{table}

\newpage
\section{Fits to the \xp\ distributions at the scan energies}
\label{app::xp_fit_for_scan}
      The fits to the \xp\ distributions at various energies are shown
      in Fig.~\ref{fig::xp_fit_for_scan_ds} for \Ds\ and
      Fig.~\ref{fig::xp_fit_for_scan_d0} for \D.
\begin{figure}[h!]
  \begin{adjustbox}{max width=\textwidth}
  \centering
  \begin{tabular}{ccccc}
    \includegraphics[width=0.20\linewidth]{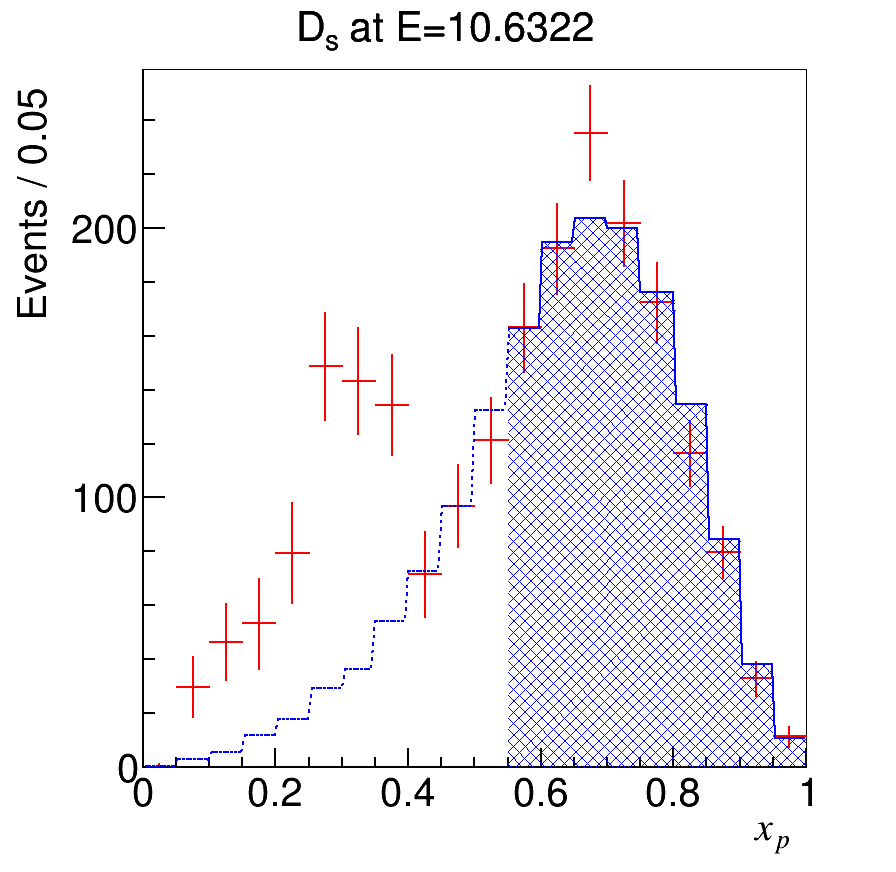} &
    \includegraphics[width=0.20\linewidth]{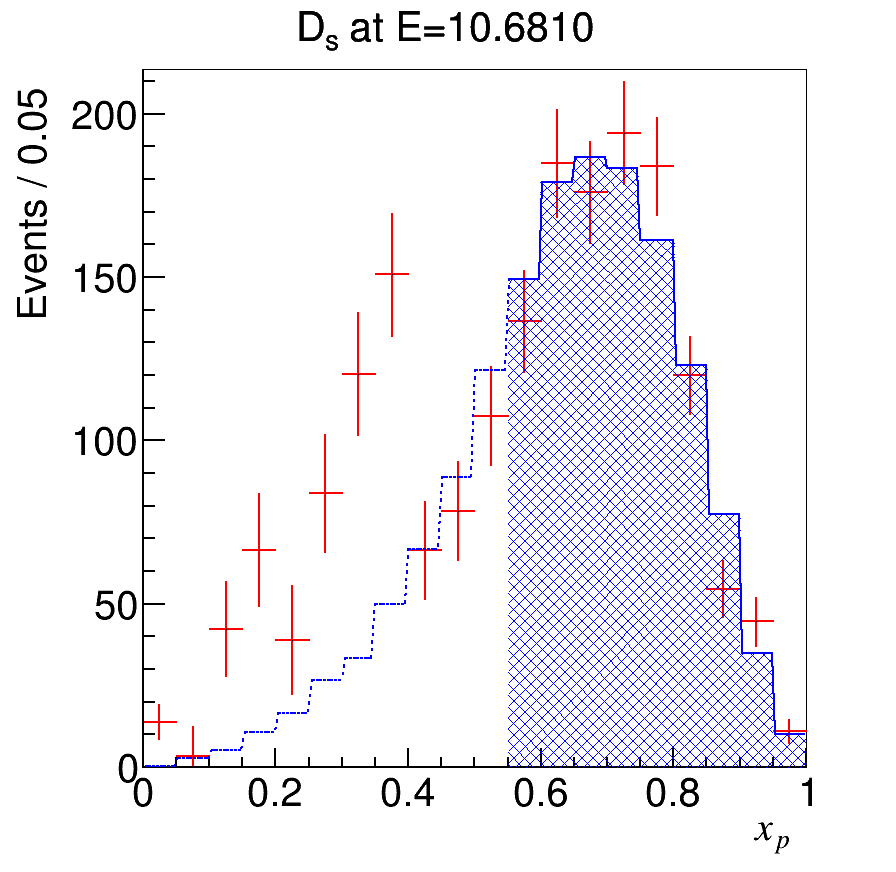} &
    \includegraphics[width=0.20\linewidth]{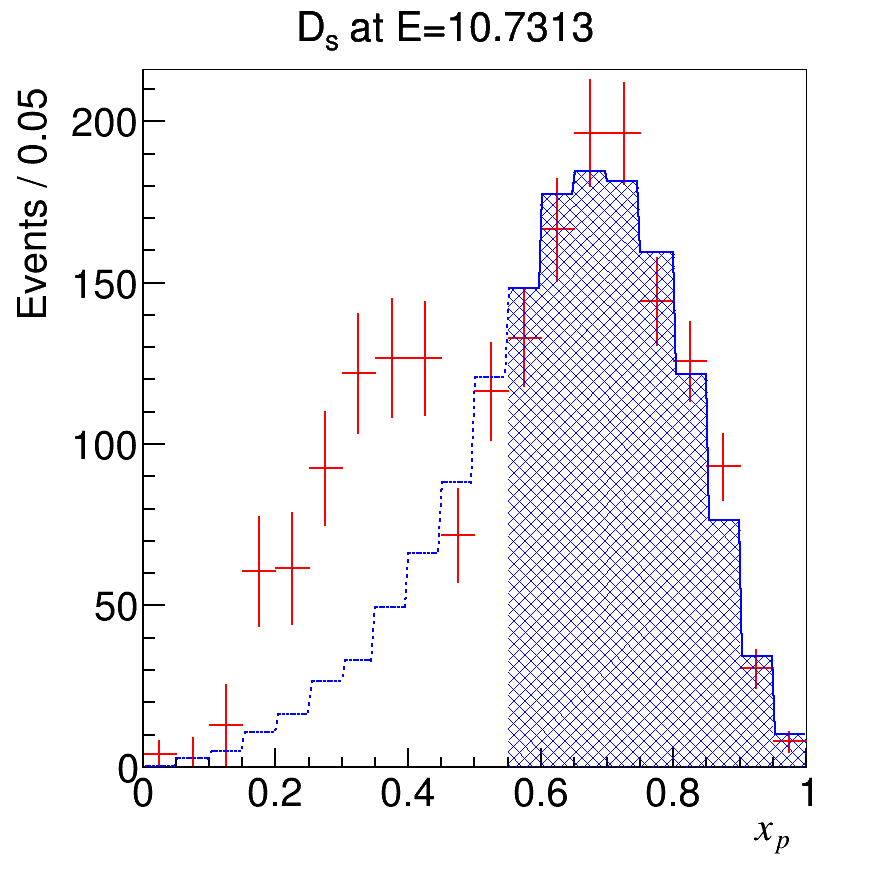} &
    \includegraphics[width=0.20\linewidth]{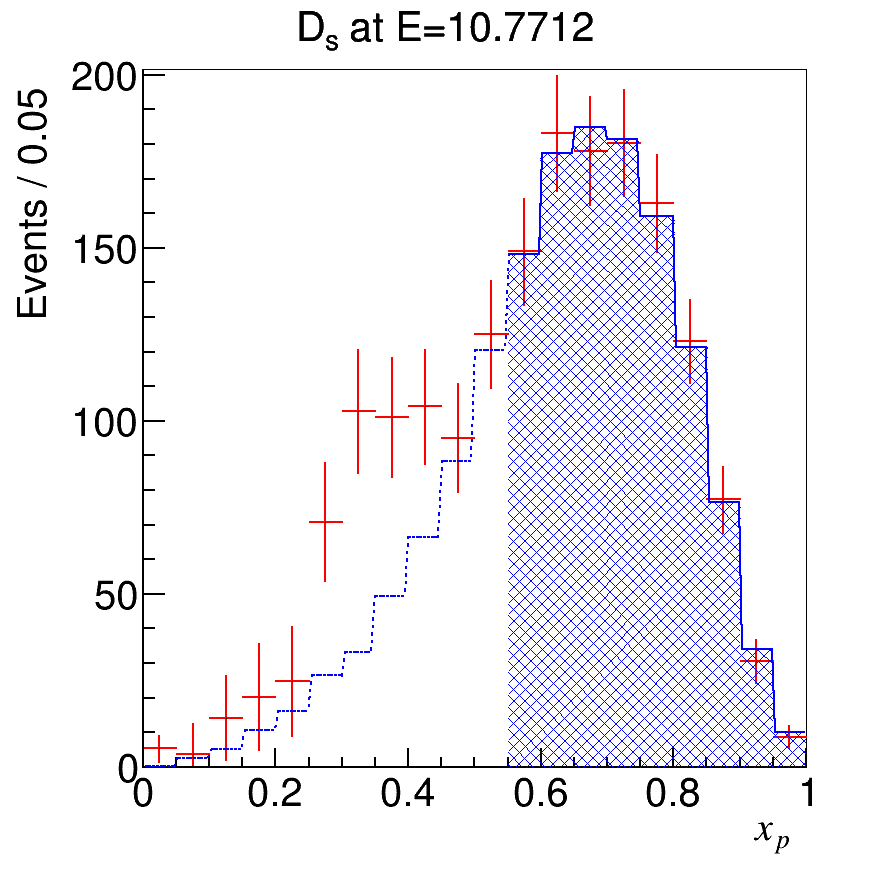} &
    \includegraphics[width=0.20\linewidth]{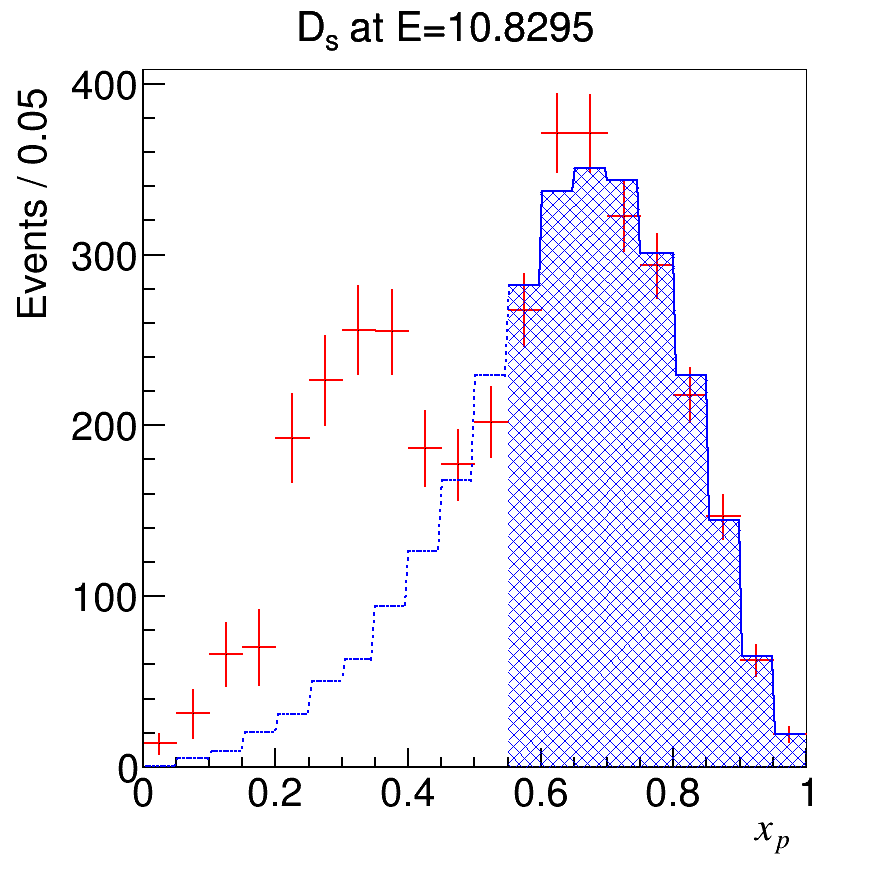} \\
    \includegraphics[width=0.20\linewidth]{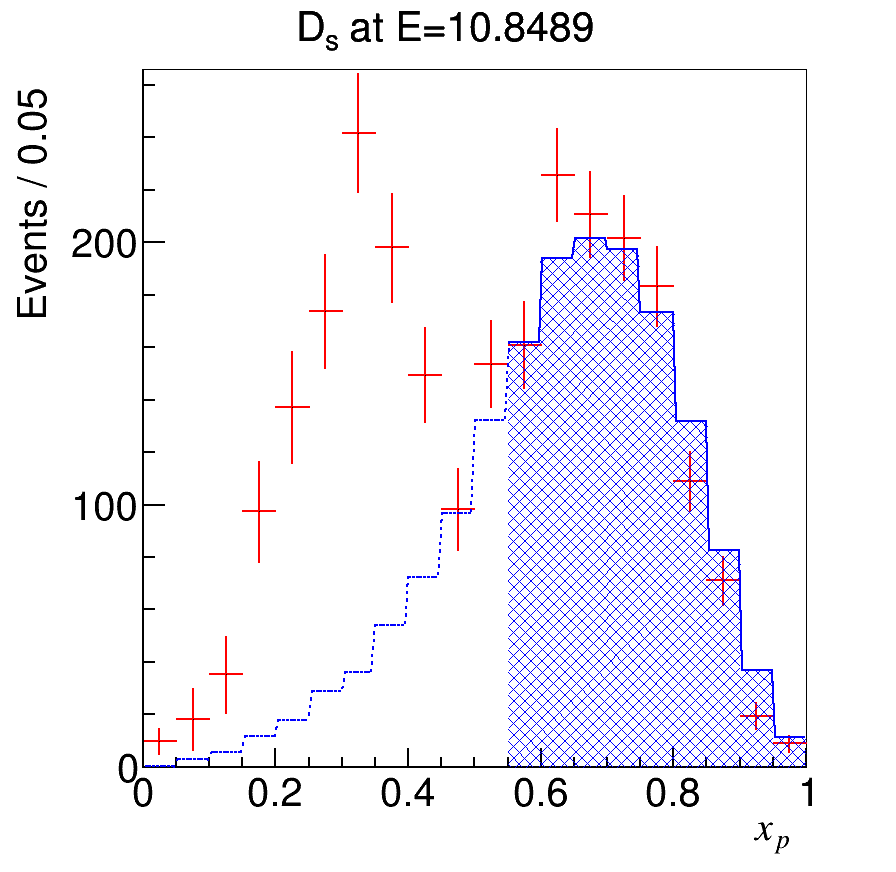} &
    \includegraphics[width=0.20\linewidth]{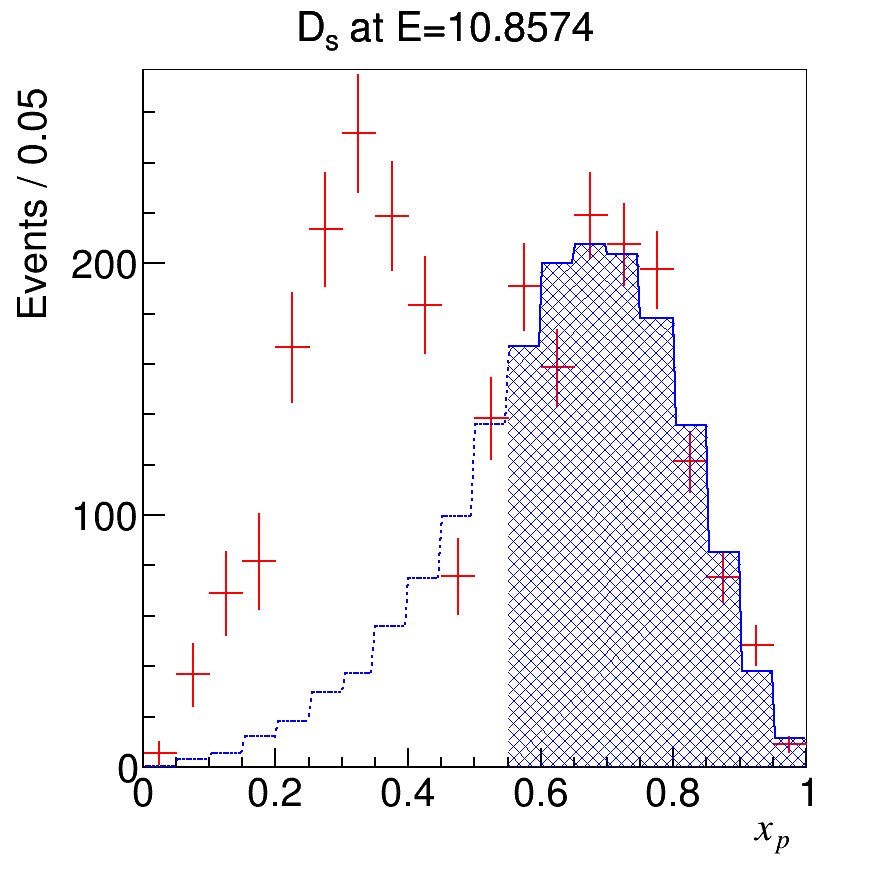} &
    \includegraphics[width=0.20\linewidth]{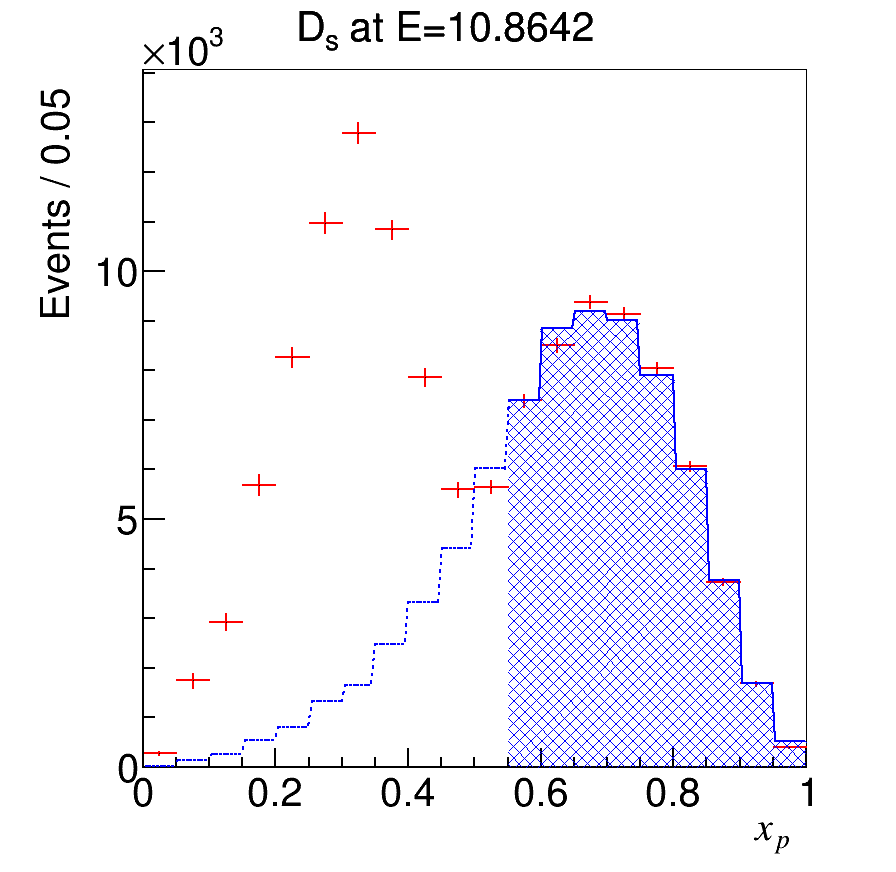} &
    \includegraphics[width=0.20\linewidth]{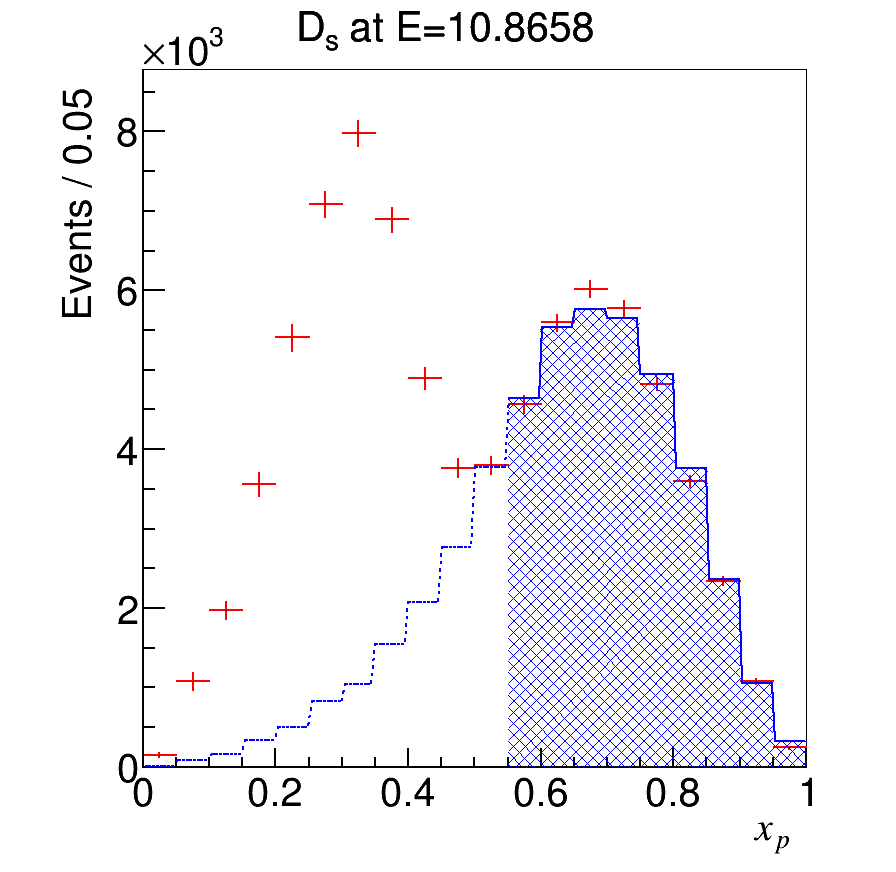} &
    \includegraphics[width=0.20\linewidth]{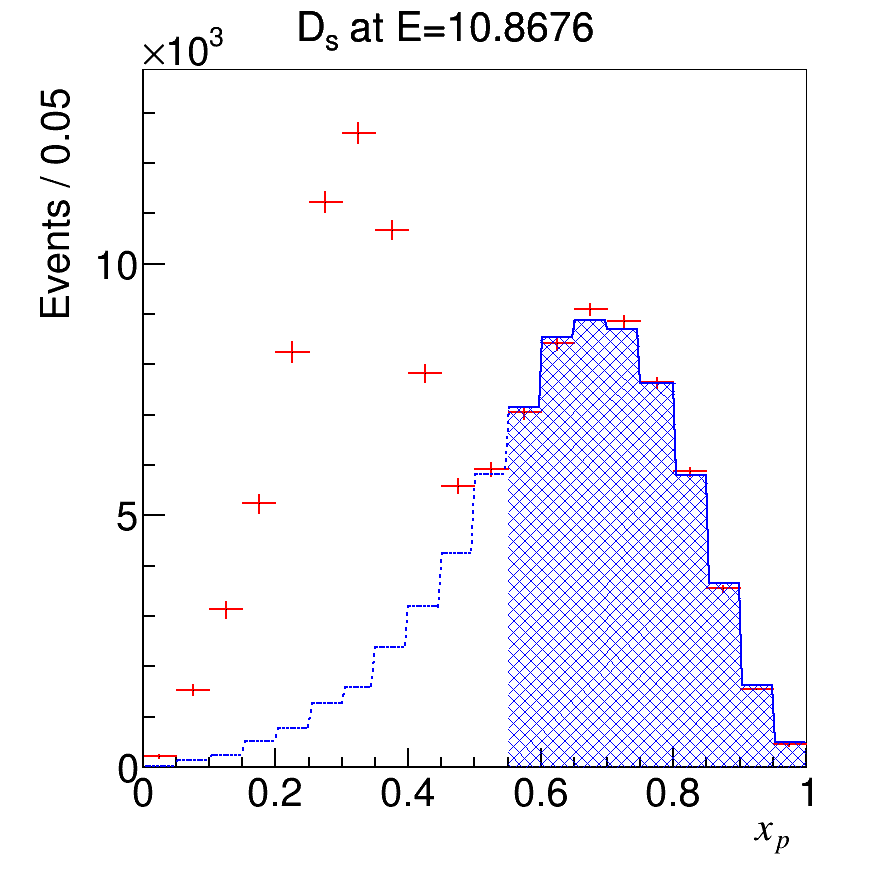} \\
    \includegraphics[width=0.20\linewidth]{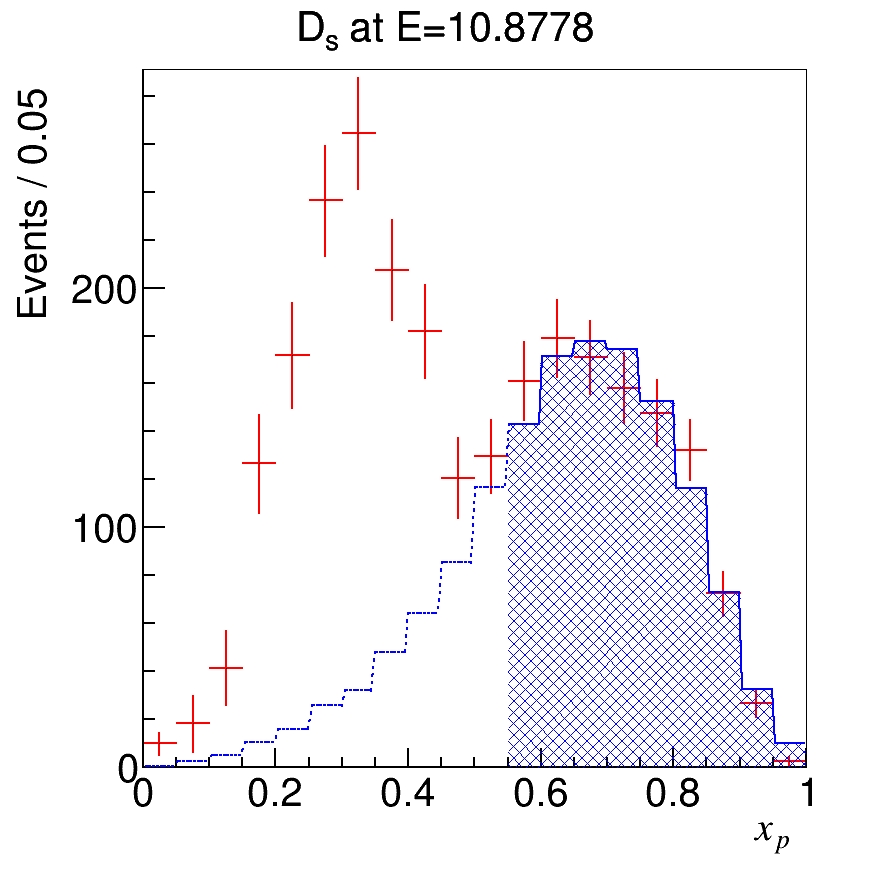} &
    \includegraphics[width=0.20\linewidth]{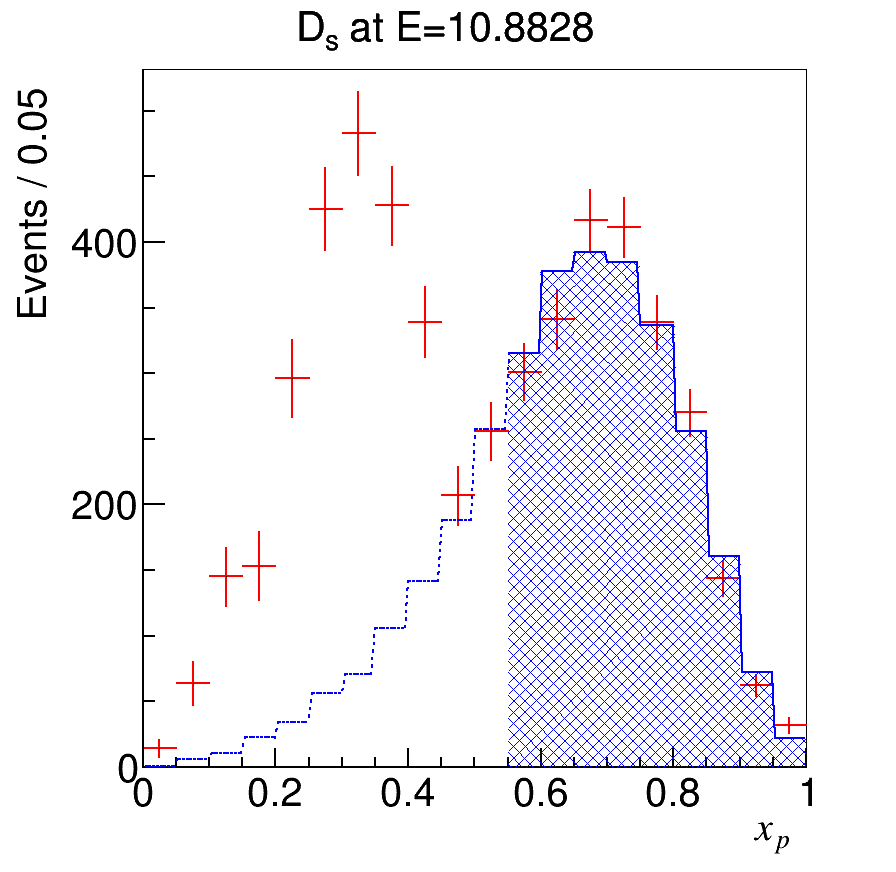} &
    \includegraphics[width=0.20\linewidth]{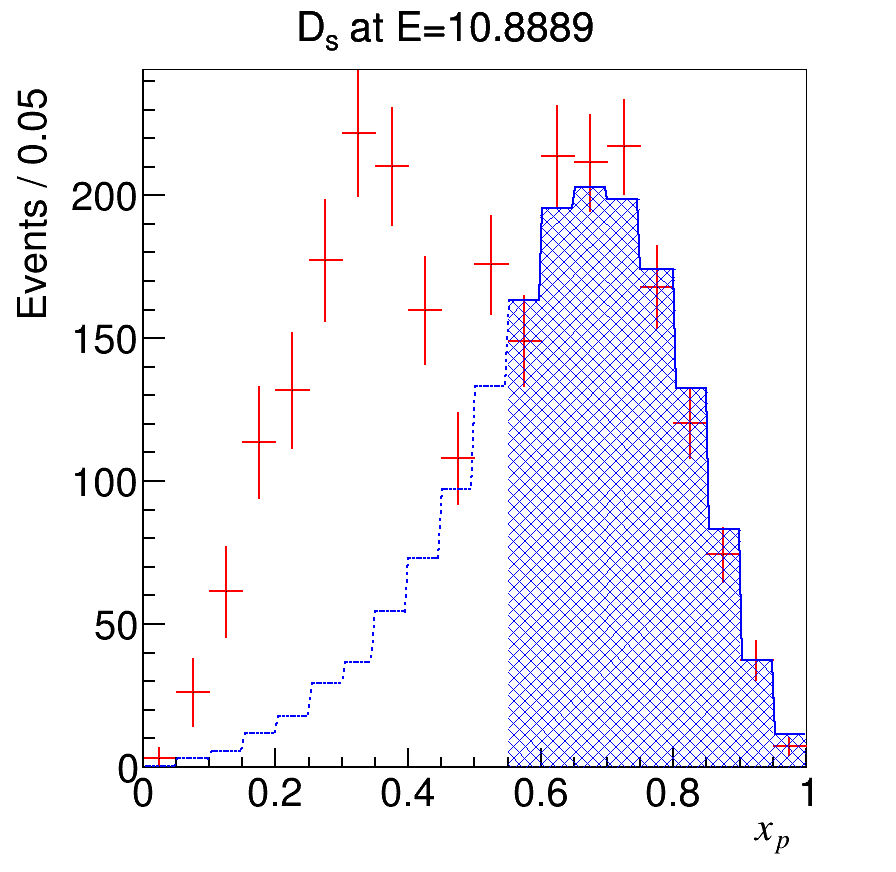} &
    \includegraphics[width=0.20\linewidth]{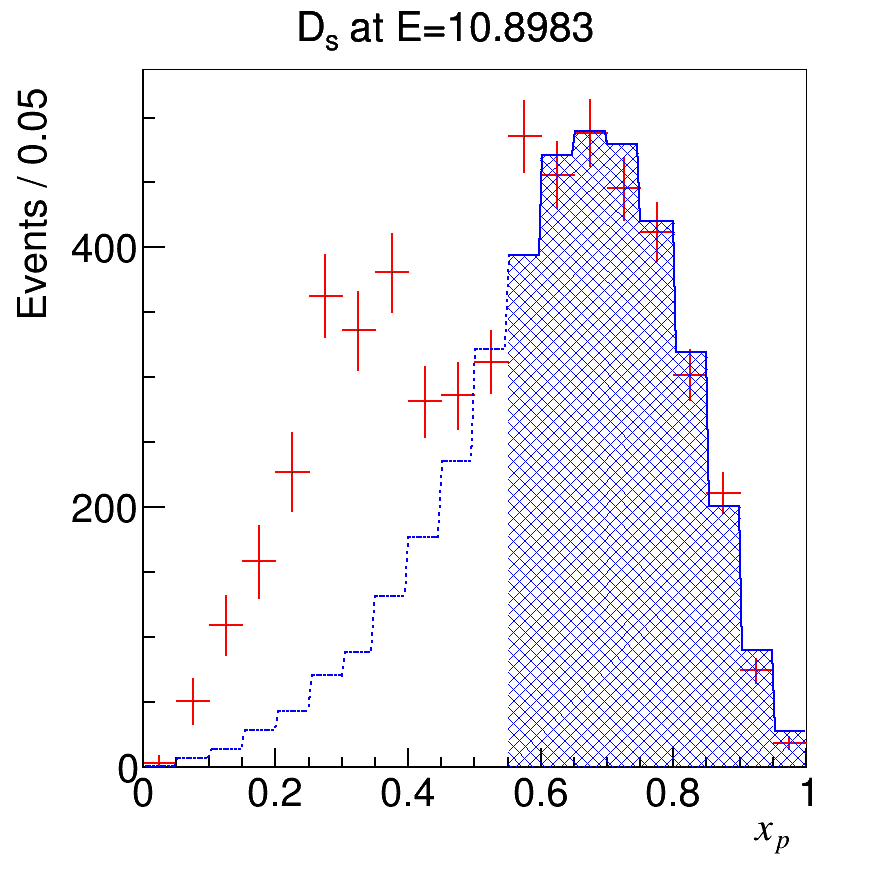} &
    \includegraphics[width=0.20\linewidth]{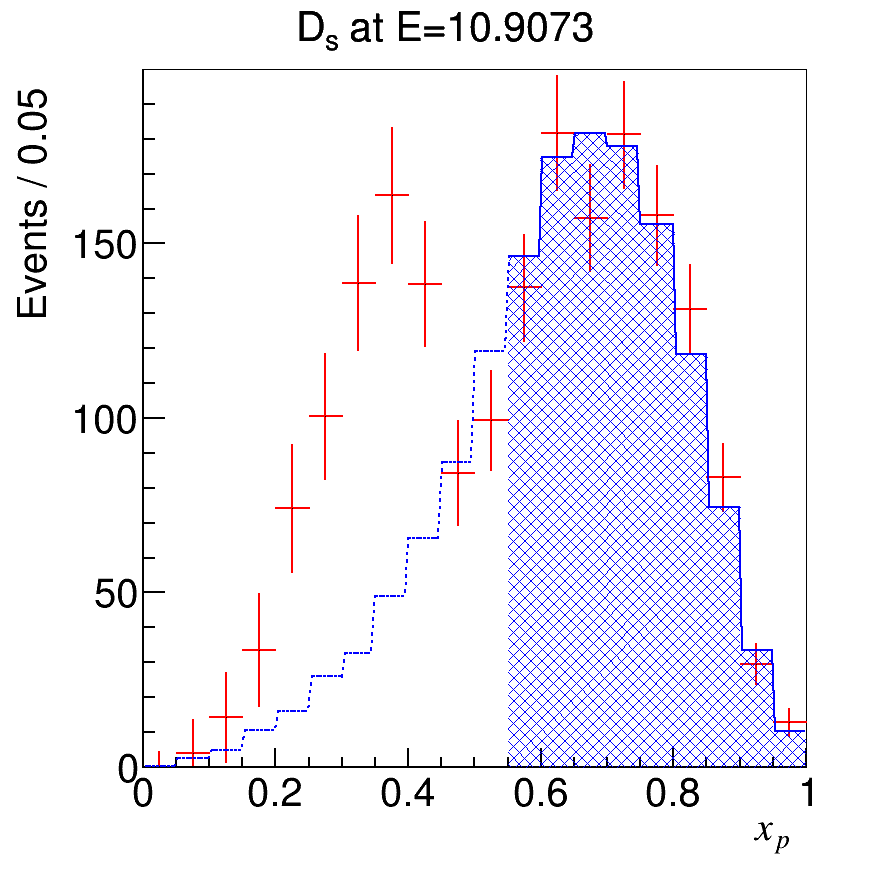} \\
    \includegraphics[width=0.20\linewidth]{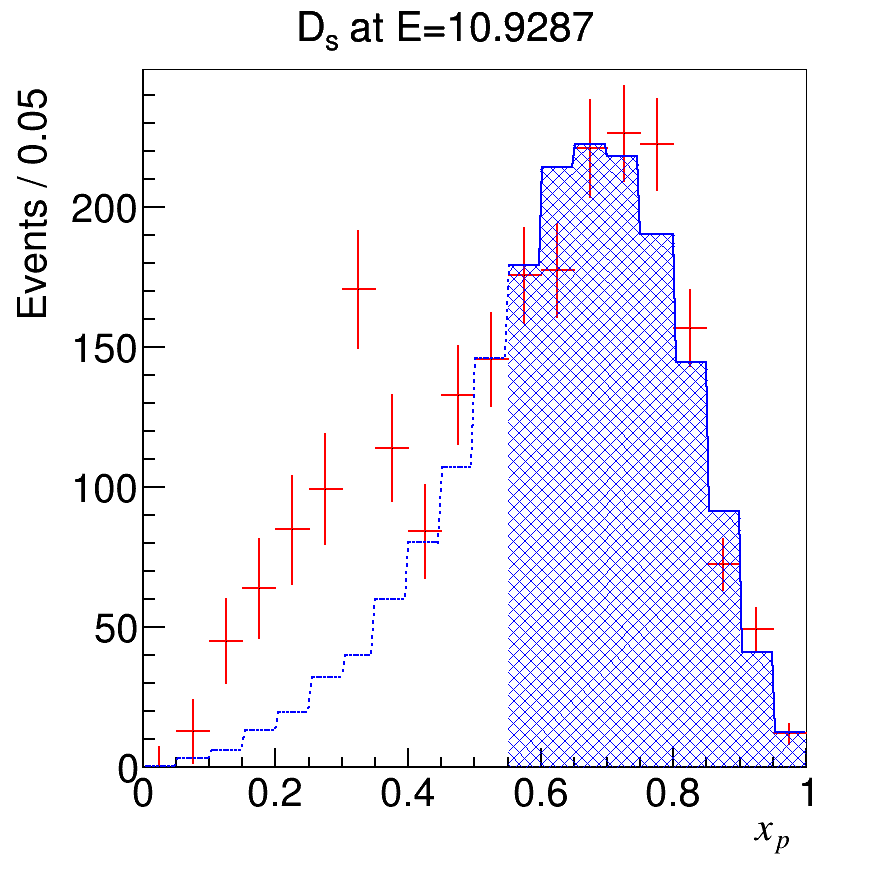} &
    \includegraphics[width=0.20\linewidth]{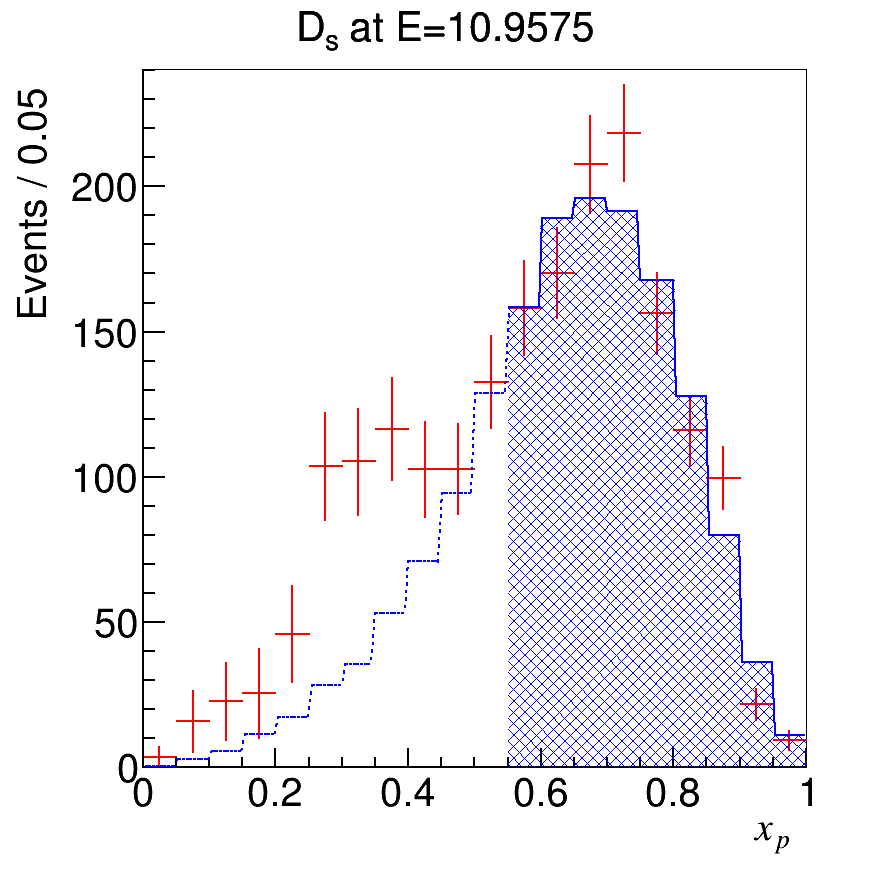} &
    \includegraphics[width=0.20\linewidth]{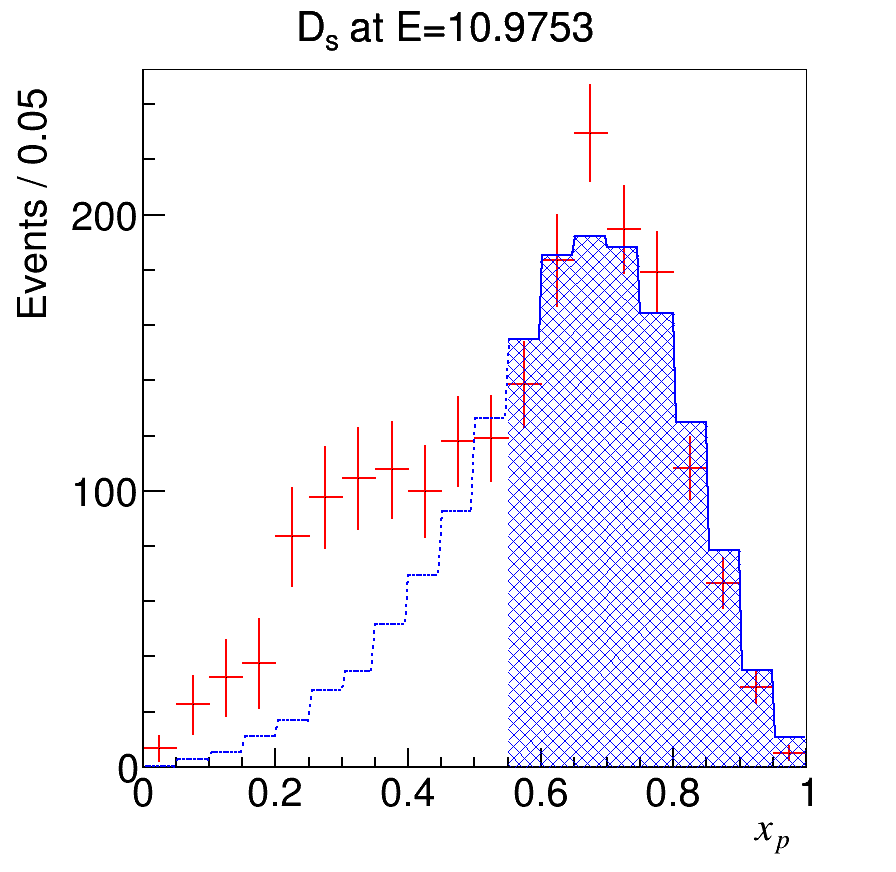} &
    \includegraphics[width=0.20\linewidth]{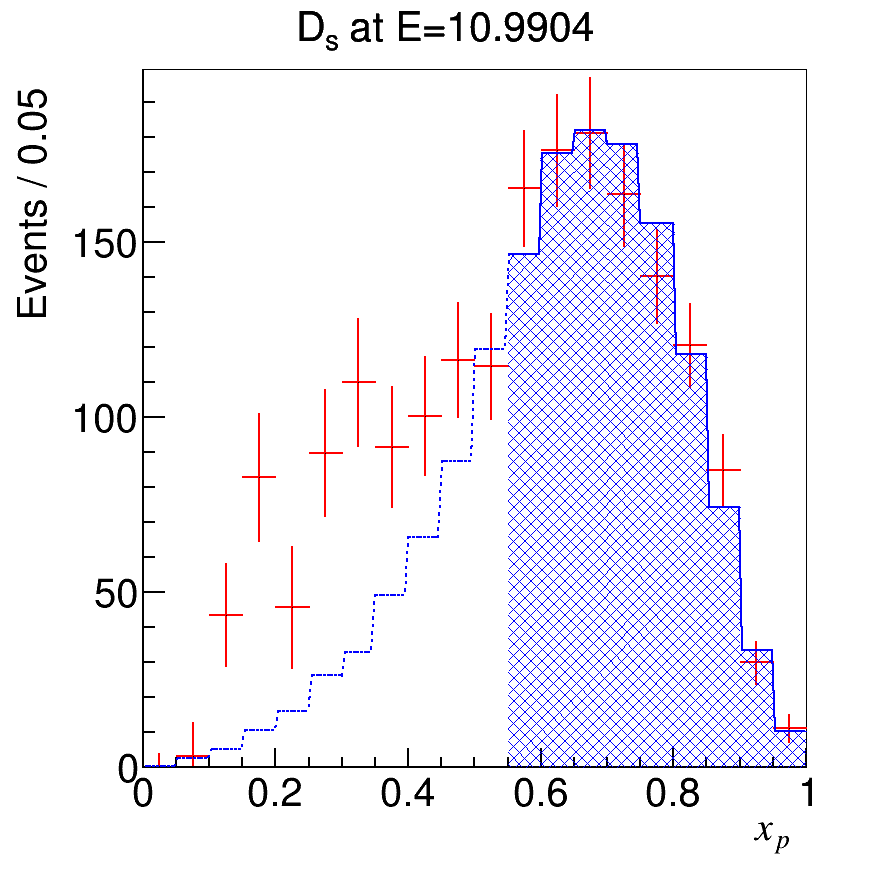} &
    \includegraphics[width=0.20\linewidth]{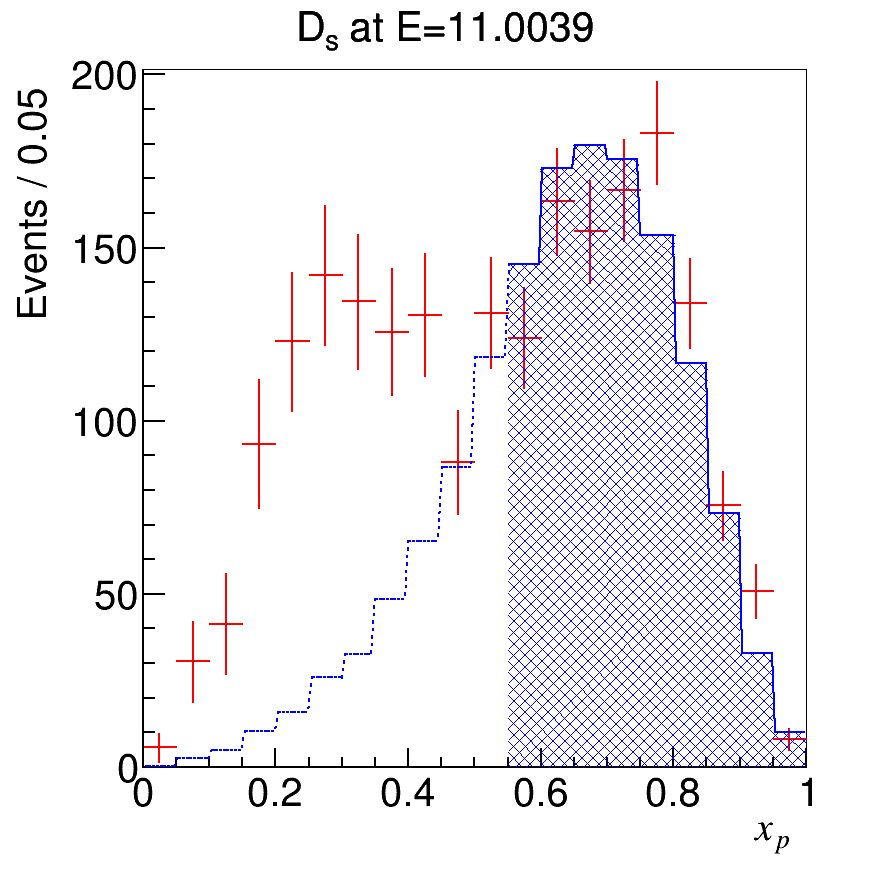} \\
    \includegraphics[width=0.20\linewidth]{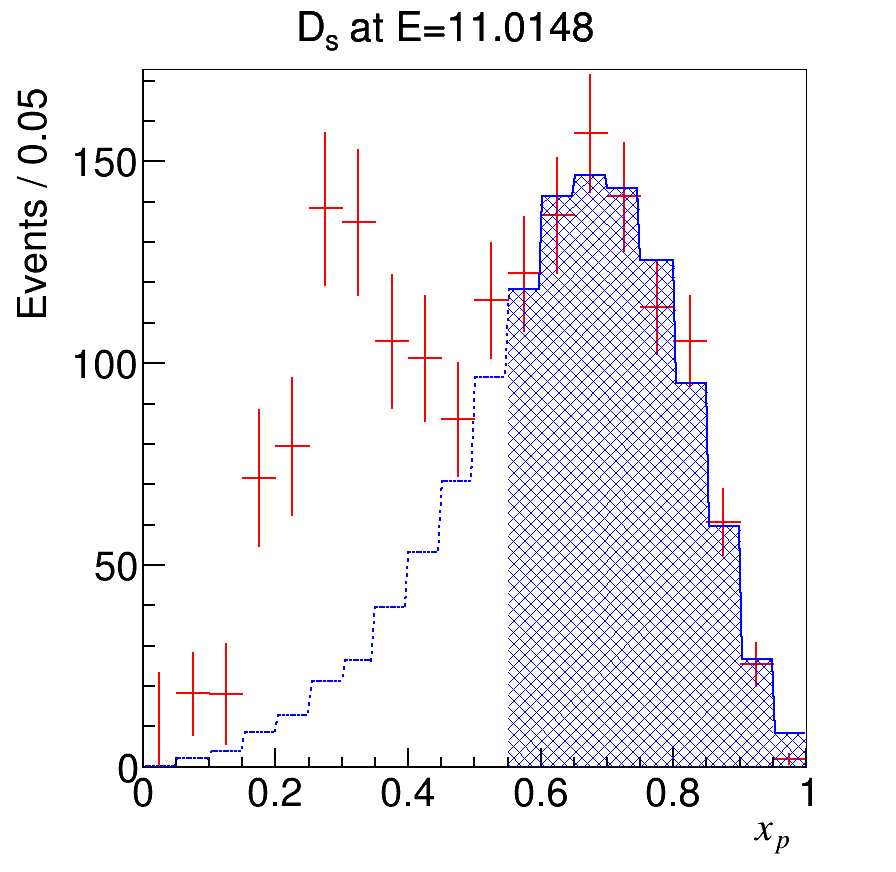} &
    \includegraphics[width=0.20\linewidth]{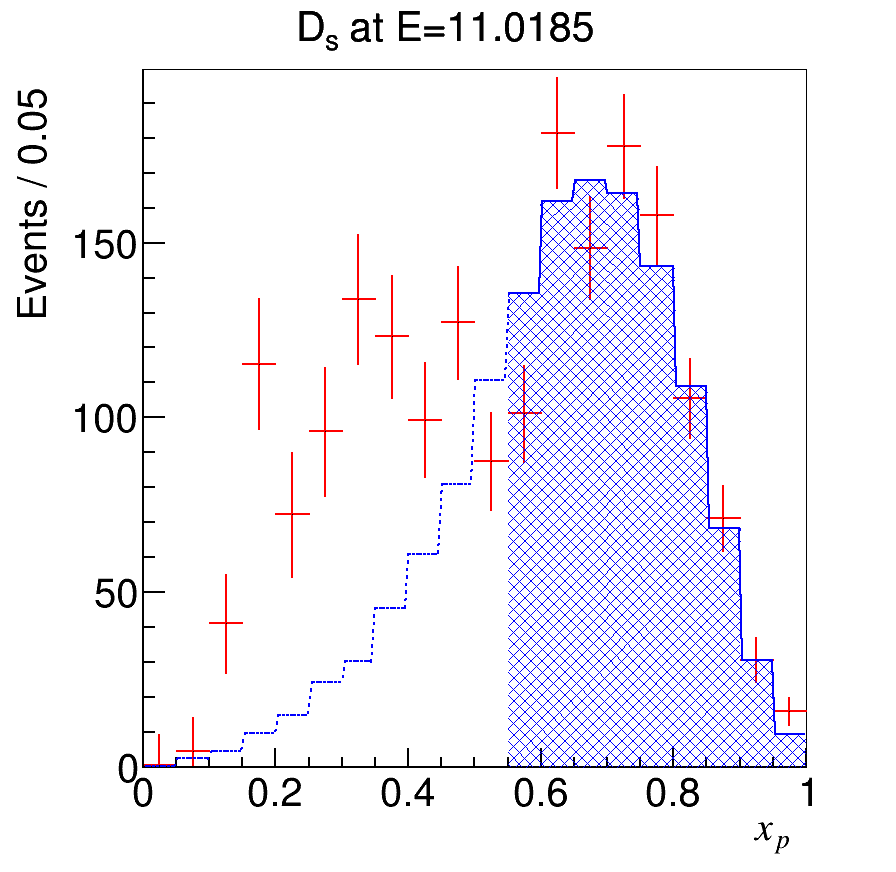} &
    \includegraphics[width=0.20\linewidth]{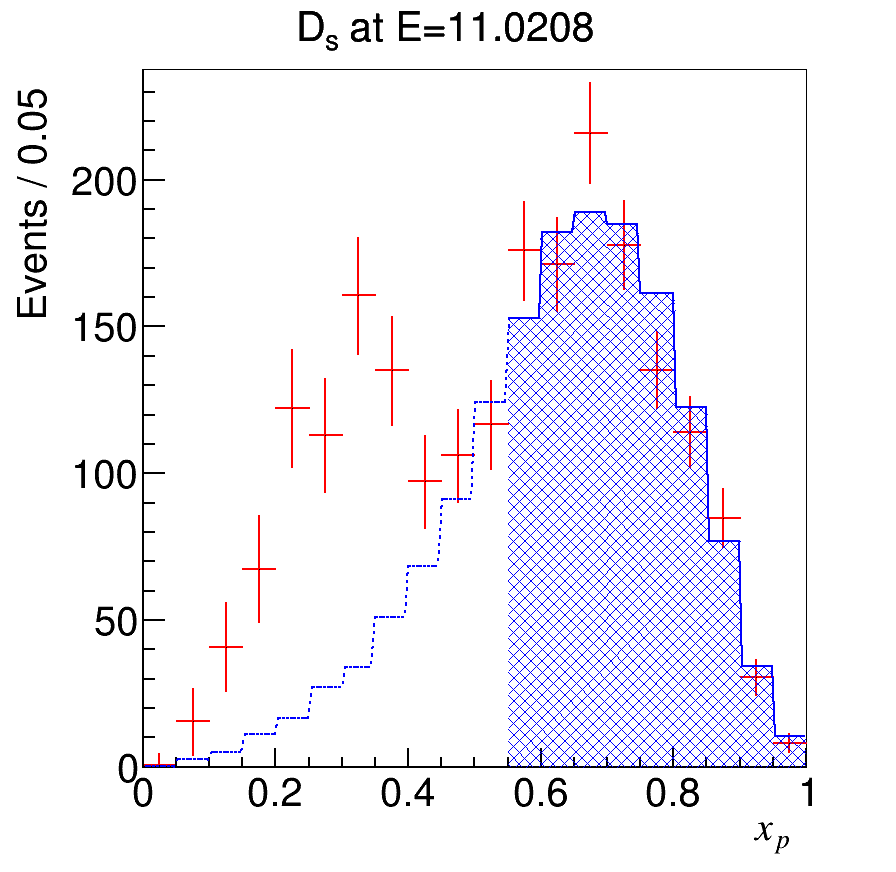} &
  \end{tabular}
  \end{adjustbox}
  \caption{The yield of \Ds\ in bins of \xp\ for the scan energies.
    Points with error bars show the data, solid hatched histograms
    show the fit results, and open dashed histograms show the
    extrapolation of the continuum component into the $b\bar{b}$
    signal region. The energy increases from left to right and from
    top to bottom. }
 \label{fig::xp_fit_for_scan_ds}
\end{figure}
\clearpage
\begin{figure}[h!]
  \begin{adjustbox}{max width=\textwidth}
 \centering
 \begin{tabular}{ccccc}
   \includegraphics[width=0.20\linewidth]{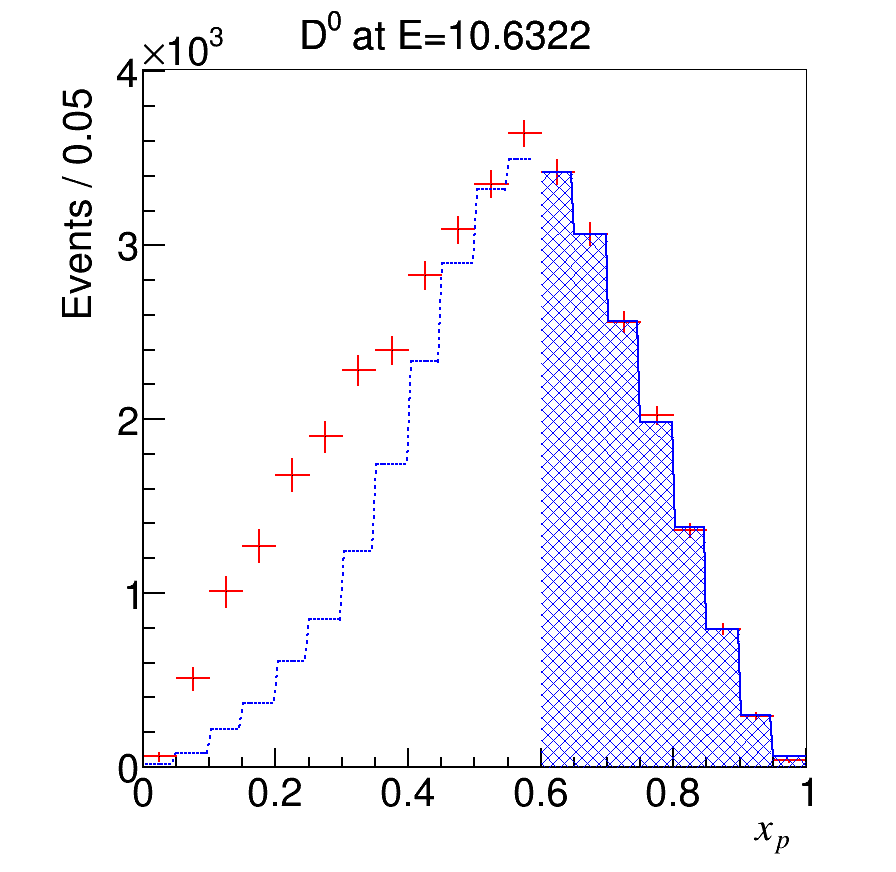} &
   \includegraphics[width=0.20\linewidth]{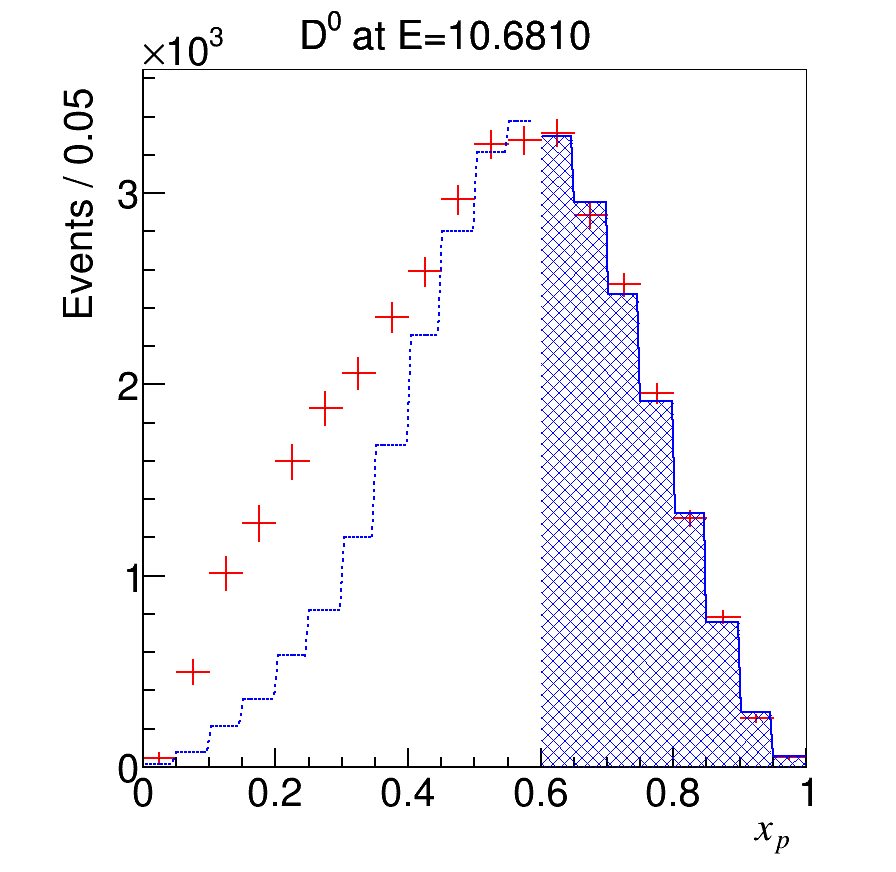} &
   \includegraphics[width=0.20\linewidth]{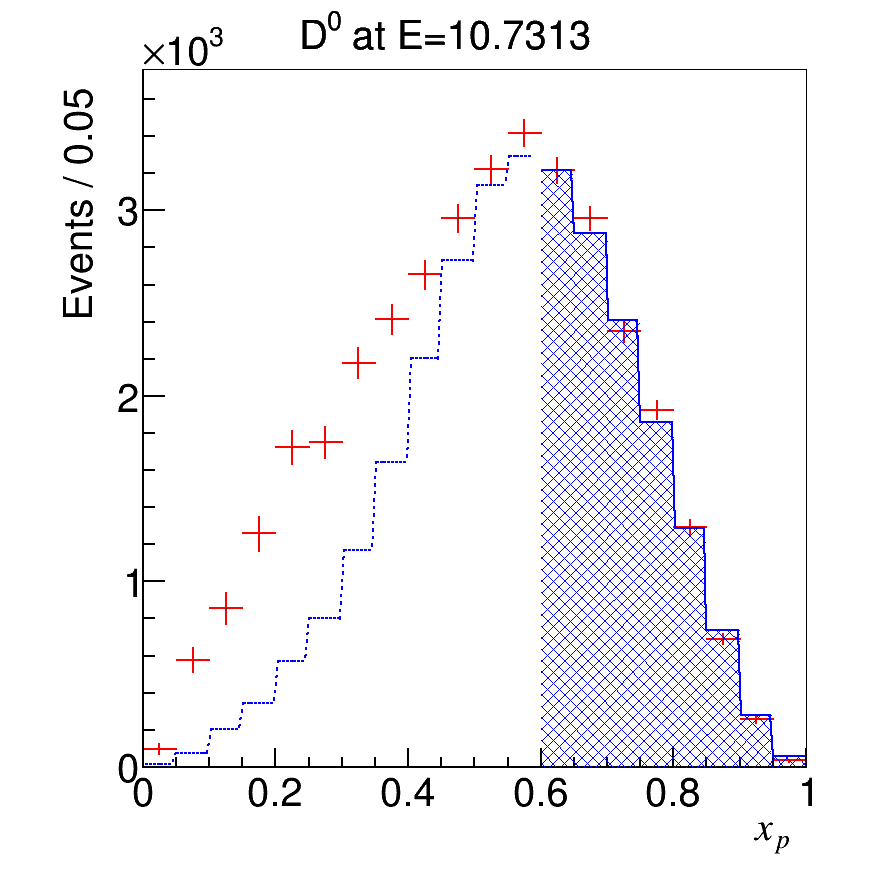} &
   \includegraphics[width=0.20\linewidth]{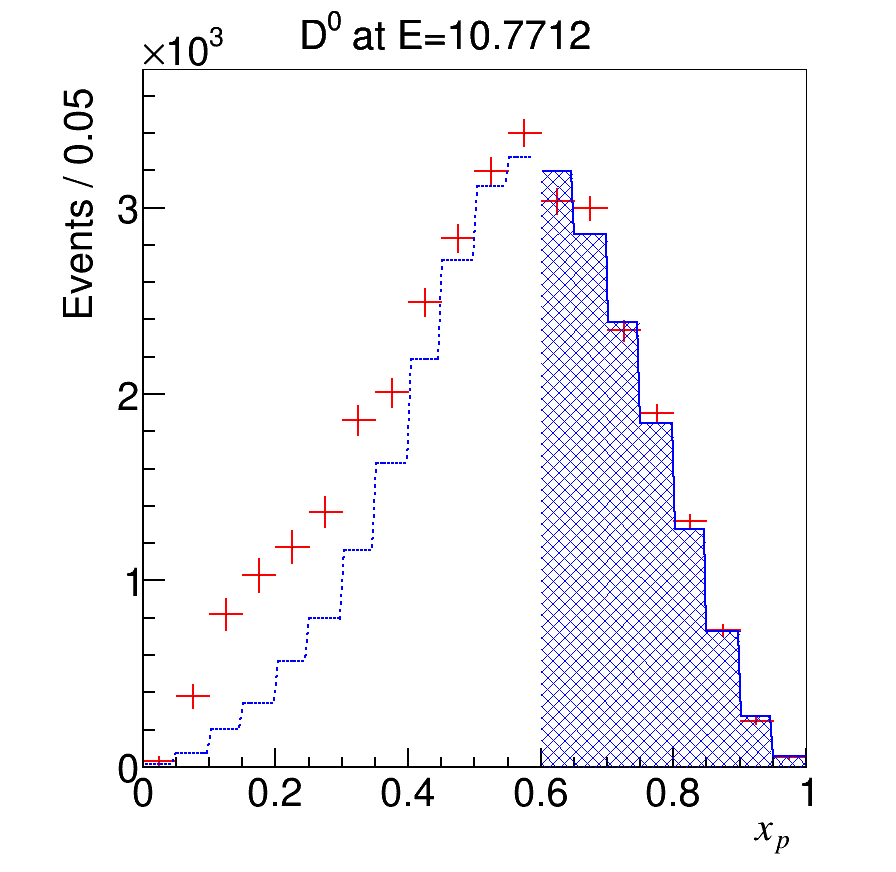} &
   \includegraphics[width=0.20\linewidth]{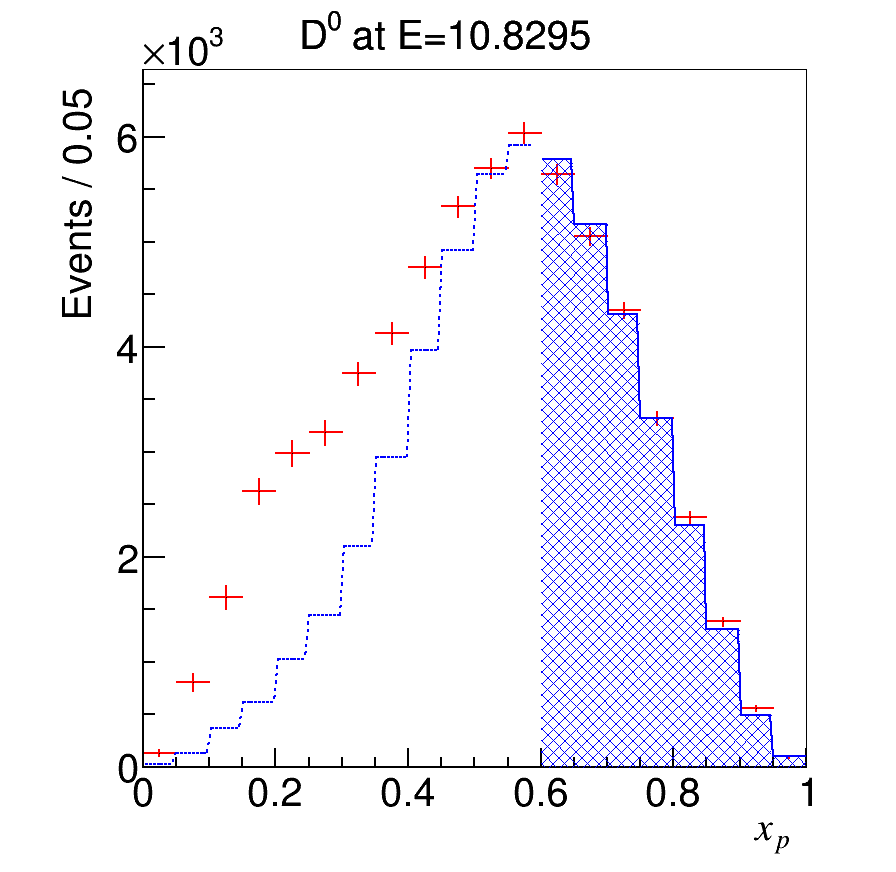} \\
   \includegraphics[width=0.20\linewidth]{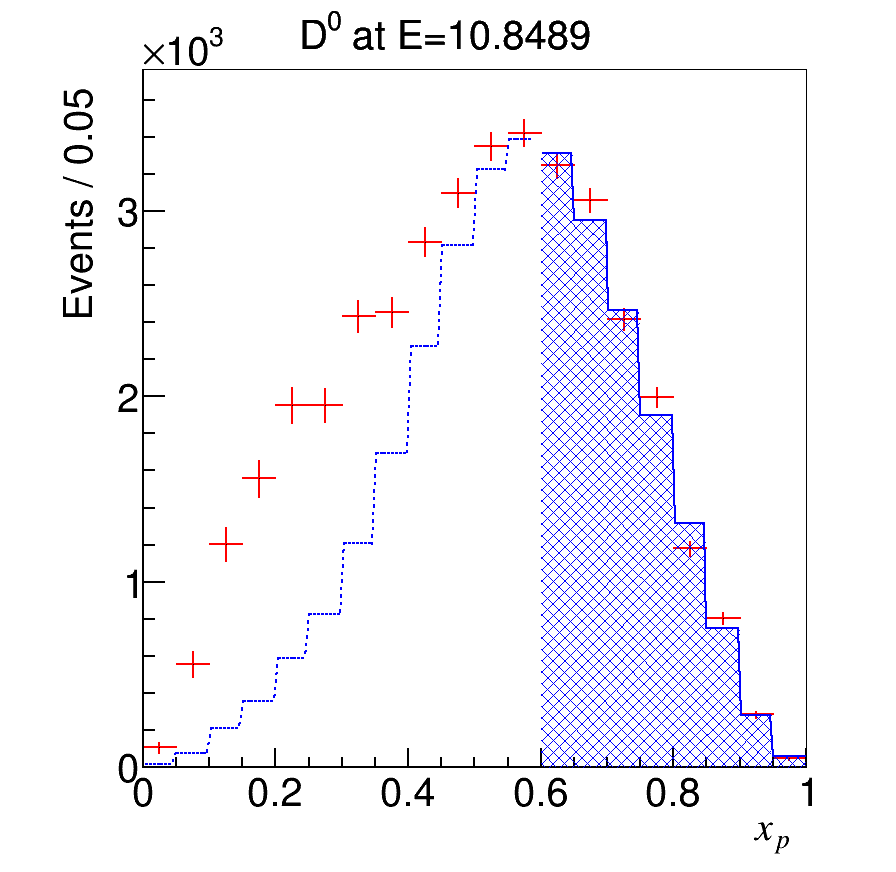} &
   \includegraphics[width=0.20\linewidth]{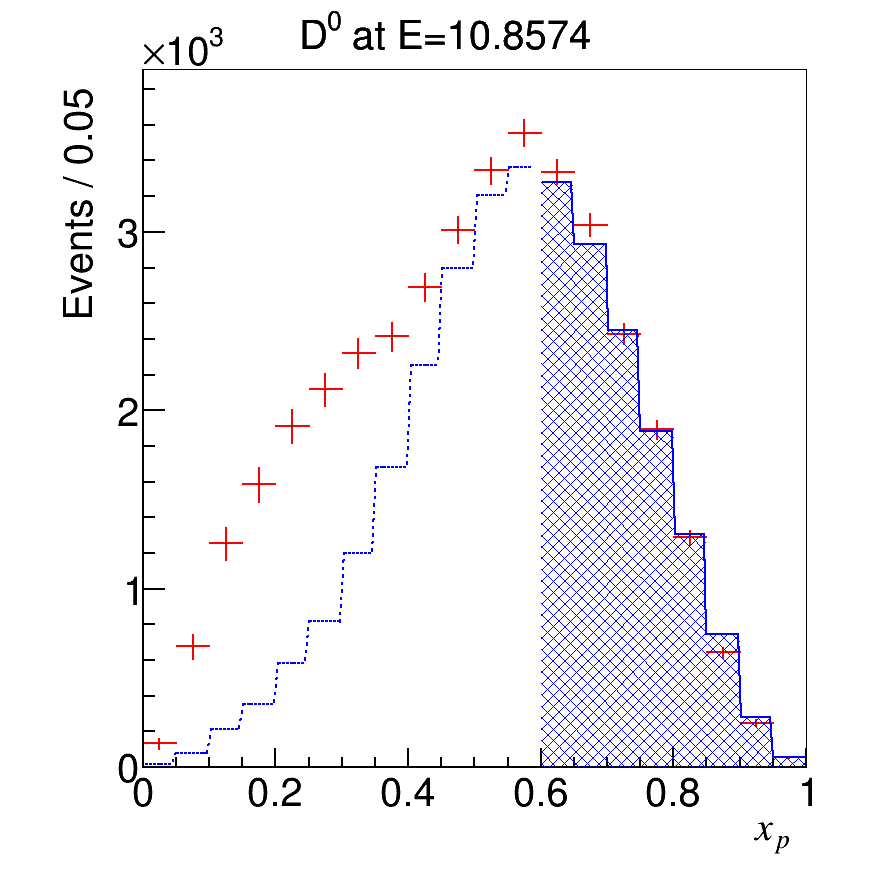} &
   \includegraphics[width=0.20\linewidth]{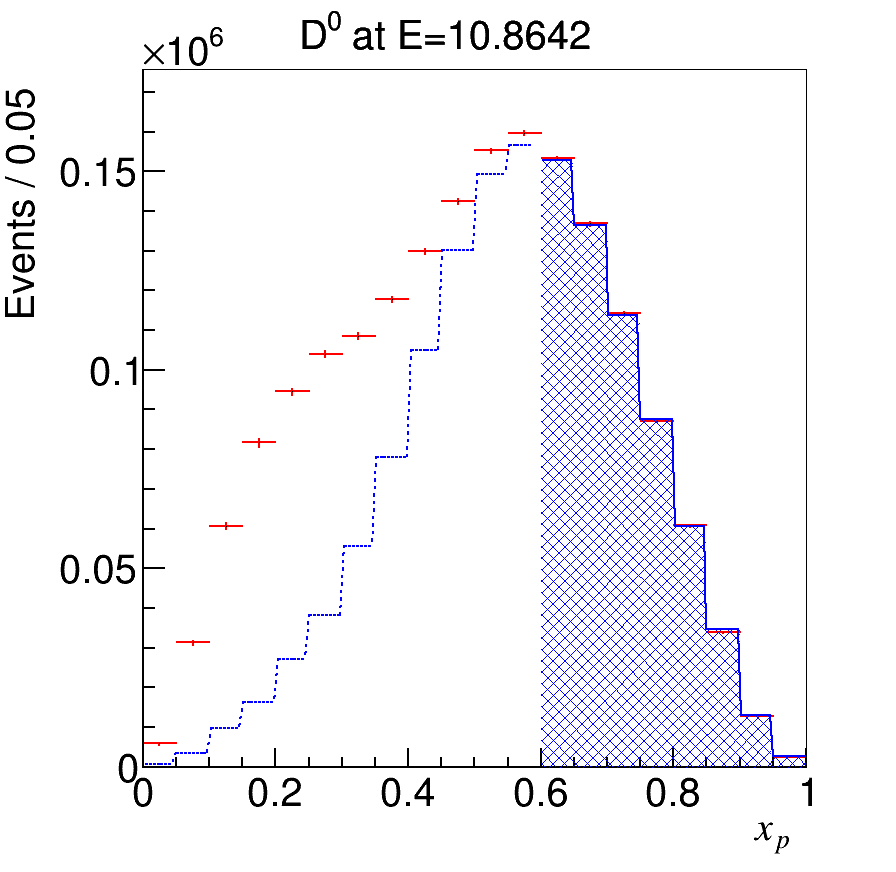} &
   \includegraphics[width=0.20\linewidth]{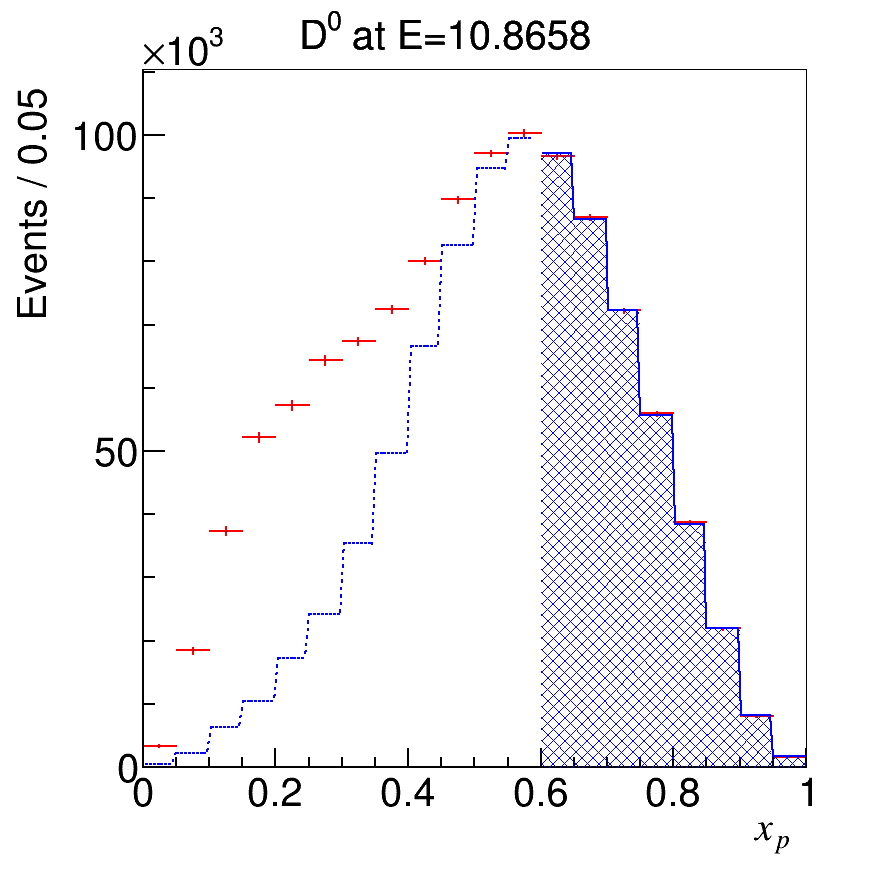} &
   \includegraphics[width=0.20\linewidth]{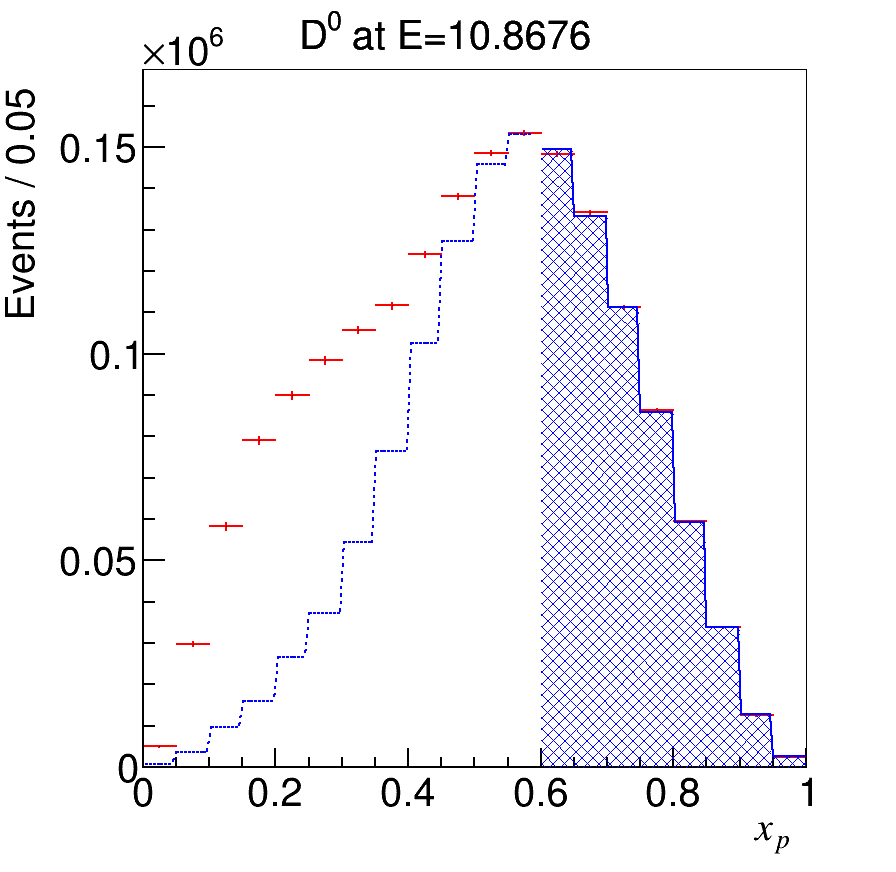} \\
   \includegraphics[width=0.20\linewidth]{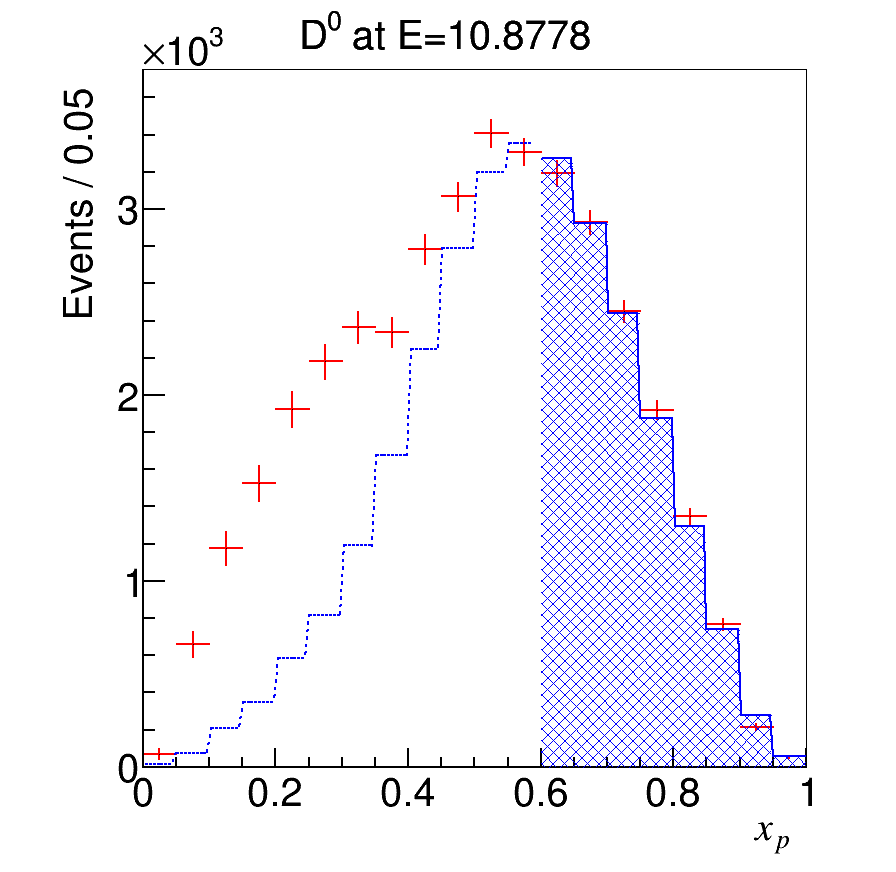} &
   \includegraphics[width=0.20\linewidth]{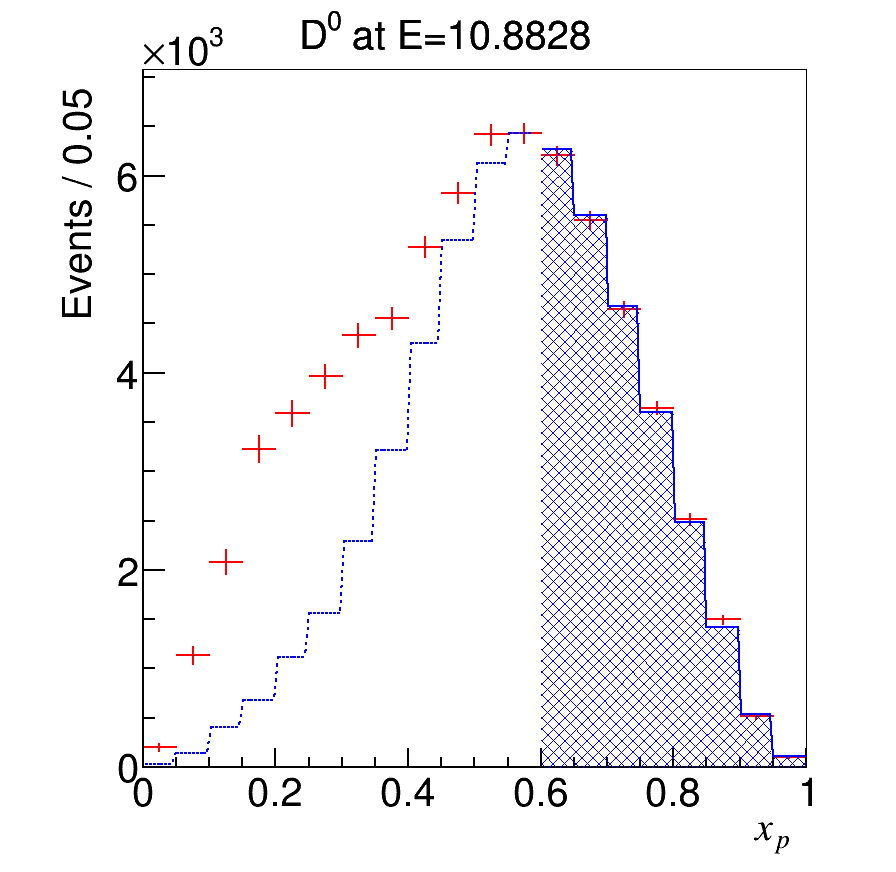} &
   \includegraphics[width=0.20\linewidth]{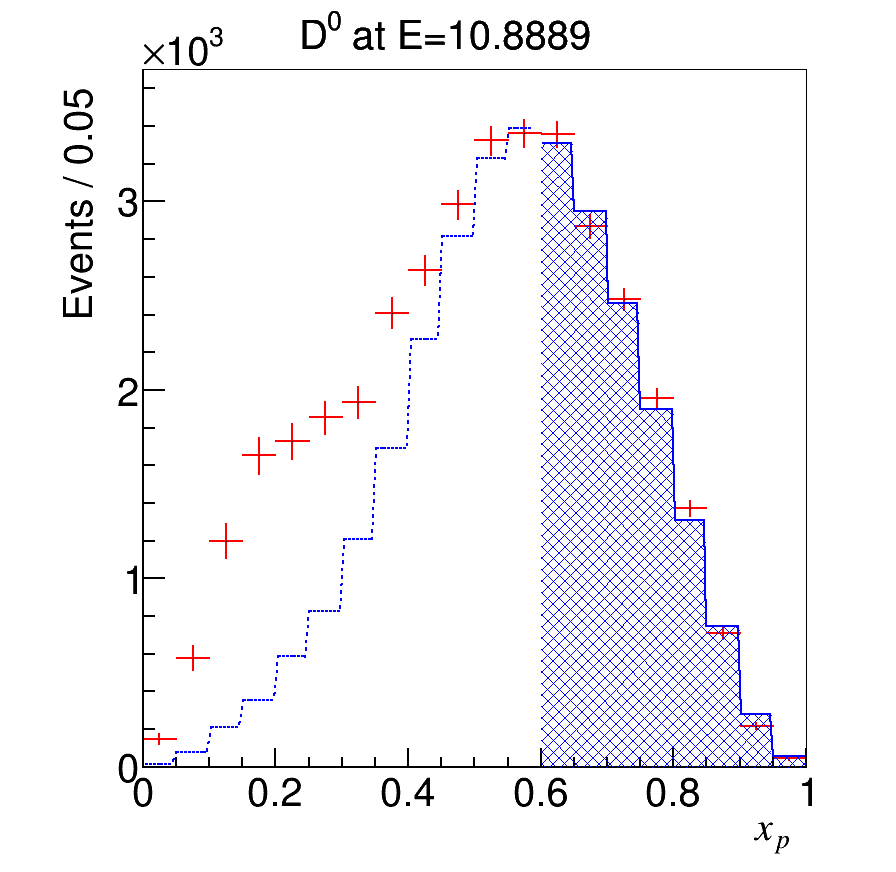} &
   \includegraphics[width=0.20\linewidth]{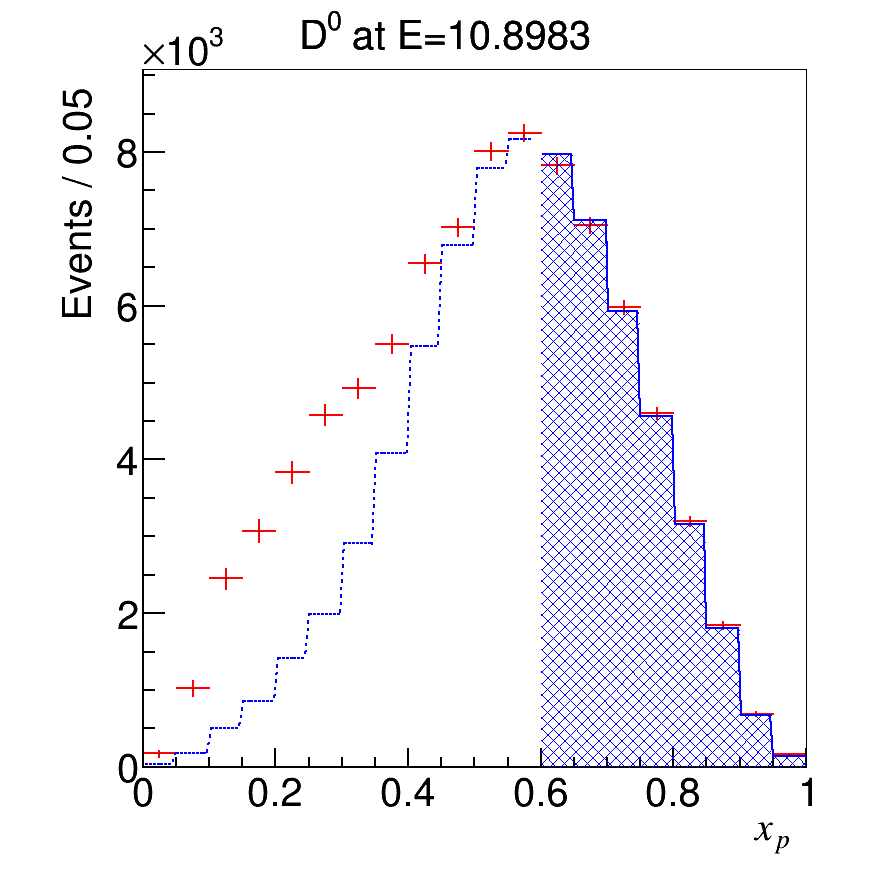} &
   \includegraphics[width=0.20\linewidth]{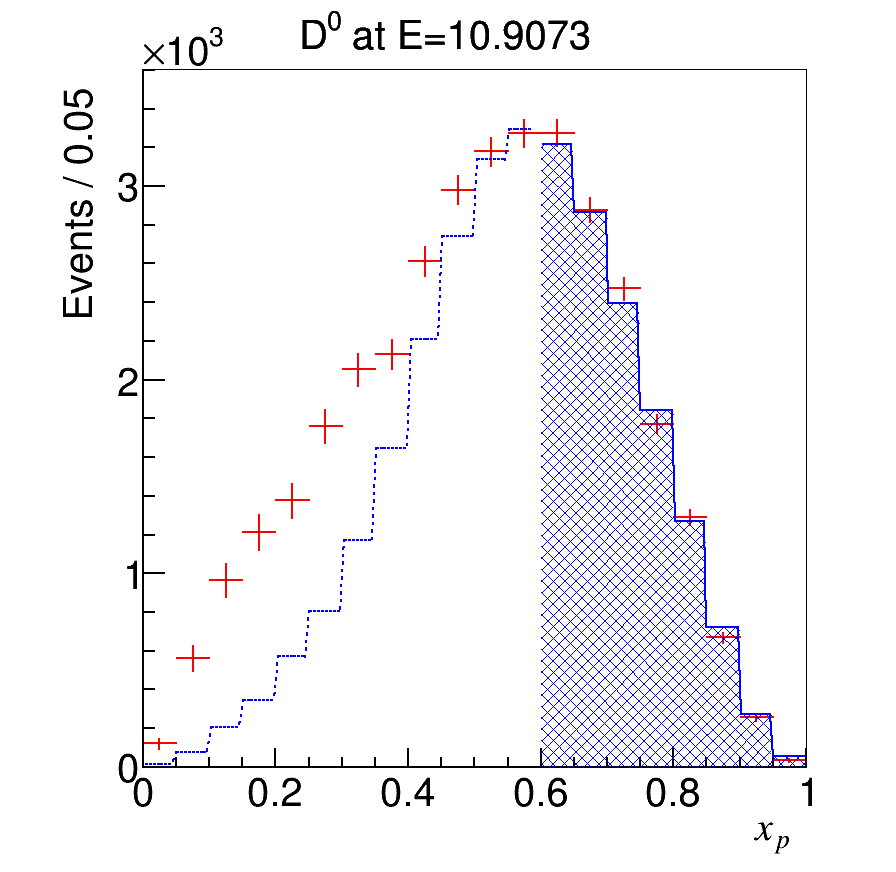} \\
   \includegraphics[width=0.20\linewidth]{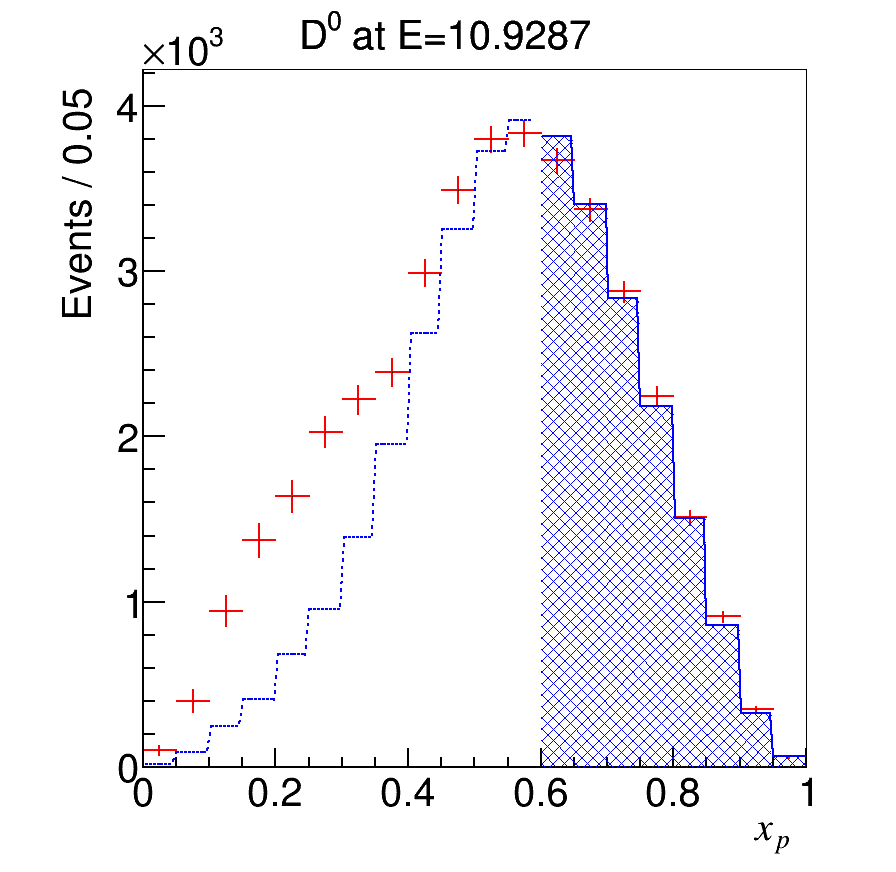} &
   \includegraphics[width=0.20\linewidth]{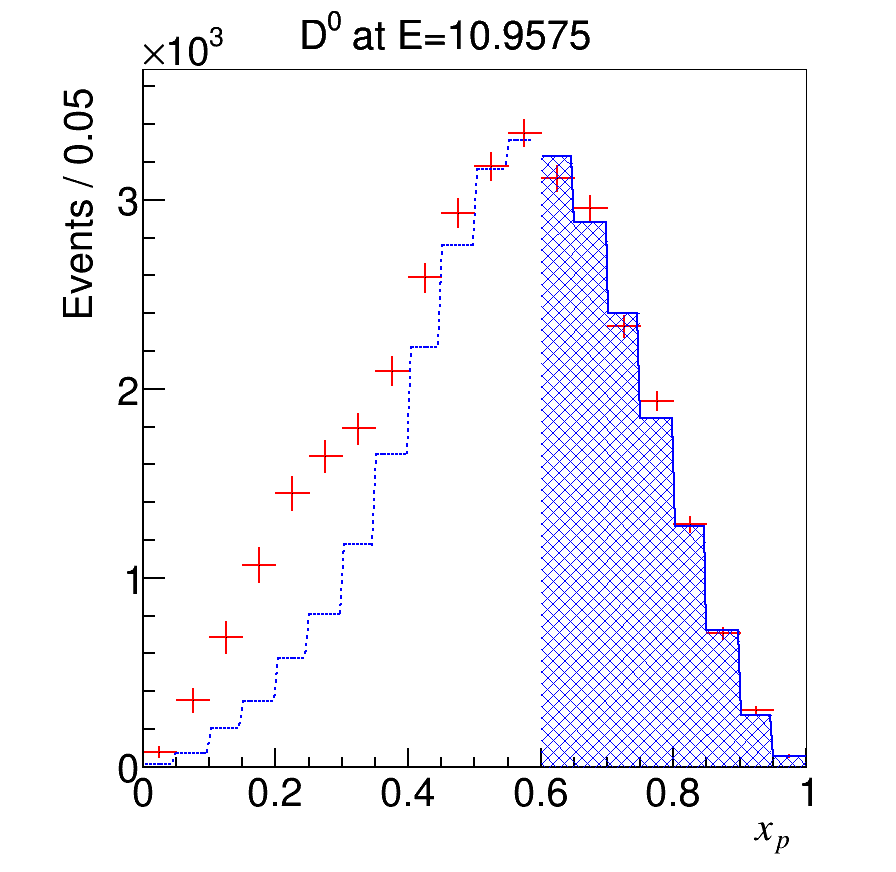} &
   \includegraphics[width=0.20\linewidth]{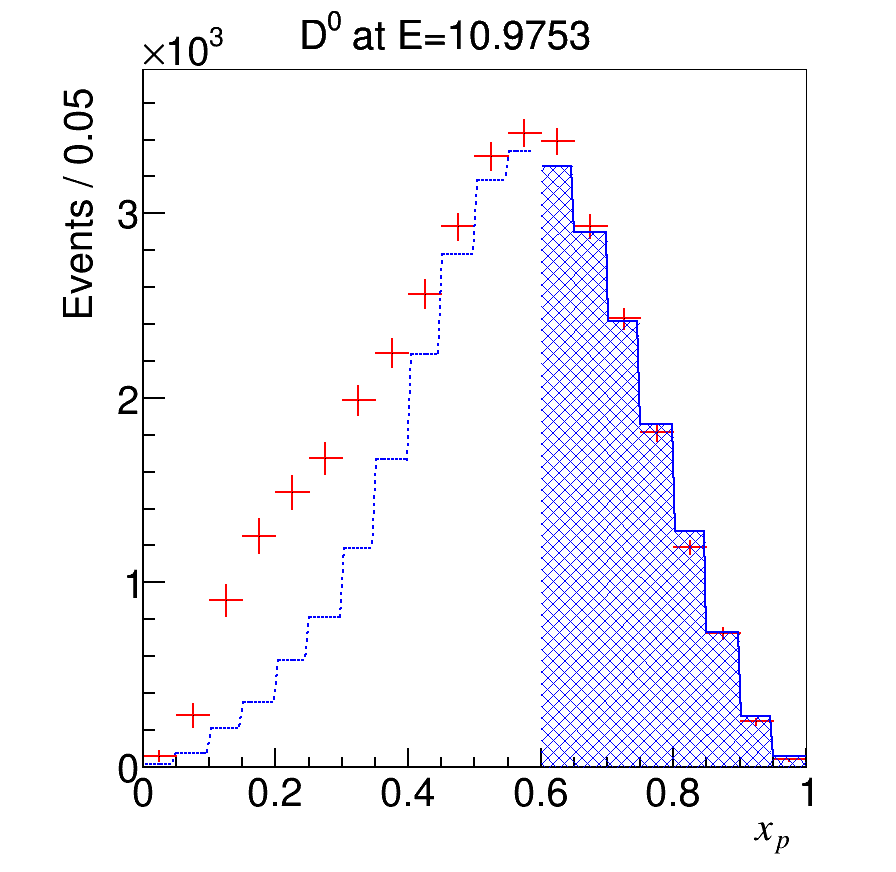} &
   \includegraphics[width=0.20\linewidth]{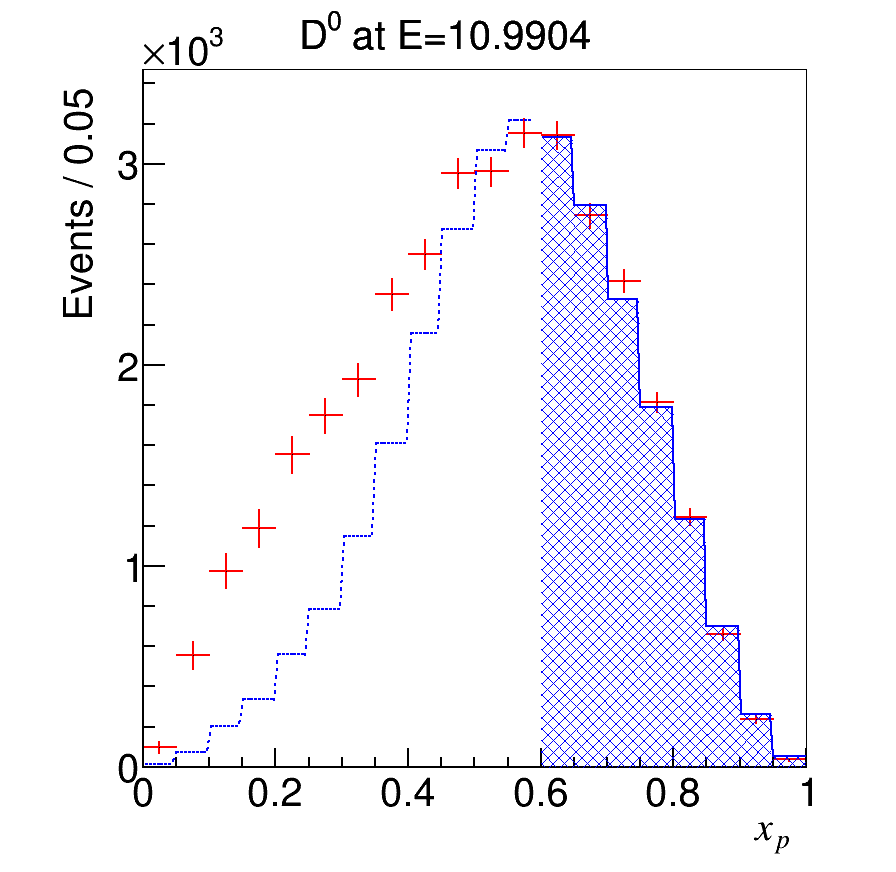} &
   \includegraphics[width=0.20\linewidth]{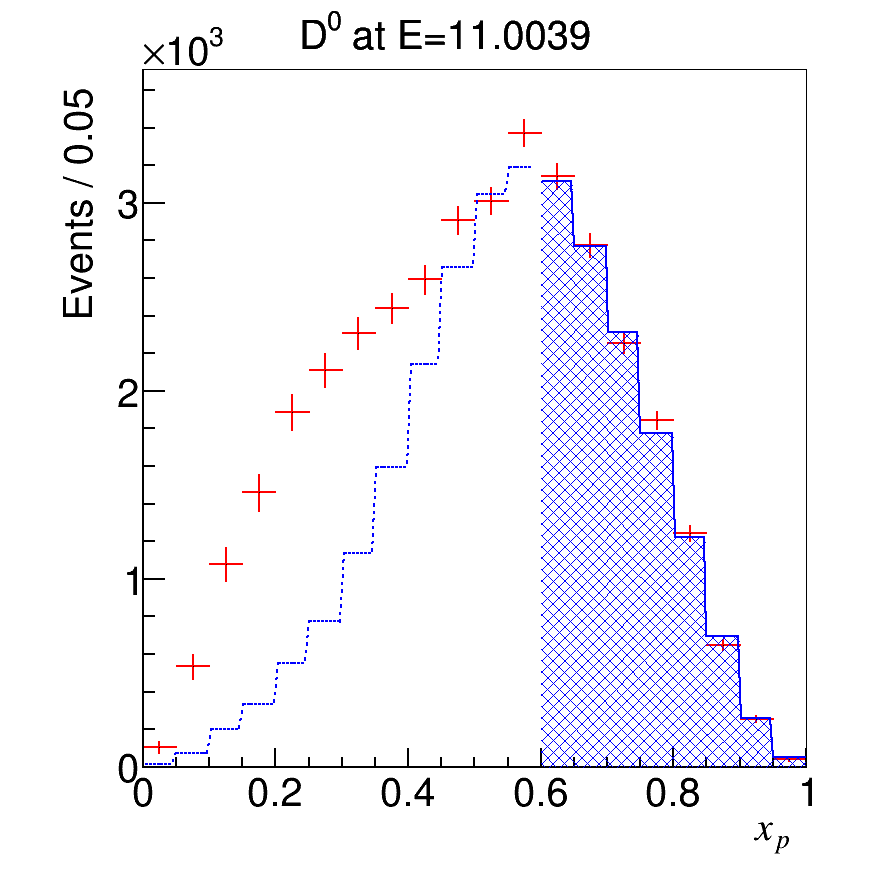} \\
   \includegraphics[width=0.20\linewidth]{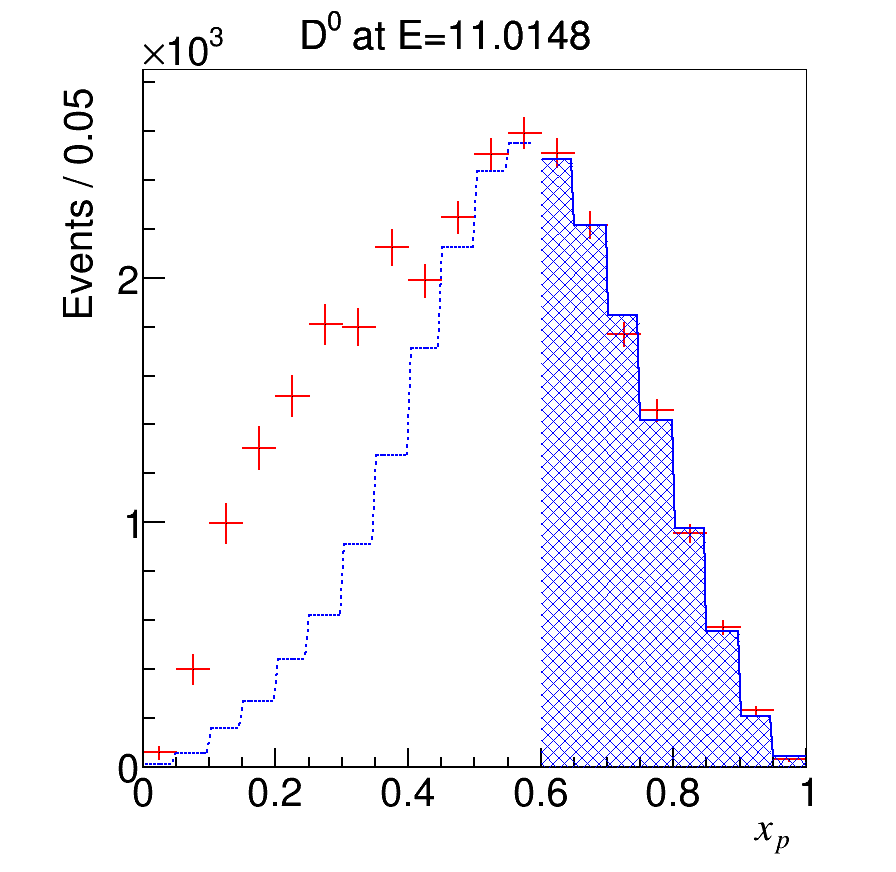} &
   \includegraphics[width=0.20\linewidth]{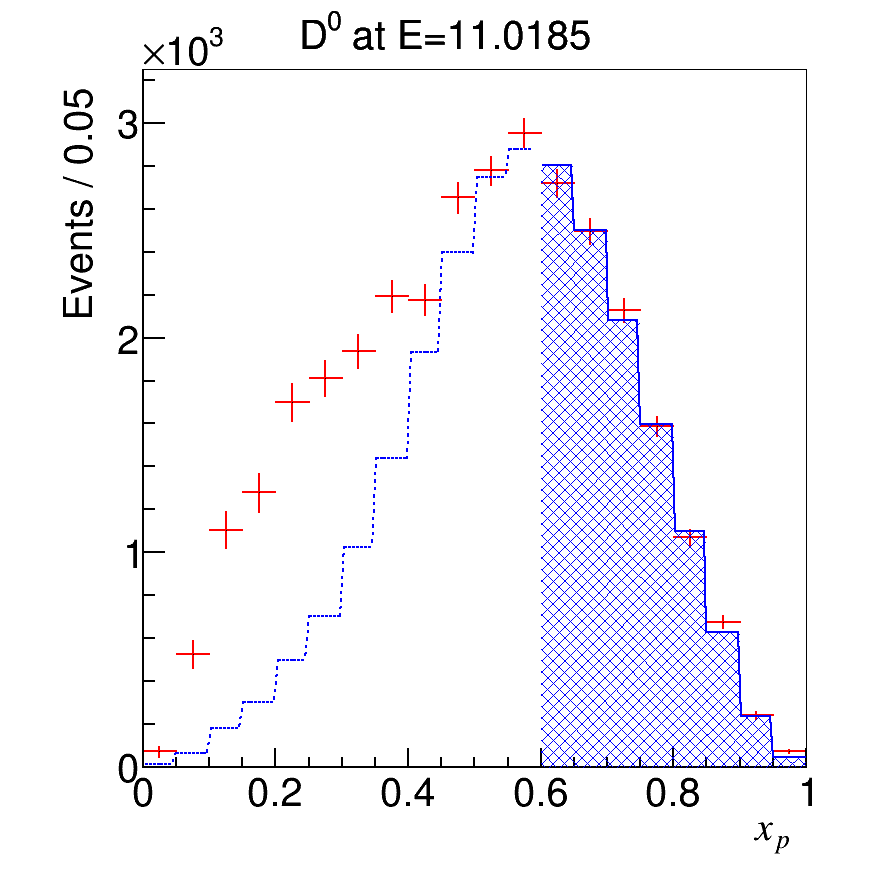} &
   \includegraphics[width=0.20\linewidth]{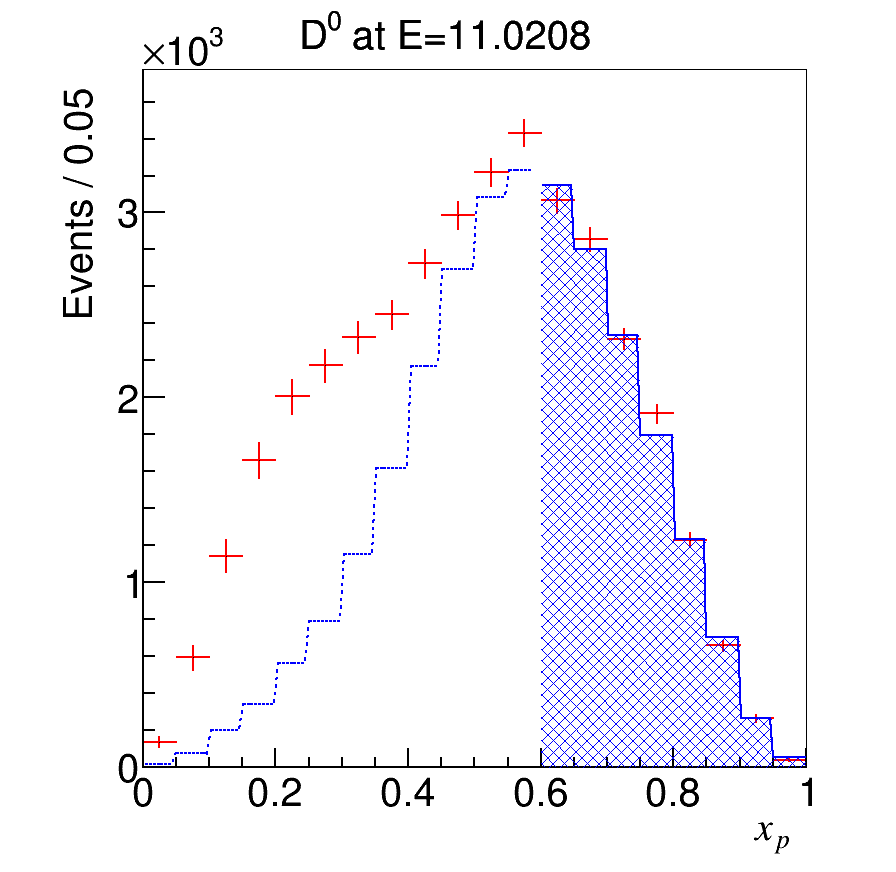} &
 \end{tabular}
 \end{adjustbox}
  \caption{The yield of \D\ in bins of \xp\ for the scan energies.
    Points with error bars show the data, solid hatched histograms
    show the fit results, and open dashed histograms show the
    extrapolation of the continuum component into the $b\bar{b}$
    signal region. The energy increases from left to
   right and from top to bottom.}
 \label{fig::xp_fit_for_scan_d0}
\end{figure}

\end{document}